\documentclass[pdftex]{aa}  
\def\hi{\relax \ifmmode {\mbox H\,{\scshape i}}\else H\,{\scshape i}\fi}
\def\hii{\relax \ifmmode {\mbox H\,{\scshape ii}}\else H\,{\scshape ii}\fi}
\def\nii{\relax \ifmmode {\mbox N\,{\scshape ii}}\else N\,{\scshape ii}\fi}
\def\oi{\relax \ifmmode {\mbox O\,{\scshape i}}\else O\,{\scshape i}\fi}
\def\oii{\relax \ifmmode {\mbox O\,{\scshape ii}}\else O\,{\scshape ii}\fi}
\def\oiii{\relax \ifmmode {\mbox O\,{\scshape iii}}\else O\,{\scshape iii}\fi}
\def\sii{\relax \ifmmode {\mbox S\,{\scshape ii}}\else S\,{\scshape ii}\fi}
\def\siii{\relax \ifmmode {\mbox S\,{\scshape iii}}\else S\,{\scshape iii}\fi}
\def\neiii{\relax \ifmmode {\mbox Ne\,{\scshape iii}}\else Ne\,{\scshape iii}\fi}
\def\ha{\relax \ifmmode {\mbox H}\alpha\else H$\alpha$\fi}                                        
\def\hb{\relax \ifmmode {\mbox H}\beta\else H$\beta$\fi}
\def\hd{\relax \ifmmode {\mbox H}\delta\else H$\delta$\fi}
\def\hg{\relax \ifmmode {\mbox H}\gamma\else H$\gamma$\fi}
\def\me{$^{-1}$}
\def\arcsec{\hbox{$^{\prime\prime}$}}

\def\deg{\hbox{$^{\circ}$}}
\usepackage{graphicx,subfigure}
\usepackage{amssymb}
\usepackage{lscape}
\usepackage{epsfig,natbib}
\bibpunct{(}{)}{;}{a}{}{,} 
\usepackage{supertabular}
\usepackage{txfonts}
\begin{document} 


\title{Central enhancement of the nitrogen-to-oxygen abundance ratio in barred galaxies}


 \author{E. Florido\inst{1,2}
         \and
         A. Zurita\inst{1,2}
          \and
         I. P\'erez\inst{1,2} 
         \and 
         E. P\'erez-Montero\inst{3}
         \and 
         P.R.T. Coelho\inst{4} 
         \and 
         D.A. Gadotti\inst{5}}

          \institute{Dpto. F\'\i sica y del Cosmos. Campus de Fuentenueva, Universidad de Granada, 18071 Granada, Spain\\
          \email{}             
          \and
          Instituto Carlos I de F\'\i sica Te\'orica y Computaci\'on, Granada, Spain  \and     
          Instituto de Astrof\'\i sica de Andaluc\'\i a, CSIC, Apartado 3004, 18080  Granada, Spain \and     
          Instituto de Astronomia, Geof\'\i sica e Ci$\hat{\rm e}$ncias Atmosf\'ericas. Universidade de S\~ao Paulo, R. do Mat\~ao, 1226, 05508-090  S\~ao Paulo, SP, Brazil    \and 
          European Southern Observatory, Casilla 19001, Santiago 19, Chile                }
   \date{}
\abstract
{Bar-induced gas inflows towards galaxy centres are recognised as a key agent for the secular evolution of galaxies. One immediate consequence of this inflow is the accumulation of gas in the centre of galaxies where it can form stars and alter the chemical and physical properties.}
{Our aim is to study whether the properties of the ionised gas in the central parts of barred galaxies are altered by the presence of a bar and whether the change in central properties is related to bar and/or parent galaxy properties.}
{We use a sample of nearby face-on disc galaxies with available SDSS spectra, morphological decomposition, and information on the stellar population of their bulges, to measure the internal Balmer extinction from the \ha\ to \hb\ line ratio, star formation rate, and relevant line ratios to diagnose chemical abundances and gas density.} 
{The distributions of all the parameters analysed (internal Balmer extinction at \hb\ (c(\hb)), star formation rate per unit area, electron density, [\nii]$\lambda$6583/\ha\ emission-line ratio, ionisation parameter, and nitrogen-to-oxygen (N/O) abundance ratio)  are different for barred and unbarred galaxies, except for the R$_{23}$ metallicity tracer and the oxygen abundance obtained from photoionisation models. 
The median values of the distributions of these parameters point towards (marginally) larger dust content, star formation rate per unit area, electron density, and  ionisation parameter in the centres of barred galaxies than in their unbarred counterparts. The most remarkable difference between barred and unbarred galaxies appears in the [\nii]$\lambda$6583/\ha\ line ratio that is, on average, $\sim$25\% higher in barred galaxies, due to an increased N/O abundance ratio in the centres of these galaxies compared to the unbarred ones. 
We analyse these differences  as a function of galaxy morphological type (as traced by bulge-to-disc light ratios and bulge mass), total stellar mass, and bulge S\'ersic index. We observe an enhancement of the differences between central gas properties in barred and unbarred galaxies in later-type galaxies or galaxies with less massive bulges. However, the bar seems to have a lower impact on the central gas properties  for galaxies with bulges above $\sim10^{10}$~M$_\odot$ or total mass M$_\star \gtrsim10^{10.8}~$M$_\odot$.}
{We find observational evidence that the presence of a galactic bar affects the properties of the ionised gas in the central parts of disc galaxies (radii $\lesssim$~0.6--2.1~kpc). The most striking  effect is an enhancement in the N/O abundance ratio. This can be interpreted qualitatively in terms of our current knowledge of bar formation and evolution, and of chemical evolution models, as being the result of  a different star formation history in the centres of barred galaxies caused by the gas inflow induced by the bar. Our results lend support to the scenario in which less massive and more massive bulges have different origins or evolutionary processes, with the gaseous phase of the former currently having a closer relation to the bars.
}
\maketitle
%

\section{Introduction}

The distribution of gas and stars in disc galaxies has a basic axisymmetric structure, which is  driven by rotation. However,  superimposed non-axisymmetric  morphological features, such as spiral arms or bars,  may have important consequences on their evolution \citep[e.g.][]{kormendy}. This secular, internally driven, evolution is thought to be dominant in recent times, as opposed to the evolution driven by galactic hierarchical merging and external gas accretion \citep[e.g.][]{review_barnes}.
Stellar bars have received a lot of attention and are seen as important agents for the secular evolution of galaxies,  because their non-axisymmetric gravitational potential redistributes the angular momentum of the gas and stars in  galactic discs \citep[e.g.][]{SW93,athanassoula03}. Gas is dissipative and, at radii well inside  corotation, loses angular momentum as a result of the gravitational torque exerted by the bar. As a result, gas is  driven towards the galaxy centre  \citep[e.g.][]{athanassoula,friedli_benz,friedli_kenni}. 
Analytical and numerical simulations also show that the effectiveness with which a bar influences  galaxy dynamics (bar strength) changes as the bar evolves \citep{athanassoula03} and, as a consequence, the bar-induced gas inflow rate changes accordingly \citep{ReganTeuben}.

According to simulations, bar-induced gas inflow can affect the central properties of galaxies in a number of ways:  it can increase the central gas and dust concentration and trigger star formation (SF) \citep[][]{friedli_benz,martinet},  produce the formation and growth of a disky pseudobulge  \citep[][]{kormendy, cheung}, and alter the central abundance and dilute the initial disc metallicity gradient \citep[][]{friedli_benz,friedli_kenni}. If the gas were to reach the central parsecs it could potentially  act as  fuel for a central massive black hole \citep[][]{shlosman} and thus form an active galactic nucleus or AGN \citep[e.g.][]{coelho,ooy}. Eventually, bars can produce central mass concentrations and, once they have accumulated sufficient mass in the centre, they weaken. For a sufficiently high mass, the bar may even  dissolve  \citep[e.g.][]{Pfenniger_Norman, friedli_benz, Bournaud_Combes}, although, in more recent simulations, bars appear more robust than initially thought  \citep[e.g.][]{shen,athanassoula05,athanassoula13}.
For a detailed description of the role of bars in secular evolution, see the review by \cite{kormendy} and references therein.

Observational evidence exists for  bar-induced gas inflows \citep[e.g.][]{regan97,zurita2004}. The inflow rate has been measured to within a range of $\sim$0.1--1 M$_\sun$ yr\me\ \citep{regan97,sakamoto}, with a higher rate being found in earlier-type galaxies \citep[][]{ReganTeuben,sheth}. However, the observational search for the theoretically predicted effects of bar inflows on the properties of the central gas-phase component of galaxies has not been successful in many cases, and conclusions are frequently contradictory.

Compared to unbarred galaxies, barred galaxies have higher central molecular mass concentrations \citep{sakamoto,sheth},  which produce a higher star formation rate (SFR)  in the centres of barred galaxies, as observed by several authors  \citep[e.g.][]{hummel90,Ho_barras,wang2012,zhou,ellison,ooy}. However, there is no agreement on the dependence of this enhancement in SFR with bar properties. For example,  \cite{wang2012} and \cite{zhou} find a positive trend between SFR and ellipticity or ellipticity-based parameters for strong bars, and \cite{ooy} find it with bar length,  but no dependency is found by other authors \citep{pompea,roussel,ellison}. Meanwhile, \cite{ellison} find an enhancement in central SFR  only for barred galaxies  with a mass greater than $10^{10}$~M$_\odot,$ and the enhancements observed by \cite{ooy} are restricted to the reddest galaxies in their sample.
Moreover,  \cite{cacho} have recently found no enhancement in SFR in barred galaxies compared with unbarred ones.

Studies of the central gas-phase metallicity yielded even more complex results.  \cite{ellison} find that there are central oxygen abundances $\sim$0.06 dex larger in barred than in unbarred galaxies at a given galactic  stellar mass, but \cite{cacho} find no difference in a similar study to the one performed by \cite{ellison}. Other authors \citep{considere,Dutil_Roy} used a smaller galaxy sample and obtained lower central oxygen abundances in barred starbursts than in normal, unbarred, galaxies. So far no relation has been found between bar strength and central oxygen abundance.

Thus, while simulations predict strong effects in the central gas properties of barred galaxies, no convincing observational confirmation has  yet  been found. The same bars that produce an enhancement in SFR  do not seem to produce the metallicity alteration that simulations predict \citep{friedli_kenni,martel}. Bar evolution, with its accompanying
bar-strength change, linked to the fact that bars can even be destroyed (meaning that currently unbarred galaxies have maybe had a bar) might be part of the reason for the current discrepancies in findings. Also, central properties might also depend on the availability of gas and on host galaxy properties in general (namely total mass or Hubble-type), which makes sample selection critical. 

The advent of the SDSS spectroscopic data has made it possible to study the effects of bars in central gas properties using large galaxy samples and covering wide ranges in host galaxy properties. Although studies in this field have contributed to interesting results, all publications, so far, have focused on central star formation. Only in a few cases has oxygen abundance also been studied \citep{ellison,cacho}. However, bar-induced gas inflows might also alter other physical properties of the central gas.  For example, if bars are able to produce higher central mass concentrations \citep{sakamoto,sheth}, higher concentrations of dust should be expected in the centres of barred galaxies, which would produce greater extinction. This greater concentration of gas and dust in the centres of barred galaxies, combined with a larger bar-induced SFR, could also lead to higher electron densities in the centres of barred galaxies \citep{Ho_barras}. The history of star-formation  in galaxy centres is likely to be different in barred and unbarred galaxies due to the gas supply induced by the bars, which might leave footprints in the  N/O abundance ratio \citep{molla,mallery07,edmunds_pagel78}.  
Consequently, it is worth exploring other properties of the ionised gas that might be modified by the effect of a bar. This is the  motivation for the work presented below. 

In this paper we  study ionised gas properties in the centres of barred and unbarred galaxies.  In addition to the SFR and oxygen abundance, we also measure the Balmer extinction, electron density, and the N/O abundance ratio. We  use a sample of nearby galaxies with available spectra from which we  remove AGN. Our galaxy sample has important advantages compared to previous samples: galaxies are face-on ($i < 26$\deg), the stellar component has already been studied by \cite{coelho}, and morphological 2D decomposition (and therefore structural information on bars, bulges and discs) is also available \citep{dimitri_morpho}.
Our  aim is to look for observational evidence that bar-induced inflows have important consequences for the central properties of the gas and, therefore, on secular evolution.

This paper is organised as follows. In Sect. \ref{sample} we describe our galaxy sample. In Sect.~\ref{spectra} we describe the procedure to measure the emission-line fluxes on which our study is based. Afterwards we remove AGN from the sample (Sect. \ref{agn}). In Sects. \ref{int_extinction}, \ref{abund}, \ref{SFR}, and \ref{edensity}, the internal extinction, central oxygen abundance, N/O ratio, SFR, and electron density are calculated for all sample galaxies and the distribution of these parameters is  analysed and compared separately for both barred and unbarred galaxies. In the following sections (\ref{morfo}, \ref{trends}, and \ref{dependence_bar}) we explore the dependence of the found differences on Hubble-type, total galaxy, and bulge mass, and possible dependencies on bar properties. In Sect. \ref{discussion}, we discuss our results, which  are then summarised in Sect. \ref{conclussions}.

\section{Galaxy sample}
\label{sample}

The galaxy sample studied here is the one used by \cite{coelho} to study the stellar populations of the bulges of barred and unbarred galaxies. This sample  is based on the one studied morphologically by \cite{dimitri_morpho}.

The sample contains all spiral face-on galaxies (axial ratio b/a $\geqslant 0.9$) in the Sloan Digital Sky Survey (SDSS) Data Release 2, with stellar masses larger than $10^{10}$~M$_\odot$,  redshift $0.02\leqslant z \leqslant 0.07$, bulge-to-total luminosity ratio below 0.043 (i.e. earlier than $\sim$Sd), and with signal-to-noise ratio in their corresponding SDSS spectra greater than or equal to 10, measured over the spectral range corresponding to the SDSS $g$-band. This sample is made up of  251 barred and 324 unbarred galaxies. 
The two sub-samples of barred and unbarred galaxies have similar total stellar mass distributions \citep{coelho}.

\section{Analysis of the spectra}
\label{spectra}

\subsection{Spectra pre-processing}
The spectra analysed in this paper come from the SDSS Data Release 7.
The spectra pre-processing comprises shifting of the original spectra to the rest frame using 
the redshifts from the SDSS database, re-sampling in steps of 1\AA, and 
Galactic extinction correction as described by \cite{cid-fernandes}. 

Afterwards the emission of the stellar component was modelled with the spectral synthesis 
code {\sc STARLIGHT} \citep{cid-fernandes}. This is done by comparing the observed SDSS spectra, on a 
pixel-by-pixel basis, to stellar population models \citep{vazdekis}.

We refer the reader to \cite{coelho} and \cite{dimitri_morpho} and references therein for 
full details on the sample selection and biases, and on the spectra pre-processing and 
modelling of the stellar component.

\subsection{Emission-line measurements}
\label{fluxes}
The emission-line fluxes were  automatically measured  over the stellar-emission-subtracted SDSS spectra of the galaxy sample.  The latter had already been processed as described in the paragraph above.

We developed a code for measuring the fluxes of the brightest emission 
lines of the optical spectra. These include [\oii]$\lambda\lambda$3727,3729, \hd, \hg, \hb, [\oiii]$\lambda$4959,
[\oiii]$\lambda$5007, [\oi]$\lambda$6300, [\nii]$\lambda$6548, \ha, [\nii]$\lambda$6584, and [\sii]$\lambda\lambda$6717,6731. All line fluxes were measured independently of each other, with no assumptions on the relative strength of any pair of lines.

Initially, the code searches for significant emission (over the continuum level)
at the rest-frame wavelength of the brightest emission lines. At this point, the 
continuum level is  determined using a linear fit of the median continuum level 
in two spectral regions on both sides of each emission line.
If the code detects an emission peak, with a signal  of at least twice the standard deviation of the continuum level, then the emission line profile is fitted with a Gaussian function, and the central wavelength and full-width at half maximum (FWHM) 
of the line are determined. Only fitted emission features yielding reasonable values\footnote{At this stage, we only considered as true detections those features having simultaneously: (a) a maximum central wavelength shift from the expected rest-frame wavelength smaller than 4.5\AA, and (b) either a FWHM within a factor or two, the FWHM of the [\oii]$\lambda\lambda$3727,3729 doublet, when the doublet was detected, or a  FWHM between 2 and 45\AA\ when [\oii]$\lambda\lambda$3727,3729 was not detected} for the central wavelength and FWHM are considered as true detections.
Afterwards, the code measures the emission-line fluxes. To do this, the code first defines
new regions dominated by continuum emission on both sides of each emission line. The size and 
distance of these continuum regions from the emission-line central wavelength depend 
on both the previously determined FWHM and on the location of neighbouring emission lines. 
The continuum level at the emission-line position is estimated through a linear fit of
the median continuum levels in the  continuum regions.
This is an important step in the procedure, as the greatest contribution to uncertainties in the line fluxes 
stems from the determination of the continuum level.
Line fluxes are then obtained from direct integration of the emission-line profiles over the
continuum level, which was determined as described above. An exception is made for  [\nii]$\lambda$6548, \ha, [\nii]$\lambda$6583,
and [\sii]$\lambda\lambda$6717,6731, for which fluxes obtained from Gaussian line-fitting were adopted 
to ensure accurate flux retrieval in cases where these sets of lines were blended.

The preliminary flux measurements were carefully inspected for unreliable measurements and/or undetected
emission lines which are clearly detectable by eye. These checks were done interactively, by visual inspection of the individual spectra and on the Gaussian and  continuum fits. We also compared our automatic flux measurements with measurements done by hand with the {\tt SPLOT} IRAF\footnote{IRAF is distributed by the National Optical 
Astronomy Observatory, which is operated by the Association of Universities for Research in Astronomy (AURA) 
under a cooperative agreement with the National Science Foundation.} task for random sets of spectra. 
The code was then modified accordingly until  automatic results were in reasonable agreement (within $\sim$20\%) 
with by-hand flux measurements.

The uncertainty of the line fluxes was estimated by propagating the uncertainty of the continuum level
emission in the line-flux calculation. Emission-line flux measurements with signal-to-noise ratios lower than three 
were rejected and set to zero.
Typically, relative errors are below $\sim$5\% for line fluxes brighter than 
$\sim$1.5$\times10^{-15}$~erg~s\me~cm$^{-2}$ for all emission lines,  except for 
[\oii]$\lambda\lambda$3727,3729, for which relative errors are  
larger ($\sim$10\%) at that flux level.
Line flux relative errors increase by up to $\sim$20\% at flux levels of $\sim3\times10^{-16}$~erg~s\me~cm$^{-2}$ for all lines except for 
[\oii]$\lambda\lambda$3727,3729, where  errors are typically larger ($\sim$25-30\%).

\begin{figure*}
\centering
\includegraphics[width=\textwidth]{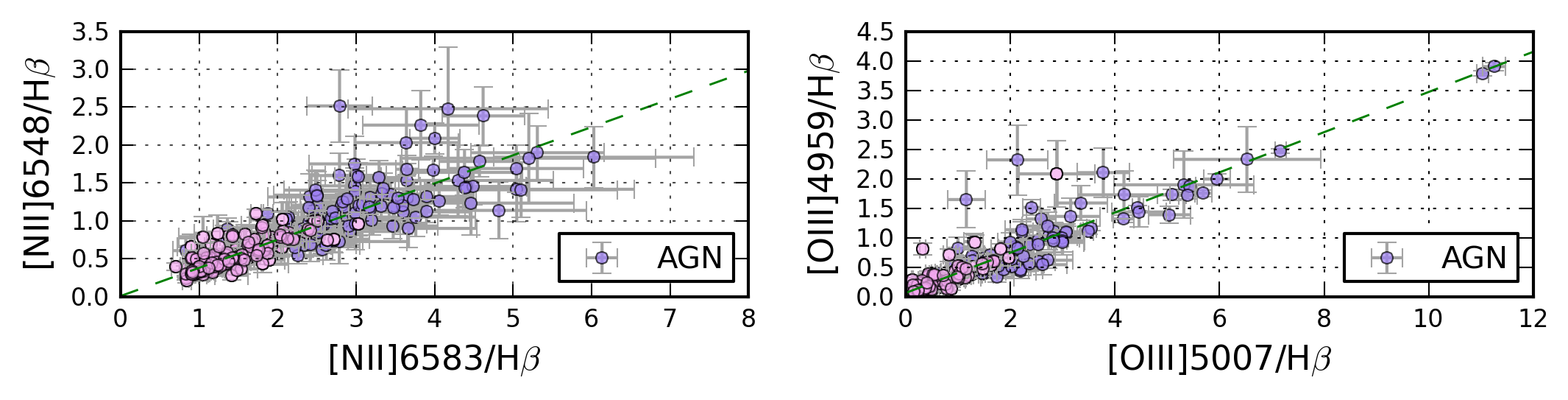}
\caption{Comparison of the measured emission-line fluxes of the two  lines in doublets [\nii]$\lambda\lambda$6548,6584 ({\em left}) and  [\oiii]$\lambda\lambda$4959,5007 ({\em right}) normalised to the \hb\ emission-line flux. Purple points indicate galaxies classified as AGN (Sect.~\ref{agn}). The dashed green line shows the best linear fit to the data, which yields a slope of  $0.37\pm0.02$ and $0.34\pm0.01$ for [\nii]$\lambda$6548/\hb\ vs  [\nii]$\lambda$6584/\hb,\ and [\oiii]$\lambda$4959/\hb\ vs  [\oiii]$\lambda$5007/\hb,\ respectively (see text in Sect.~\ref{fluxes} for details).}
\label{measurements_accuracy}
\end{figure*}

We have empirically checked the quality of our emission-line measurements by comparing the emission-line ratios, [\nii]$\lambda$6548/[\nii]$\lambda$6584 and  [\oiii]$\lambda$4959/[\oiii]$\lambda$5007, to their theoretical values. This is done in Fig.~\ref{measurements_accuracy}. The emission-line fluxes of both doublets correlate  with each other, with correlation coefficients of 0.97 and 0.88 for the [\oiii] and [\nii] doublets, respectively. The fitted slopes are $0.34\pm0.01$ and $0.37\pm0.02$  for [\oiii]$\lambda$4959 vs [\oiii]$\lambda$5007, and [\nii]$\lambda$6548 vs [\nii]$\lambda$6584, respectively, which are close to the theoretical value (i.e. 1/3) \citep{osterbrock}. The dispersion in the  [\nii]$\lambda$6548 vs [\nii]$\lambda$6584 is larger than that in the [\oiii] doublet, especially for higher values of [\nii]$\lambda$6584/\hb. In general, the galaxies with larger deviations from the best linear fits are also those with larger error bars. This fact reinforces our flux error estimates, which seems to represent the real uncertainties in the measurements well.
Data points with the largest deviation from the best fit are normally AGNs (especially for the [\nii] doublet). We only concentrate on non-AGN galaxies in this paper and, therefore, our results will not be affected by these larger uncertainties.

\subsection{Comparison with emission-line fluxes from other databases}
\label{comp_fluxes}
We compared our emission-line flux measurements with those available in some public databases. The most popular 
public databases with emission-line fluxes are the MPA-JHU\footnote{http://www.mpa-garching.mpg.de/SDSS/DR7/} release 
of spectrum measurements (from SDSS DR7) and the OSSY\footnote{http://gem.yonsei.ac.kr/$\sim$ksoh/wordpress/} database \citep{ossy}. 
As in our case, these databases also used the 7th release of the SDSS spectra.
MPA-JHU line fluxes of extended sources are normalised to match the photometric fibre magnitude in the $r$-band. We  removed 
this normalisation from the MPA-JHU data. 

\begin{figure*}
\centering
\includegraphics[width=\textwidth]{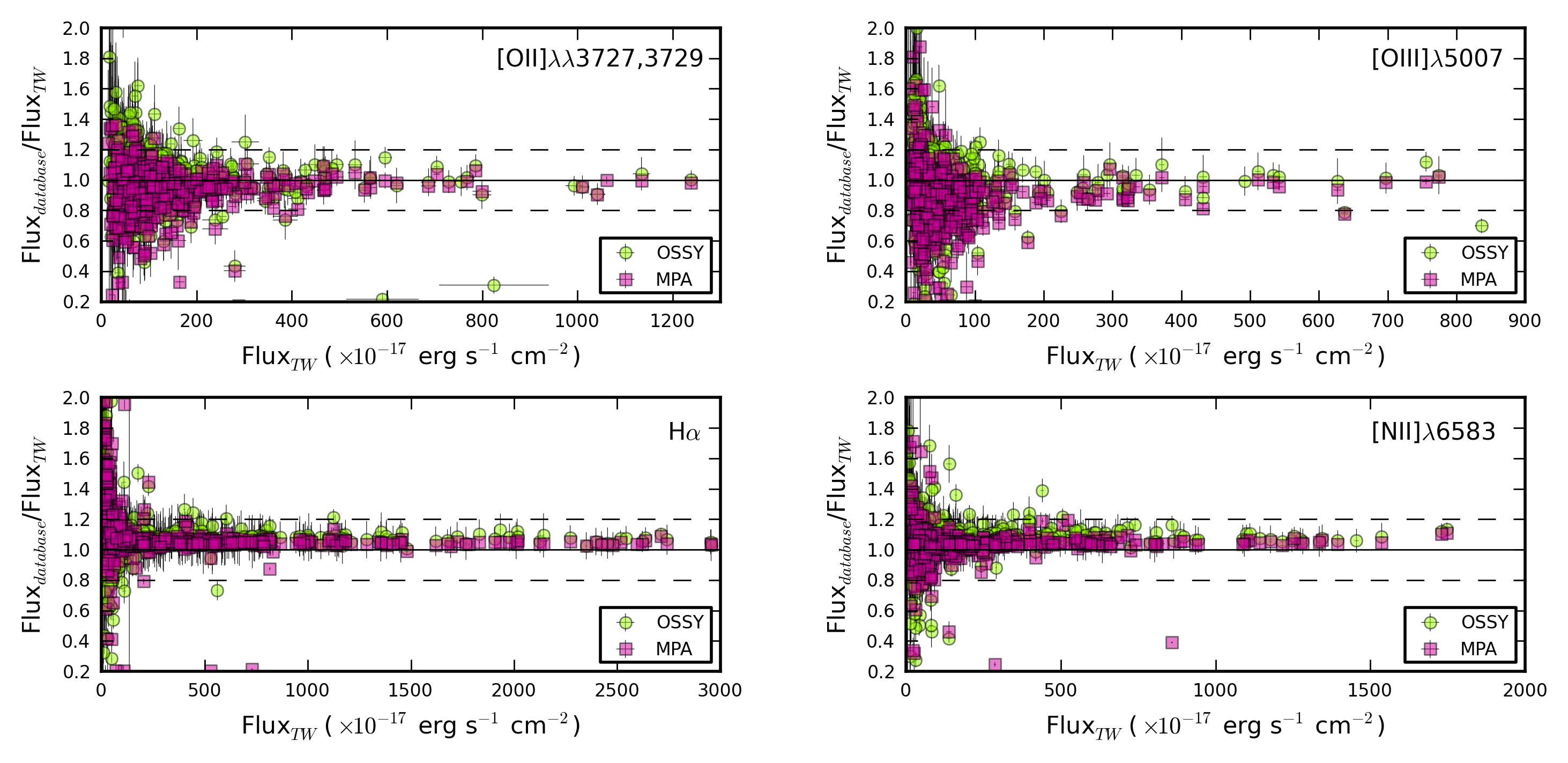}
\caption{Ratio of emission-line fluxes available in public databases to the fluxes measured in this work (Flux$_{TW}$), for the OSSY (green circles) 
and MPA-JHU (pink squares) databases, as a function of the emission-line fluxes measured in this work for the  [\oii]$\lambda\lambda$3726,3729 (top left), 
[\oiii]$\lambda$5007 (top right), H$\alpha$ (bottom left), and [\nii]$\lambda$6583 (bottom right) emission-lines. All fluxes are corrected for Galactic extinction (see Sect.~\ref{comp_fluxes} for details).}
\label{lineas}
\end{figure*}

Figure~\ref{lineas} shows the comparison of the [\oii]$\lambda\lambda$3727,3729, [\oiii]$\lambda$5007, \ha, and [\nii]$\lambda$6583 
emission-line fluxes, measured by these databases, with our measurements from the spectra of our galaxy sample (Sect.~\ref{sample}).
The three sets of data represented in Fig.~\ref{lineas} have been corrected for foreground Galactic reddening (but not for internal extinction). 
Our emission-line fluxes match fairly closely  those from the OSSY and MPA-JHU databases. In general
the fluxes match within $\sim$20\% for lines brighter than $\sim$10$^{-15}$~erg~s\me~cm$^{-2}$, except for [\oii]$\lambda\lambda$3727,3729, where differences are within  $\sim$30\%. For weaker lines, the scatter increases considerably, as a result of larger relative 
errors in the measurements.

For emission lines brighter than $\gtrsim$3$\times$10$^{-15}$~erg~s\me~cm$^{-2}$, our fluxes match with the OSSY and MPA-JHU  measurements to within $\sim$4-6\%, except for [\nii]$\lambda$6583 and \ha\ as measured by OSSY, which are, on average, $\sim$8\% higher than our measurements. Our measurements of [\nii]$\lambda$6583 and \ha\ are also systematically lower than those from MPA-JHU, but only by $\sim$4\% on average. For [\oii]$\lambda\lambda$3727,3729 and [\oiii]$\lambda$5007, however, our measurements are systematically $\sim$5\% higher than MPA-JHU measurements, and are a better match with the OSSY measurements (within $\sim$2-3\%). These systematic differences can, in part (but not completely), be explained by the use of different extinction laws for the correction of the foreground Galactic extinction. Our data was corrected using the \cite{schlegel98} maps and the \cite{ccm} extinction curve \citep[see][]{cid-fernandes}. The  MPA-JHU data have been corrected using \cite{0Donnell}. Differences of approximately $\sim$3\% at [\oii]$\lambda\lambda$3727,3729 are expected for typical  values of A$_V$$\sim1-1.5$~mag, given the differences in the extinction curves. The extinction curve used by OSSY authors is not specified.

In any case, differences between our observed fluxes and the ones available in the OSSY and MPA-JHU databases are small and comparable to measurement errors, even taking into account that the stellar emission subtraction, the flux measurement procedures, and the foreground  Galactic reddening correction differ in the three sets of data.

\section{Nuclear classification and removal of AGN}
\label{agn}

\begin{figure*}
\centering
\subfigure{
\label{BPTbar}
\includegraphics[width=0.7\columnwidth]{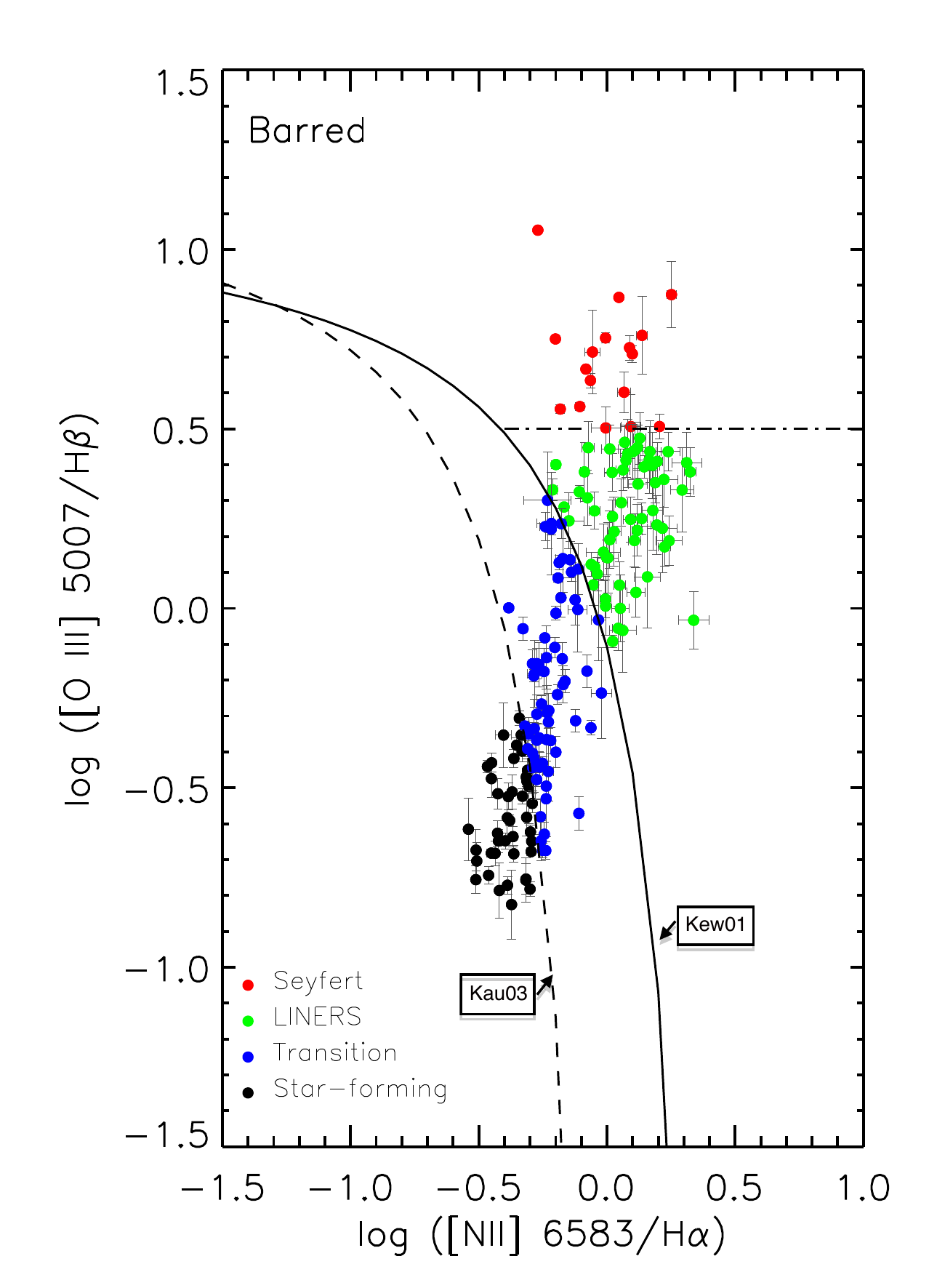}}
\subfigure{
\label{BPTnobar}
\includegraphics[width=0.7\columnwidth]{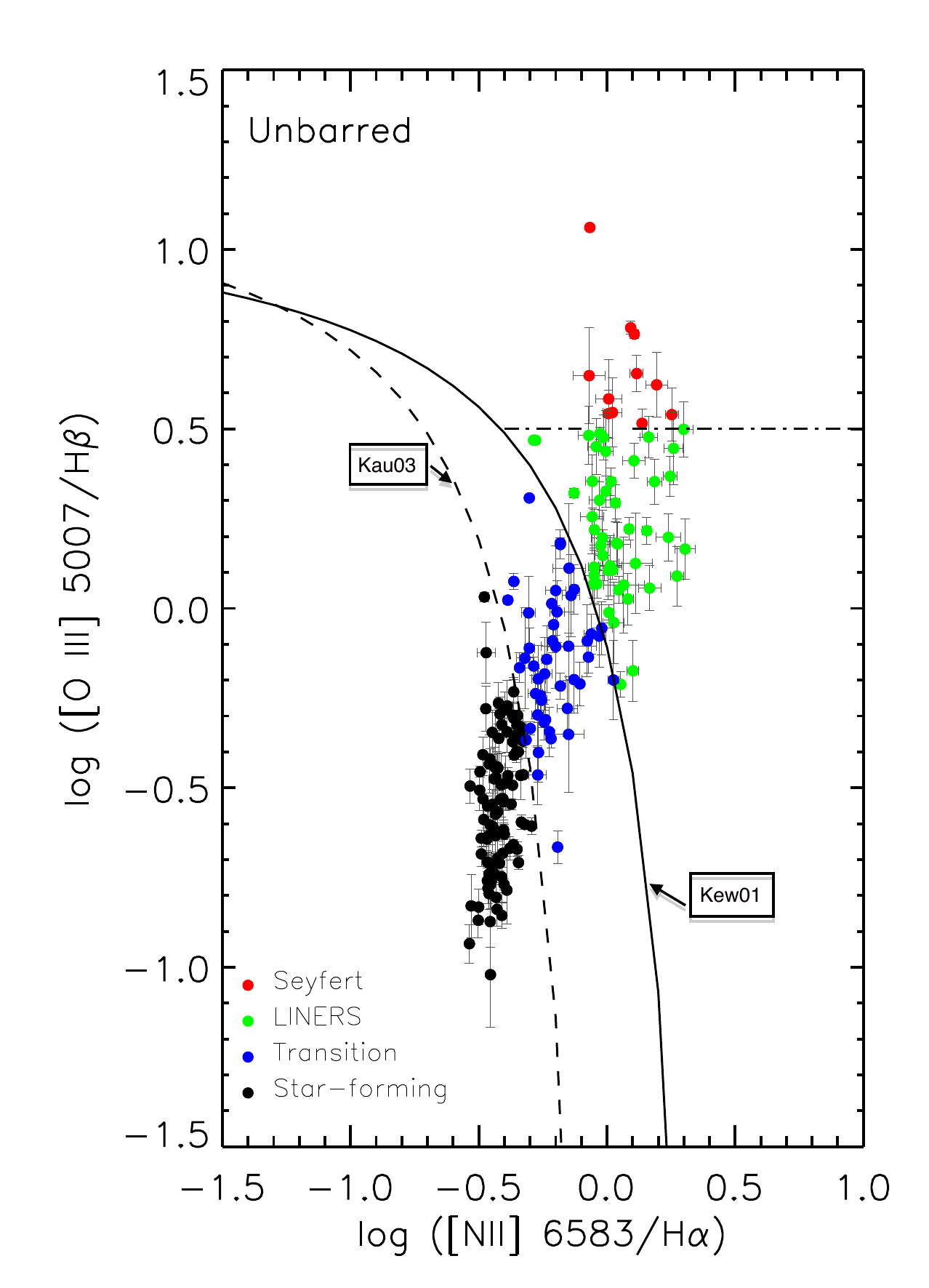}}
\caption{Diagnostic diagrams to classify emission-line galaxies according to their dominant ionisation mechanism for the barred ({\em left}) and unbarred ({\em right}) galaxies of the sample. The solid and dash lines separate the region of star-forming galaxies from the region of AGN according to \cite{kewley01} and \cite{kauffmann2003_AGN}, respectively. The horizontal dot-dashed line separates the AGN region into Seyfert and LINER galaxy regions \citep{kewley06}.}
\end{figure*}

Our initial galaxy sample (see Sect.~\ref{sample}) includes active and normal galaxies. We  used standard diagnostic diagrams based on bright emission-line ratios \citep[also known as BPT diagrams, after][]{bpt} to separate the galaxies where the dominant mechanism of ionisation is UV radiation from massive young stars (i.e. star-forming galaxies) from those galaxies being ionised by an active galactic nucleus (AGN).  In Figs.~\ref{BPTbar} and  \ref{BPTnobar}, we show the diagnostic diagrams based on our measured (non-extinction corrected\footnote{The [\oiii]$\lambda$5007/\hb\ vs [\nii]$\lambda$6584/\ha\ is the least reddening dependent BPT plot. After internal extinction correction as described in Sect.~\ref{int_extinction}, we have checked that our classification was correct. BPT diagrams shown in Sect.~\ref{discussion} are based on extinction-corrected emission-line fluxes.}) ratios [\oiii]$\lambda$5007/\hb\ vs [\nii]$\lambda$6584/\ha, for the barred and unbarred sub-samples of galaxies.

The solid line separates star-forming galaxies from AGN according to the predictions of the photoionisation models of \cite{kewley01}, while the dashed line sets this separation empirically \citep{kauffmann2003_AGN}. Both lines do not match, and the number of AGN predicted by \cite{kauffmann2003_AGN} is larger than the theoretical prediction. We refer to the galaxies located between the two lines (i.e. classified as "star-forming" according to \citealt{kewley01} and as AGN according to \citealt{kauffmann2003_AGN}) as transition objects. These are also known as  composite objects and might be a population of objects where ionisation is partly produced by recent star-formation and partly due to an AGN \citep{kewley06}, but they might also include photoionised nuclei with high nitrogen abundance \citep{pmc09}. For this reason we  used the \cite{kewley01} criteria to remove AGN from our sample.
The horizontal dot-dashed line  separates the Seyfert region from the Low-Ionisation Nuclear-line Regions area \citep[LINERs;][]{kewley06}.
  
According to this classification into star-forming, transition and AGN galaxies, the percentage of AGN in the sub-samples of barred and unbarred galaxies is 31\% and 18\%, respectively. In other words,  there are $\sim1.7$ times more AGN in barred than in unbarred galaxies. This factor remains when we consider the \cite{kauffmann2003_AGN} criteria, i.e. AGN \cite[according to][]{kewley01} plus transition objects are 1.7 times more frequent in the sub-sample of barred galaxies than in the unbarred galaxy sub-sample\footnote{\cite{coelho} used the AGN catalogue from \citealt[][http://www.mpa-garching.mpg.de/SDSS/DR4/Data/agncatalogue.html]{kauffmann2003_AGN}, to separate AGN galaxies from the whole sample. Their figures differ from the ones used in this work, in which classification comes from our emission-line measurements. However, in both cases AGN are $\sim1.7$ times more frequent in barred than in unbarred galaxies.}.  The fraction of AGN in barred galaxies increases when we consider galaxies with low-mass bulges. AGN are about twice as frequent in barred galaxies as in unbarred galaxies, if we only consider bulges with mass lower than $10^{10.1}$M$_\odot$.

Below, we  consider all the galaxies of the sample, except those classified as AGN. We will refer to them as  non-AGN galaxies, and this sample includes star-forming galaxies according to the criteria of \cite{kauffmann2003_AGN}, transition objects, and unclassified galaxies (normally due to the non-detection of relevant emission lines involved in the BPT diagram).  Our effective sample of non-AGN galaxies has 173 and 265  barred and unbarred galaxies, respectively.

\begin{table}
\caption{\label{tablaAGN} Classification of the galaxy sample.}
\centering
\begin{tabular}{|l|c|c|c|}
\hline\hline     
Type         & whole   & barred      & unbarred    \\
             & sample  & subsample & subsample \\
\hline 
Star-forming&  139    & 43          & 96   \\
Transition   &  113    & 66          & 47   \\
LINER        &  109    & 61          & 48   \\
Seyfert      &   28    & 17          & 11   \\
Unclassified &  186    & 64          & 122  \\
\hline
\end{tabular}
\tablefoot{Number of active and non-active galaxies in the global sample and in the sub-samples of barred and unbarred galaxies, according to the BTP diagrams in Figs.~\ref{BPTbar} and \ref{BPTnobar} (see Sect.~\ref{agn} for details).}
\end{table}

\begin{figure}
\centering
\includegraphics[width=1.04\columnwidth]{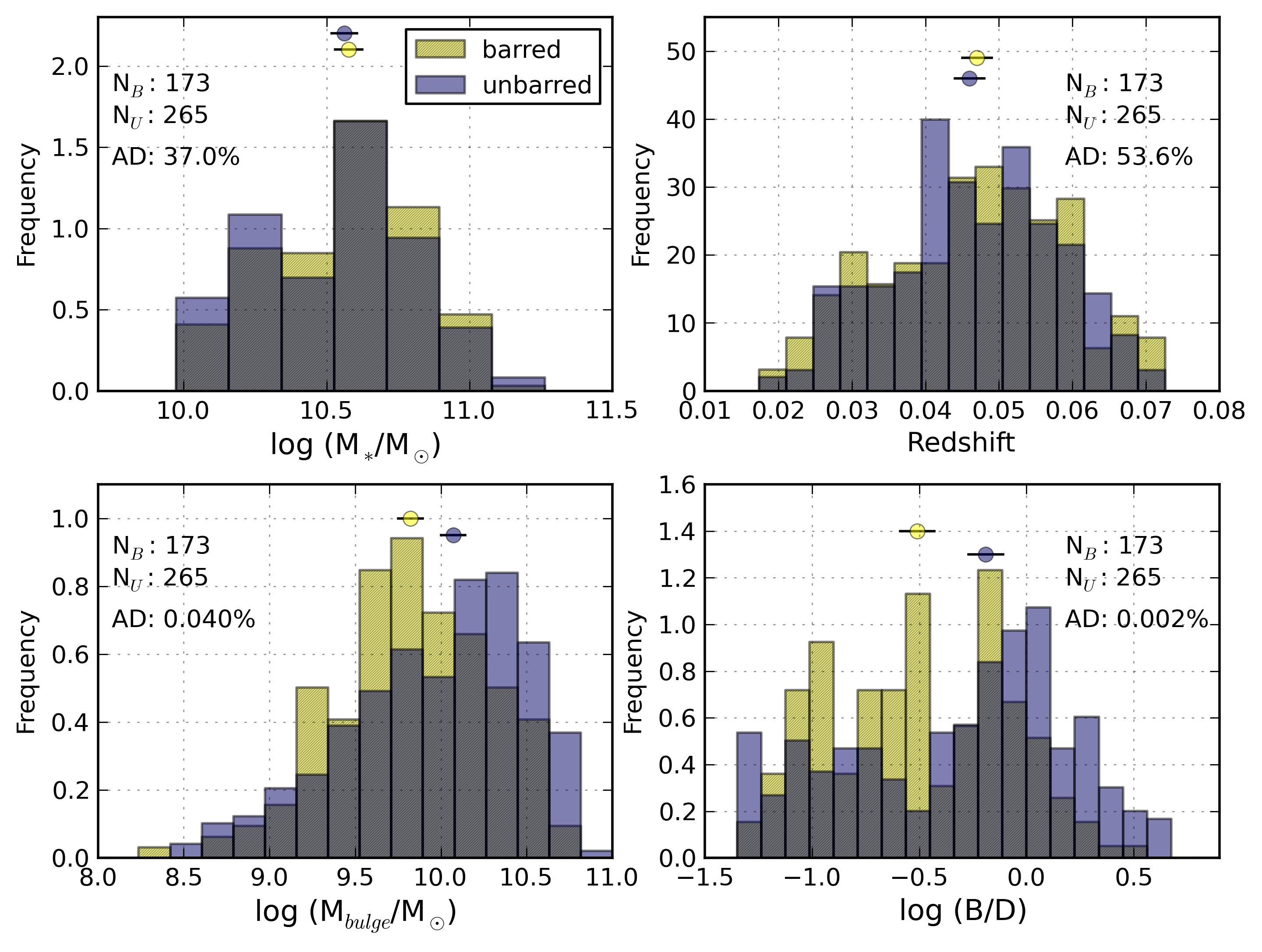}
\caption{Histograms showing the distribution of decimal logarithm of the total galaxy stellar mass, redshift, logarithm of bulge mass and  logarithm of the B/D light ratio in the $i$ band for barred (hatched yellow) and unbarred (purple) galaxies separately, for all the non-active galaxies of the sample. The number of galaxies in each sub-sample (N$_B$ and N$_U$) is indicated, together with the two-sample Anderson-Darling test $P$-values (AD, expressed in \%). The yellow and purple circles in the upper side of the panels indicate the median value of the barred and unbarred distributions, respectively. The horizontal error bar covers the 95\% confidence interval (estimated as $1.57 \times IQR / \sqrt{N}$, where $IQR$ is the interquartile range, or 1st quartile subtracted from the 3rd quartile, and $N$ the number of data points) for the corresponding median value.}
\label{sample_noAGN}
\end{figure}
\begin{figure}
\centering
\includegraphics[width=1.04\columnwidth]{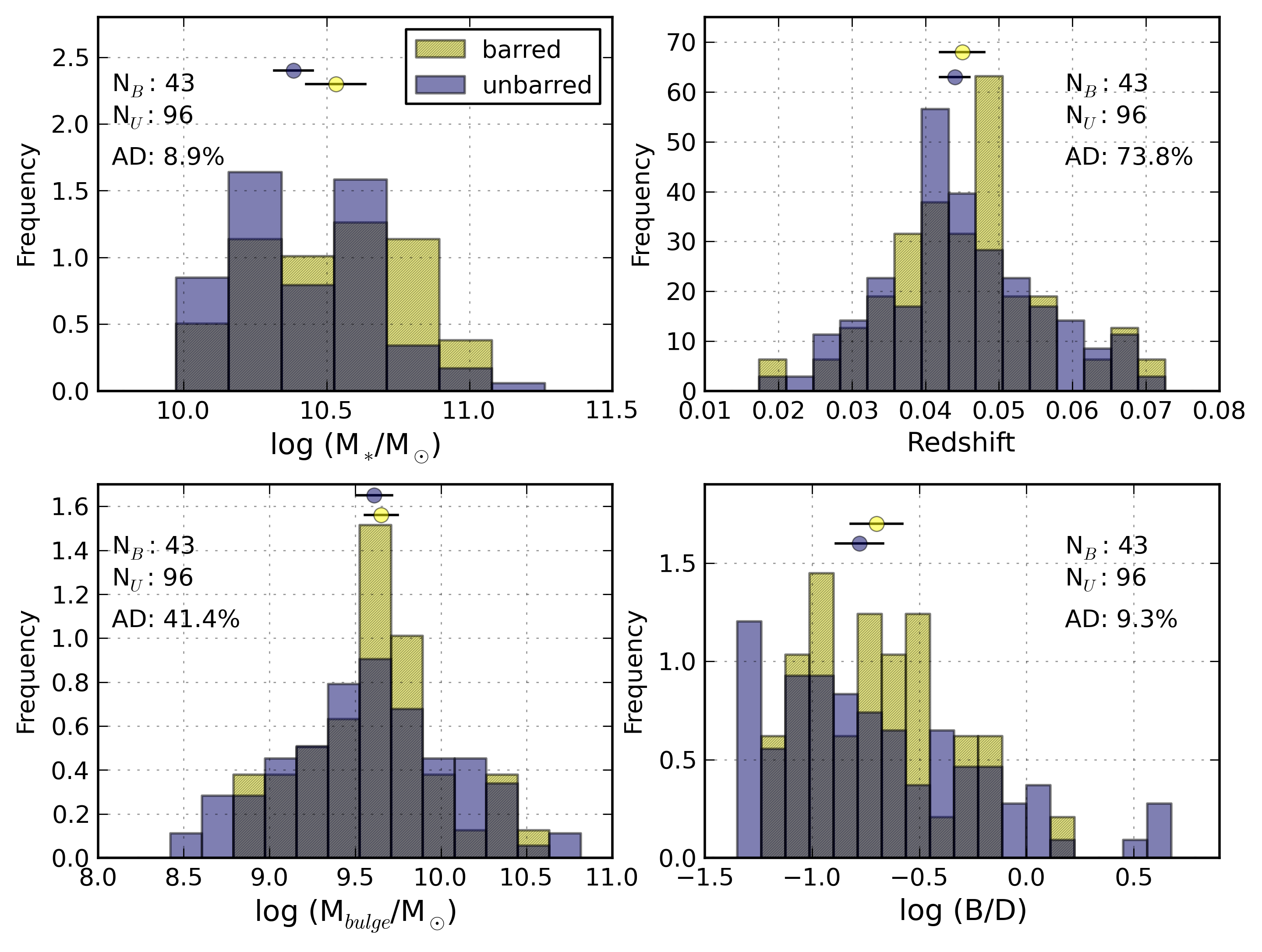}
\caption{Same as Fig.~\ref{sample_noAGN} but only for galaxies classified as "star-forming" according to \cite{kauffmann2003_AGN}. N$_B$ and N$_U$ are the number of barred and unbarred galaxies, respectively, and AD is the two-sample Anderson-Darling test $P$-value (expressed in \%).} 
\label{sample_SF}
\end{figure}

\subsection{The characterisation of non-AGN and star-forming sub-samples}
\label{subsample-nonAGN}
Figure~\ref{sample_noAGN} shows a comparison between the  total galaxy stellar mass, redshift, bulge mass, and bulge-to-disc (B/D) light ratio ($i$-band) distributions for the sub-samples of barred and unbarred galaxies. The total stellar masses are taken from \cite{kauffmann03mass} and are based on fitting to stellar spectral features (the 4000\AA\ break and the Balmer absorption line index H$\delta_A$). The bulge masses come from  \cite{coelho} (based on bulge luminosity and mass-to-light ratio in the $i$ band) and the B/D light ratio comes from morphological decomposition performed by \cite{dimitri_morpho}.
The total galaxy stellar mass and  redshift distributions are similar for barred and unbarred galaxies. To confirm this statistically we  used the  $k$-sample Anderson-Darling (hereinafter A-D) test \citep[see][]{AD,stephens74}\footnote{The $k$-sample Anderson-Darling test is based on the Anderson-Darling rank statistics \citep{AD54} for testing homogeneity of samples with possibly different sample sizes and unspecified distributions. The Anderson-Darling test is, in turn, a modification of the Kolmogorov-Smirnov test. The $k$-sample version of the test is more recommended than $k$-sample Kolmogorov-Smirnov for small sample sizes \citep[e.g.][]{hou}. See also discussion at  https://asaip.psu.edu/Articles/beware-the-kolmogorov-smirnov-test.}.  It tests the null hypothesis that the two samples are drawn from the same parent distribution. The output of the A-D test is a $P$-value or significance level at which the null hypothesis can be rejected. For the total stellar mass and redshift, we find $P$-values of  $\sim$37\% and 54\%, respectively. These values are much greater than the 5\% threshold value normally adopted  below which the result is statistically significant and the null hypothesis can be rejected.
The  distributions of  bulge mass and B/D light ratio are, however, different between barred and unbarred galaxies when all non-AGN galaxies are considered. Median values and distribution shapes are different and the A-D tests confirms this difference, with $P$-values well below 0.1\%.
The bulges of barred galaxies are less massive  than in unbarred galaxies and B/D light ratios are also lower for barred galaxies, even though both sub-samples of barred and unbarred galaxies have the same total stellar mass distributions. This has already been pointed out by \cite{coelho}.

If we consider the sub-sample containing only galaxies classified as star-forming, Fig.~\ref{sample_SF} and the A-D test results indicate that the  distributions of total stellar mass, redshift, bulge mass, and  B/D light ratio are no different for barred and unbarred galaxies,  since $P$-values  are higher than 5\% in all cases.


\section{Internal extinction}
\label{int_extinction}
The emission-line fluxes were corrected for internal extinction from the Balmer decrement using 
the \ha\ to \hb\ emission-line flux ratio. In the absence of internal dust extinction, the
\ha\ to \hb\ line flux ratio, (F$_{H\alpha}$/F$_{H\beta}$)$_{intr}$ is equal to 2.86, for typical electron 
temperatures and densities in star-forming regions (T$_e\sim$10000~K 
and n${_e}\sim$100~cm$^{-3}$), in the  case B of the recombination theory \citep{osterbrock}.
The presence of dust in the interstellar medium increases this line ratio as a result of differential 
extinction. The amount of attenuation can be quantified with the parameter $c(\hb)$, the 
internal extinction at the \hb\ emission line
\begin{equation}
(F_{H\beta})_{\rm obs} = (F_{H\beta})_{\rm intr}\, 10^{-c(H\beta)}
,\end{equation}
where the sub-indexes "obs" and "intr" stand  for observed and intrinsic fluxes, respectively.
The \ha\ to \hb\ emission-line ratio together with an assumed extinction curve $f(\lambda)$ can be used to obtain $c(\hb)$
\begin{equation}
c({\rm H}\beta) = \frac{-1}{f(\ha)} \left[\log{\left(\frac{F_{H\alpha}}{F_{H\beta}}\right)_{\rm obs}} -\log{\left(\frac{F_{H\alpha}}{F_{H\beta}}\right)_{\rm intr}}\right]
, \end{equation}where $f(\ha)$ is the reddening function at \ha\ normalised to \hb\ (i.e. $f(\hb)=0$).
We  employed the \cite{seaton} reddening law with the \cite{howarth} parametrisation\footnote{The relation of $f(\lambda)$ with the function X(x) of the \cite{howarth} parametrisation is 
$f(\lambda)=\frac{X(x)}{X(x_{H\beta})}-1$, where $x=1/\lambda$ ($\lambda$ in microns), with $x_{H\beta}$ the 
corresponding value for the \hb\ spectral line. In this parametrisation, the extinction in magnitudes at a given wavelength $\lambda$ is A$_\lambda = X(x)\, E(B-V)$.}, assuming $R_V=3.1$. With this combination of reddening law and $R_V$, the extinction in magnitudes at \ha\ is A$_{H\alpha}$ = 1.515 $\times c(\hb)$. 
The intrinsic \ha\ to \hb\ line flux ratio was assumed  to be 2.86. We are aware that in the central regions of galaxies the metallicity might be oversolar, with T$_e$ lower than 10000~K, and, therefore, this ratio may be higher than 2.9. Assuming an increased value of the \ha\ to \hb\ line flux ratio would decrease all $c(\hb)$ values by 0.02, 0.06, and 0.1 for (F$_{H\alpha}$/F$_{H\beta}$)$_{intr}$, equal to 2.9, 3.0, and 3.1, respectively.

\begin{figure}
\centering
\includegraphics[width=1.04\columnwidth]{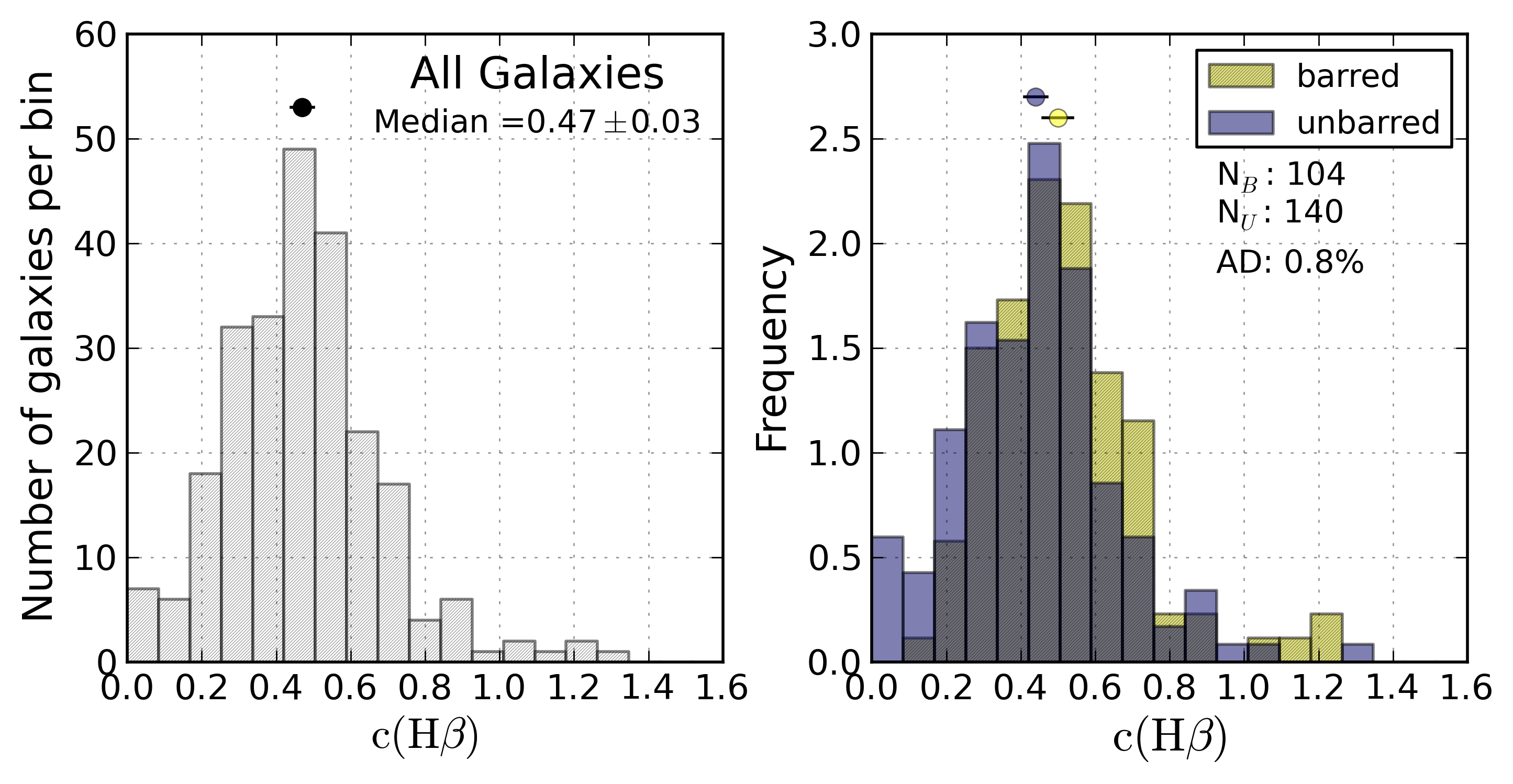}
\caption{\label{distribucionC}{\bf Left.} Histogram of the values of the Balmer extinction at the \hb\ emission line, $c(\hb)$, 
for all galaxies of the sample with $c(\hb)>$0.0, except those classified as AGN as explained in Sect. \ref{agn}. 
{\bf Right.} Same as for the left, except for barred (hatched yellow) and unbarred (purple) galaxies separately. The number of galaxies 
in each sub-sample, and the $P$-value from the A-D test for the two distributions, are shown below the figure legend.}
\end{figure}

The histogram with the values of the Balmer extinction $c(\hb)$ for all galaxies of the sample (except AGN) is in the left-hand panel of Fig.~\ref{distribucionC}. Values occur in the range 0-1.4, being the median value of the distribution $(0.47\pm0.03)$. We assign an internal extinction of zero when the observed Balmer ratio F$_{H\alpha}$/F$_{H\beta}$  is lower than the theoretical value. These objects are not considered in Fig.~\ref{distribucionC}. The  distribution obtained is very similar in shape and values to that found by \cite{stasinska04} for 10854 spectra from the First Data Release of the SDSS.
\begin{figure}
\centering
\includegraphics[width=0.9\columnwidth]{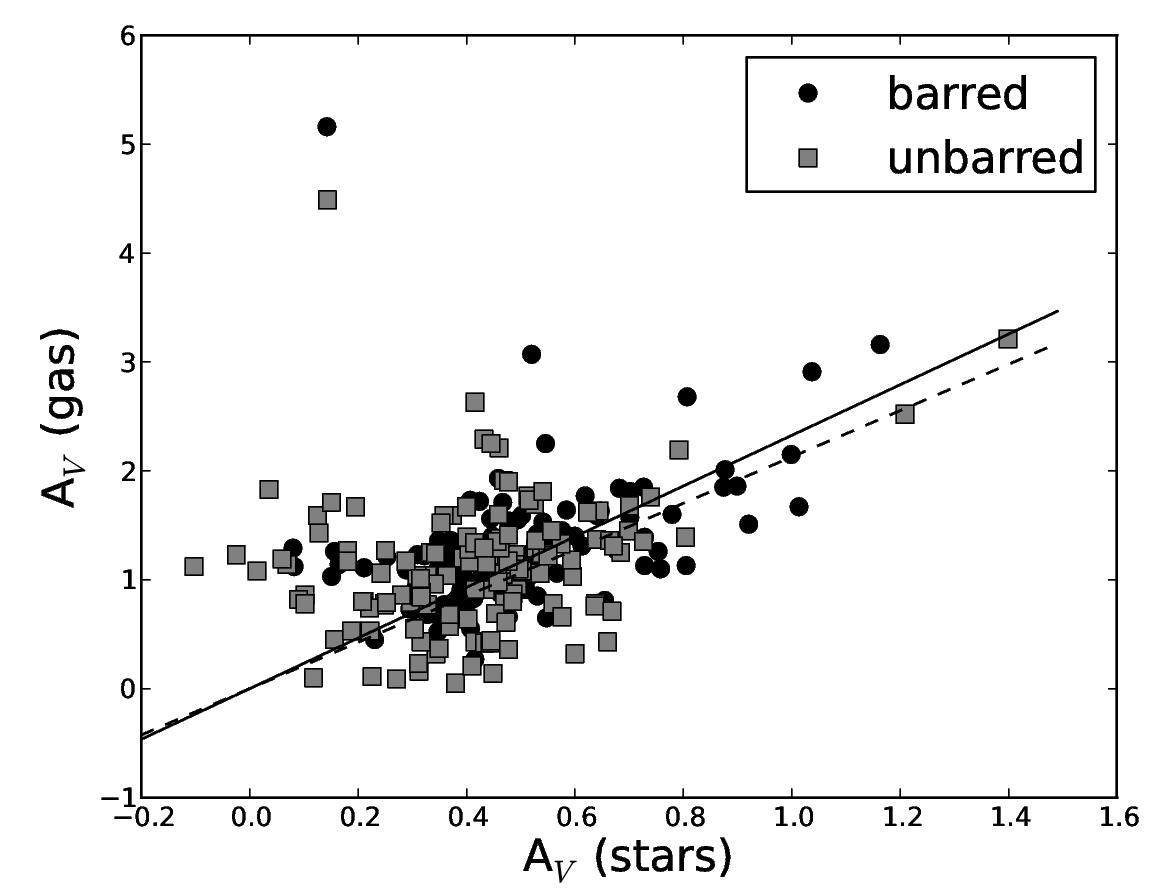}
\caption{\label{AvminAv}$V$-band extinction in magnitudes,  $A_{V,g}$, as obtained from the \ha\ to \hb\ Balmer decrement,  versus the stellar continuum $V$-band extinction ($A_{V,s}$), as derived from the spectral synthesis-fitting with STARLIGHT for all non-AGN galaxies. The straight solid and dashed lines show the relations between $A_{V,g}$ and $A_{V,s}$ found by \cite{calzetti2000} and \cite{kreckel}, respectively.}
\end{figure}

Dust affects both the stellar continuum emission and the nebular emission. Figure~\ref{AvminAv} shows the $V$-band extinction in magnitudes derived from the gas emission lines  ($A_{V,g}$)  versus the stellar continuum $V$-band extinction (A$_{V,s}$), as derived from the  spectral synthesis fitting with STARLIGHT (see Sect.~\ref{sample}).  We should point out that the \cite{ccm} extinction curve was used in STARLIGHT, and we used \cite{seaton} for the gas component. However, both extinction curves give roughly the same extinction in the $V$-band (Seaton gives on average a higher extinction rate by $\sim$0.8\% in that spectral range).
The dispersion is high, but it is clear from the figure that both quantities are correlated, and that the extinction derived from emission lines is typically about twice the extinction derived from the observed stellar continuum. This is in agreement with other author results, such as \cite{calzetti2000} and \cite{kreckel}, who find A$_{V,s}$=(0.44 $\pm$ 0.03) A$_{V,g}$ and A$_{V,s}$=(0.470 $\pm$ 0.006) A$_{V,g}$ for \hii\ regions and starburst galaxies,  respectively. These relations are overplotted in Fig.~\ref{AvminAv} and are compatible with our data in the range of extinction observed ($A_{V,g}$ from 0 to 3~mag).

\begin{figure*}
\centering
\includegraphics[width=1.0\textwidth]{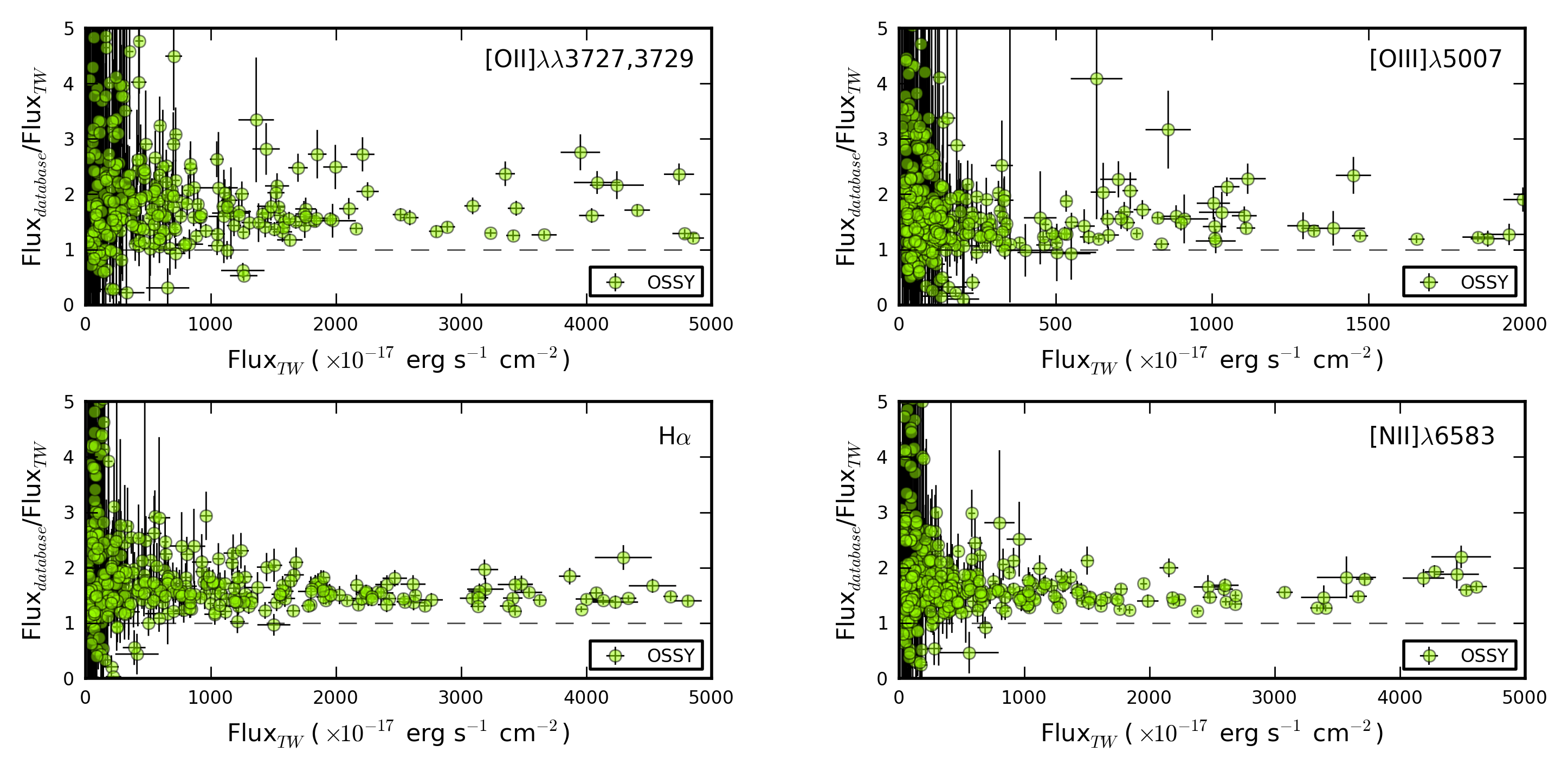}
\caption{\label{flujosOSSY}Ratio of internal extinction corrected emission-line fluxes from the OSSY database to those measured in this work, as a function of the latter for the  [\oii]$\lambda\lambda$3726,3729 (top left), 
[\oiii]$\lambda$5007 (top right), H$\alpha$ (bottom left) and [\nii]$\lambda$6583 (bottom right) emission lines. All fluxes have been corrected for Galactic extinction and internal extinction (see Sect.~\ref{int_extinction} for details).}
\end{figure*}

All measured emission lines were corrected for internal dust extinction using the derived values for $c(\hb)$. The corrected  emission-line fluxes have been compared with the corresponding values provided in one of the most popular public databases: OSSY \citep{ossy}. Figure~\ref{flujosOSSY} shows the comparison of extinction-corrected fluxes for some of the brightest emission lines. As  can be seen, OSSY fluxes are typically larger than our extinction-corrected line fluxes by a factor of about 1.6 ($\sim$1.8 for [\oii]). Also, the OSSY colour excess tabulated values for the gas and the stellar continuum (called E(B-V)\_gas and E(B-V)\_star, respectively) do not maintain the relation  mentioned above (Fig.~\ref{AvminAv}). However, we would like to point out here that our observed (uncorrected) emission-line fluxes  correspond extensively with OSSY quoted observed fluxes (see Sect.~\ref{comp_fluxes} and Fig.~\ref{lineas}). 

A table containing extinction-corrected  line fluxes measured in this paper is available at the CDS. The table contains the following information: Column 1 gives the SDSS plate used to collect the spectrum; Column 2 the modified Julian Date of the observation night (mjd); Column 3 the SDSS fibre number; Columns 4 to 25 the extinction-corrected emission-line fluxes normalised to \hb=100 and associated uncertainties; Columns 25 and 27 give the \hb\ flux in units of $10^{-17}$erg~s\me~cm$^{-2}$ and the Balmer extinction at \hb, respectively; and Columns 26 and 28  give their corresponding uncertainties.

\subsection{The distribution of $c(\hb) $ for barred and unbarred galaxies} 

The right-hand panel of Fig.~\ref{distribucionC} shows the distribution of $c(\hb)$ for barred and unbarred galaxies separately. The median value of $c(\hb)$ is only marginally larger in barred galaxies (0.50$\pm$0.04) than in unbarred (0.44$\pm$0.03), but the A-D test indicates that the two distributions  differ, with a $P$-value of 0.8\%, which is well below 5\%. We find the same result when we only consider star-forming galaxies (see Table~\ref{stats}).

Previous work on the Balmer extinction at \hb\ on large SDSS data samples \citep[e.g.][]{stasinska04}  shows that $c(\hb)$ decreases from early- to late-type galaxies. Additionally,  $c(\hb)$ also depends on total galaxy luminosity, being higher for the most luminous galaxies (and probably more massive). Our sub-sample of barred non-AGN galaxies is similar to the unbarred sub-sample in terms of total stellar mass (Fig.~\ref{sample_noAGN}), but it is biased towards lower bulge mass and lower B/D light ratios compared with the sub-sample of unbarred galaxies. If we assume that the results obtained by \cite{stasinska04} for integrated spectral properties of galaxies are applicable to the inner kiloparsec of galaxies, we would predict a lower average of $c(\hb)$ in barred than in unbarred galaxies. However we find just the opposite, although differences in median values are only marginal (within error bars).

The sub-samples of barred and unbarred star-forming galaxies (Fig.~\ref{sample_SF}) are equivalent in B/D light ratios, and bulge mass and their distributions of $c(\hb)$ are also different ($P$-value 0.6\%), being larger in barred than in unbarred galaxies by $\sim$ 0.07~dex, approximately the same difference as between the non-AGN barred and unbarred sub-samples.
A larger central dust extinction might be related to a larger dust mass surface density \citep{kreckel} and, therefore, the observed difference in  $c(\hb)$, although marginal,  could be due to different central dust mass concentrations possibly resulting from the transfer of material towards the galaxy centres, induced by the bars.

\begin{table*}
\caption{\label{stats}  Statistics for barred and unbarred non-AGN and star-forming galaxies.}
\centering
\begin{tabular}{l|c|cc|cc||c|cc|cc}
\hline \hline   
             &\multicolumn{5}{c||}{{\bf Non-AGN}} &\multicolumn{5}{c}{{\bf Star-forming}}\\
\hline
                     & $P$-value\tablefootmark{a} & \multicolumn{2}{c|}{barred}&  \multicolumn{2}{c||}{unbarred}& $P$-value\tablefootmark{a}  & \multicolumn{2}{c|}{barred}&  \multicolumn{2}{c}{unbarred}\\
                     &                          & median             & N     &     median           & N      &           &  median             & N    &  median          & N        \\
\hline
$c(\hb)$             & 0.8\%                    & 0.50$\pm$0.04      &104    & 0.44$\pm$0.03        &140     & 0.6\%     &  0.53$\pm$0.04      & 43   & 0.46$\pm$0.04    & 96   \\
$\log$~$R_{23}$       & 64.0\%                   & 0.31$\pm$0.04      &105    & 0.31$\pm$0.04        &130     & 0.3\%     &  0.19$\pm$0.04      & 41   & 0.22$\pm$0.04    & 86   \\
N2                   & 0.0009\%                 & -0.28$\pm$0.02     &110    &-0.38$\pm$0.02        &152     & 0.02\%    & -0.37$\pm$0.03      & 43   & -0.43$\pm$0.01   & 96   \\
log~$\Sigma_{SFR}$    & 0.07\%                   &  -1.1$\pm$0.1      &112    &  -1.3$\pm$0.1        &155     & 0.1\%     &  -0.9$\pm$0.1       & 43   & -1.2$\pm$0.1     & 96    \\     
{[\sii]$\lambda$6717/[\sii]$\lambda$6731}&0.05\%&  1.25$\pm$0.03     &104    &  1.31$\pm$0.03        &127     & 0.7\%    &  1.28$\pm$0.03      & 40   & 1.33$\pm$0.03    & 84   \\ 

$\log$~(N/O)          &0.01\%                     &  -0.49$\pm$0.03    &105    &-0.58$\pm$0.03       &143    & 0.03\%    &  -0.49$\pm$0.04     & 40   & -0.58$\pm$0.03   & 91 \\
12+$\log$~(O/H)       &4.3\%                      &   8.68$\pm$0.01    &107    & 8.69$\pm$0.01       &151    & 7.4\%     &   8.69$\pm$0.01     & 40   &  8.70$\pm$0.01   & 92 \\
$\log$~U              &5.3\%                      &  -3.08$\pm$0.02    &107    &-3.14$\pm$0.02       &151    & 2.5\%     &  -3.09$\pm$0.03     & 40   &  -3.18$\pm$0.02  & 92 \\

\hline
\end{tabular}
\tablefoot{Median values and errors (95\% confidence interval of the median) of the distributions of: $c(\hb)$, $R_{23}$ and N2 gas-phase metallicity indicators, star formation rate per unit area, [\sii]$\lambda$6717/[\sii]$\lambda$6731 line ratio,  oxygen abundance and N/O ratio, and logarithm of the ionisation  parameter. All figures refer to the centres of barred and unbarred galaxies in the sub-samples of non-AGN  and pure star-forming galaxies (Sect.~\ref{agn}). N is the number of galaxies in each sub-sample. The $P$-values of the $k$-sample A-D test for the comparison of distributions for barred and unbarred galaxies are shown for all cases.\\
\tablefoottext{a} A-D test $P$-value or approximate significance level at which the null hypothesis (the two samples are drawn from the same population) can be rejected. Usually, significance levels lower than 5\% are requested to reject the null hypothesis.}\\
\end{table*}

\section{Central oxygen abundance and N/O abundance ratio}
\label{abund}
The faint auroral emission lines are not generally detected in our optical spectra and, consequently, the electron temperature ($T_e$) cannot be calculated. Therefore, it is not  possible to determine abundances from the $T_e$-based or direct method and the oxygen abundance can only be inferred through empirical or theoretical calibrations of nebular strong-line flux ratios, the strong-line methods \citep[see][for in-depth discussions on different methods]{perezmontero_diaz05, ls12}.
We have used two of the most widely used tracers of gas oxygen abundance, $R_{23}\equiv$ ([\oiii]$\lambda\lambda4959,5007$+[\oii]$\lambda\lambda3727$)/\hb, and N2 $\equiv \log($[\nii]$\lambda6583/\ha$).  The $R_{23}$ parameter was first proposed by \cite{pagel79}, but there are many different empirical and theoretical calibrations of this parameter \citep[e.g.][]{P01,pt05,m91,kewley02,kobulnicky}. The N2 was first proposed by \cite{Storchi-Bergmann} and has been extensively used, especially for intermediate  and high redshift studies. This is because the lines involved are  readily observable using the current generation of telescopes, and the line ratio is insensitive to reddening corrections due to the small wavelength separation between the two involved lines. Some of the most common calibrations of N2 include \cite{DTT02}, \cite{pp04}, \cite{nag06} or \cite{pmc09}.

\begin{figure}
\centering
\includegraphics[width=0.52\textwidth]{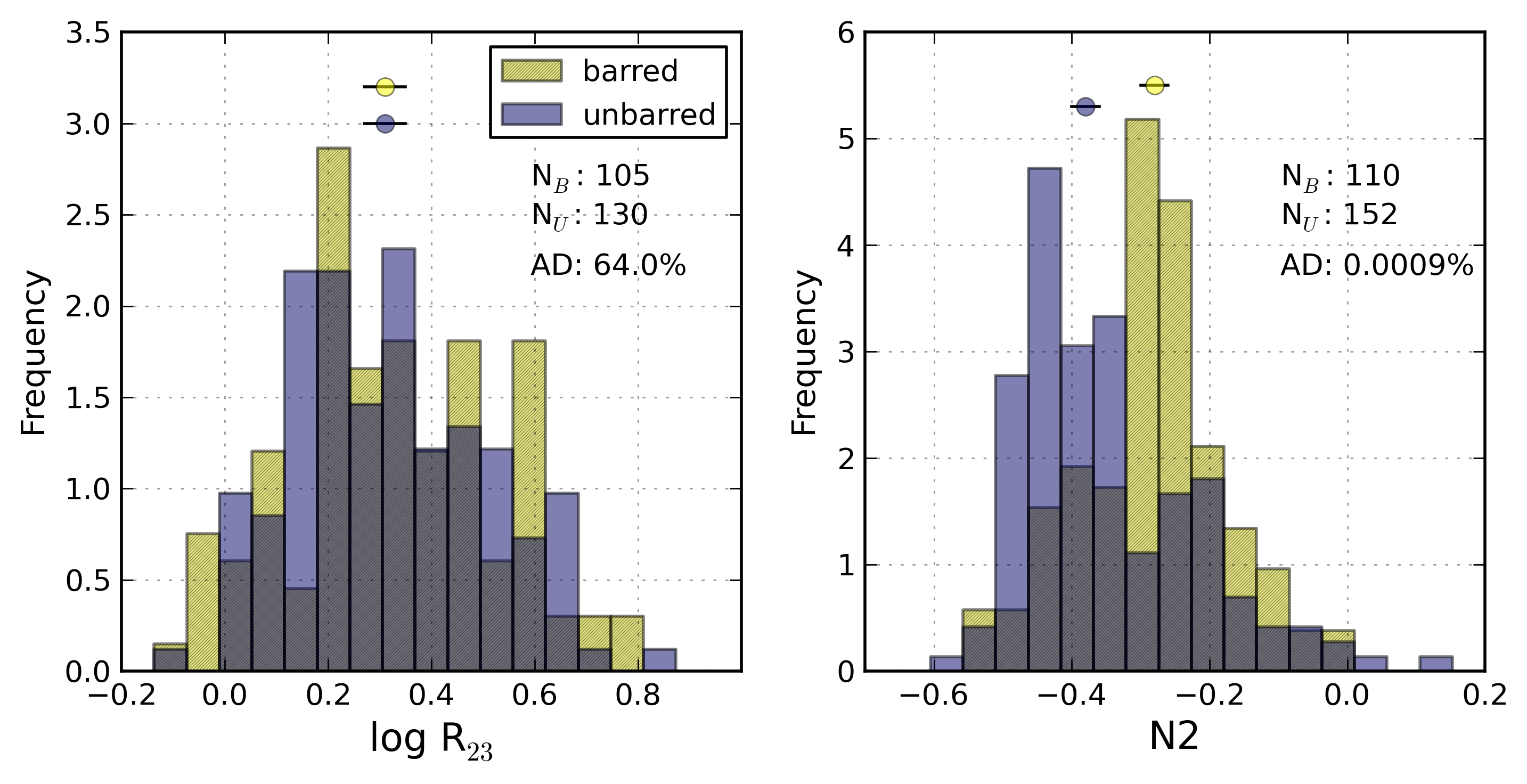}
\caption{\label{R23_N2}Histograms of the values of the oxygen abundance indicators  $R_{23}$=([\oiii]+[\oii])/\hb\ ({\em left}) and N2 = log([\nii]$\lambda 6583/\ha)$ ({\em right}) for all non-AGN barred and unbarred galaxies in the sample. The number of objects in each sub-sample and the $P$-values from the A-D test are shown in each panel.}
\end{figure}

Our aim is to investigate potential differences between barred and unbarred galaxies, and therefore what follows  only compares strong-line oxygen abundance indicators, without applying any of the existing calibrations. Figure~\ref{R23_N2} shows the comparison of the distribution of values for the $R_{23}$ and N2 parameters for barred and unbarred non-AGN galaxies. Barred and unbarred galaxies seem to be indistinguishable in terms of the median value and distribution of the $R_{23}$ parameter. There is, however, a striking difference in N2 where barred galaxies have enhanced log([\nii]$\lambda$6583/\ha) compared with unbarred galaxies by 0.10~dex (see Table~\ref{stats}), with both displaying significantly different distributions  according to the A-D test ($P$-value below 0.001\%). The two distributions of N2 are also different if we only consider star-forming galaxies ($P$-value 0.002\%), and barred galaxies also have a larger average N2 than unbarred galaxies. However,  the barred and unbarred difference in median values shortens to $\sim$0.06~dex  in the star-forming sub-sample, which is, in any case, much larger than the uncertainties in the median values.

It is well known that there is a positive mass-metallicity relation for spirals \cite[e.g.][]{tremonti,bothwell}. We will analyse in detail the dependence of N2 and $R_{23}$ with total stellar mass in Sect.~\ref{trends}. However, it should be noted that the mass distributions of our barred and unbarred sub-samples are identical and therefore the observed difference in N2 might indicate a real and strong physical difference in the centres of barred galaxies compared  with unbarred galaxies. 

\begin{figure}
\centering
\includegraphics[width=0.3\textwidth]{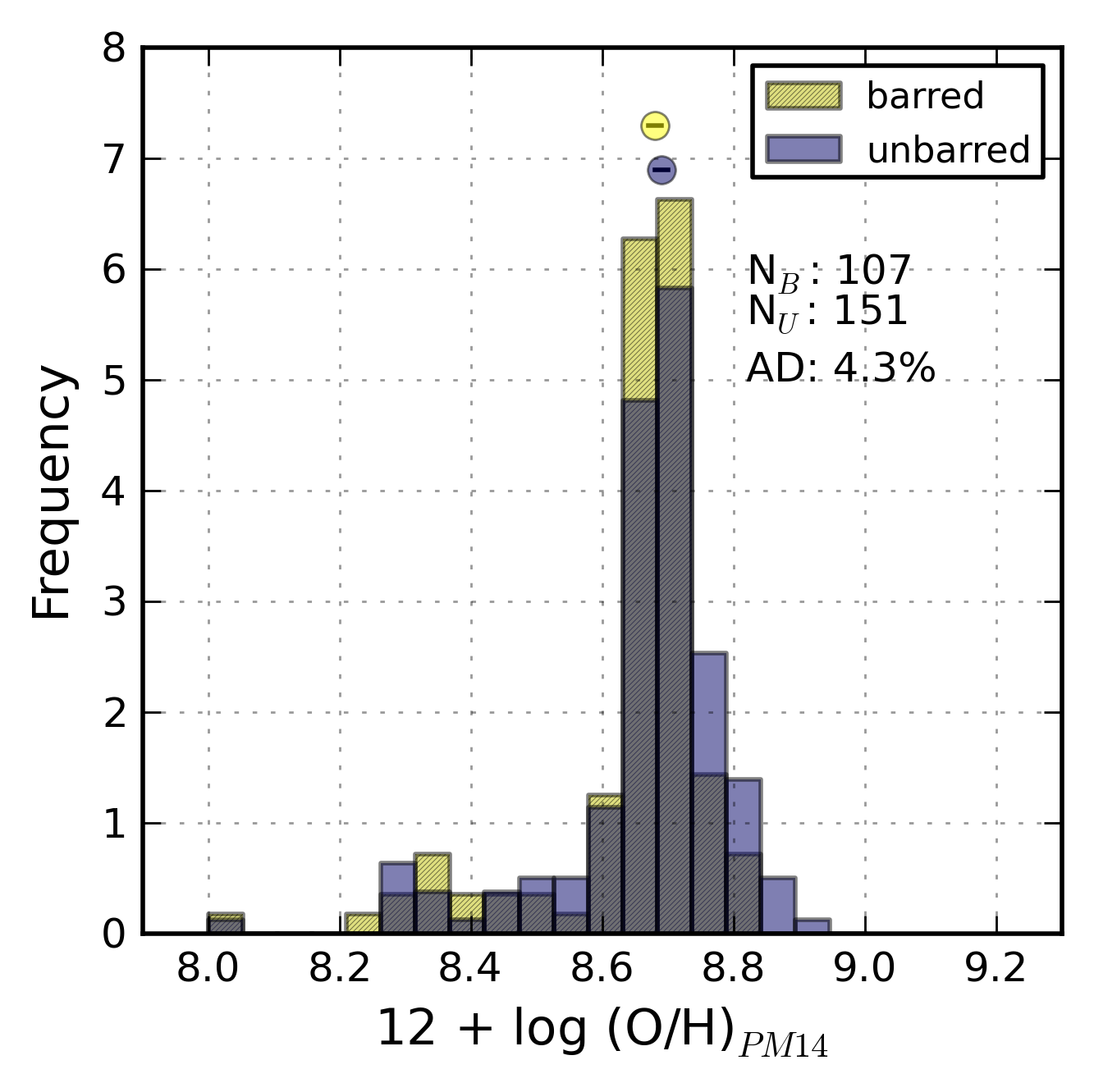}
\includegraphics[width=0.3\textwidth]{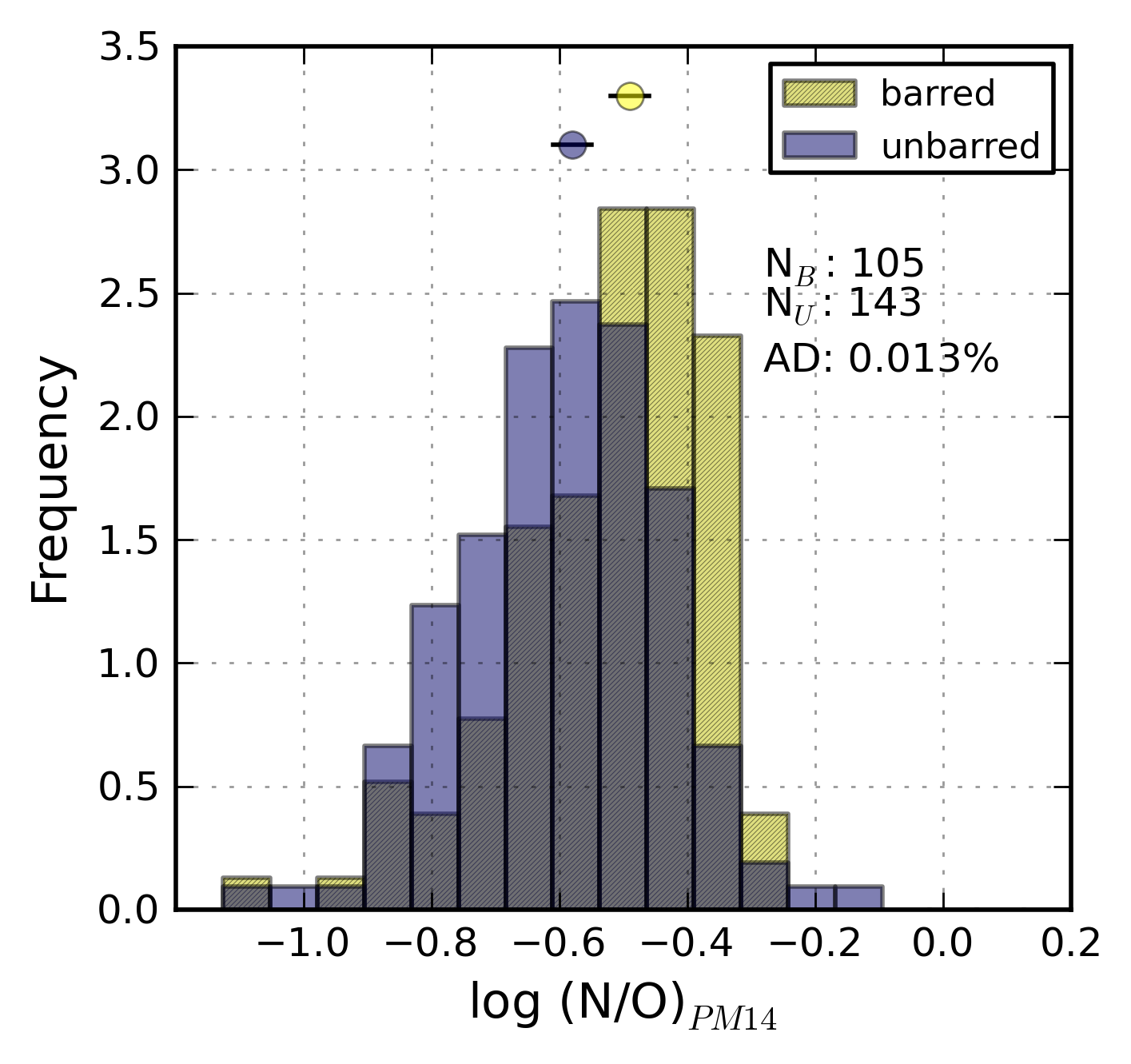}
\includegraphics[width=0.3\textwidth]{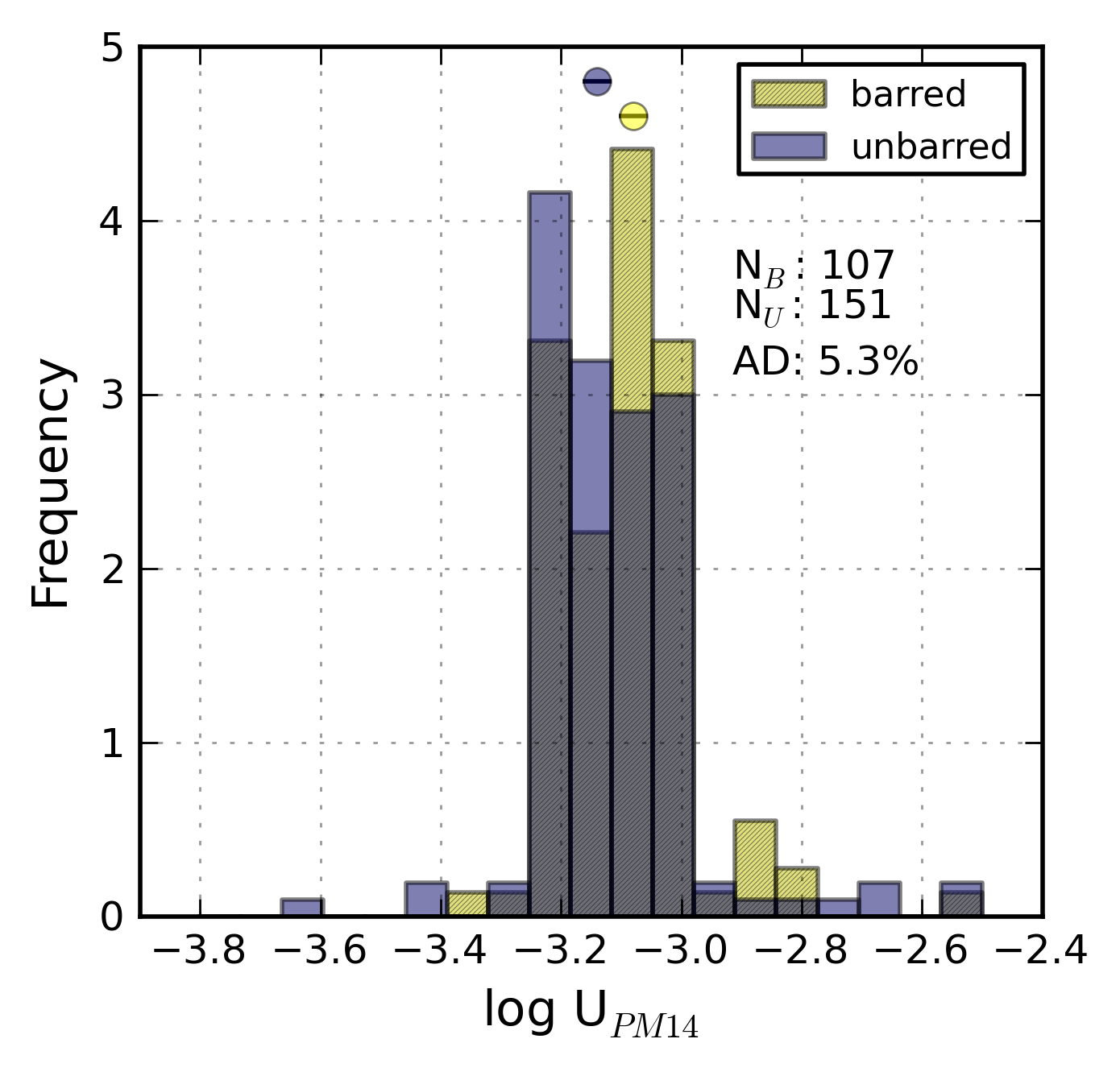}
\caption{\label{O_abund}Histograms of the oxygen abundance ({\em top}), N/O ratio ({\em middle}) and ionisation parameter  ($U$, {\em bottom}),  as calculated with the \cite{epm14} method for all non-AGN barred and unbarred galaxies in the sample. The number of objects in each sub-sample and the $P$-values from the A-D test are shown below the legend. Only non-AGN galaxies are considered.}
\end{figure}

Given that there are no differences between barred and unbarred galaxies in terms of the  $R_{23}$ parameter, any empirical calibration of this parameter yields no difference in oxygen abundance between barred and unbarred galaxies. The opposite happens with calibrations that make use of N2, except for the N2 empirical calibration by \cite{pmc09}, which includes a correction to account for the N/O ratio. We have estimated $\log$~(N/O) from the empirical calibration of the N2S2 ratio (=$\log($[\nii]$\lambda6583$/[\sii]$\lambda\lambda6717,6731$) from the same authors. Afterwards, 12 + $\log$(O/H) is calculated. We have also estimated $\log$~(N/O), 12 +$\log$(O/H) and the ionisation parameter\footnote{Regarding the ionisation parameter values derived from the grid of photoionisation models from \cite{epm14}, it is important to stress that, in absence of emission-line ratios sensitive to the electron temperature, an empirical law between $\log$~U and 12+$\log$(O/H) is assumed (i.e. lower U for higher O/H). Nevertheless, in the grid a certain range of variation in $\log$~U is allowed for each O/H value that exceeds the expected and reported $\log$~U variations in our sample of studied objects. As such, we think that the resulting $\log$~U values give an accurate idea of the variations.}, U, from the new grids of photoionisation models by \cite{epm14}. The results are shown in Table~\ref{stats} and in Fig.~\ref{O_abund}.
The barred and unbarred sub-samples of  galaxies have the same average oxygen abundance within errors, 12+$\log$(O/H) = 8.69$\pm$0.01, and the distributions are indistinguishable from each other: we obtain A-D $P$-values of 4.3\% and 7.4\%  when we consider non-AGN or star-forming galaxies, respectively. This agrees with recent results from \cite{cacho} for a similar sample of barred and unbarred nearby  disc galaxies from the \cite{nair} catalogue. However, our result apparently conflicts with \cite{ellison}, who report a larger oxygen abundance (by $\sim$0.06~dex) in barred  galaxies compared to unbarred galaxies, in relation to a galaxy sample also extracted from \cite{nair}. In  Sect.~\ref{discussion} we discuss these results and try to explain the possible sources of discrepancy between their results and ours.

However, as expected from the N2 values,  we find a statistically significant and interesting difference in the central N/O abundance ratio found in barred and unbarred galaxies. The median $\log$~(N/O) is 0.09~dex larger in barred than in unbarred galaxies, for both the non-AGN  and the star-forming samples. This difference is three times larger than the uncertainties in the median values ($\sim$0.03 dex). The A-D test confirms that the barred and unbarred distributions are different (mid-panel in Fig.~\ref{O_abund}), with $P$-values of 0.01\% and 0.03\% for the  non-AGN  and the star-forming samples, respectively. To our knowledge, this is the most obvious and largest difference in physical properties of the gas so far observed between barred and unbarred galaxies. 

\section{Star formation rate}
\label{SFR}
We have estimated the star formation rate (SFR) in the galaxy centres from the \ha\ extinction-corrected emission within the 3\arcsec\ SDSS fibres and the \cite{kennicutt98} conversion factor\footnote{The stellar component had already been subtracted from the original SDSS spectra and, therefore, we do not need to correct for stellar absorption under the Balmer lines at this point.}. The distances to the galaxies have been estimated from the redshift  given by SDSS and a Hubble constant of 72 km~s\me~Mpc\me.

Given the redshift range of the galaxy sample, the SDSS 3\arcsec\ diameter fibre corresponds to a projected size ranging from $\sim$1.2 to 4.2~kpc. Although the redshift range is small and the redshift distributions of barred and unbarred galaxies are similar,  we have derived star formation rates per unit area, $\Sigma_{SFR}$, to average out possible dependencies on galaxy distance.

\begin{figure}
\centering
\includegraphics[width=0.75\columnwidth]{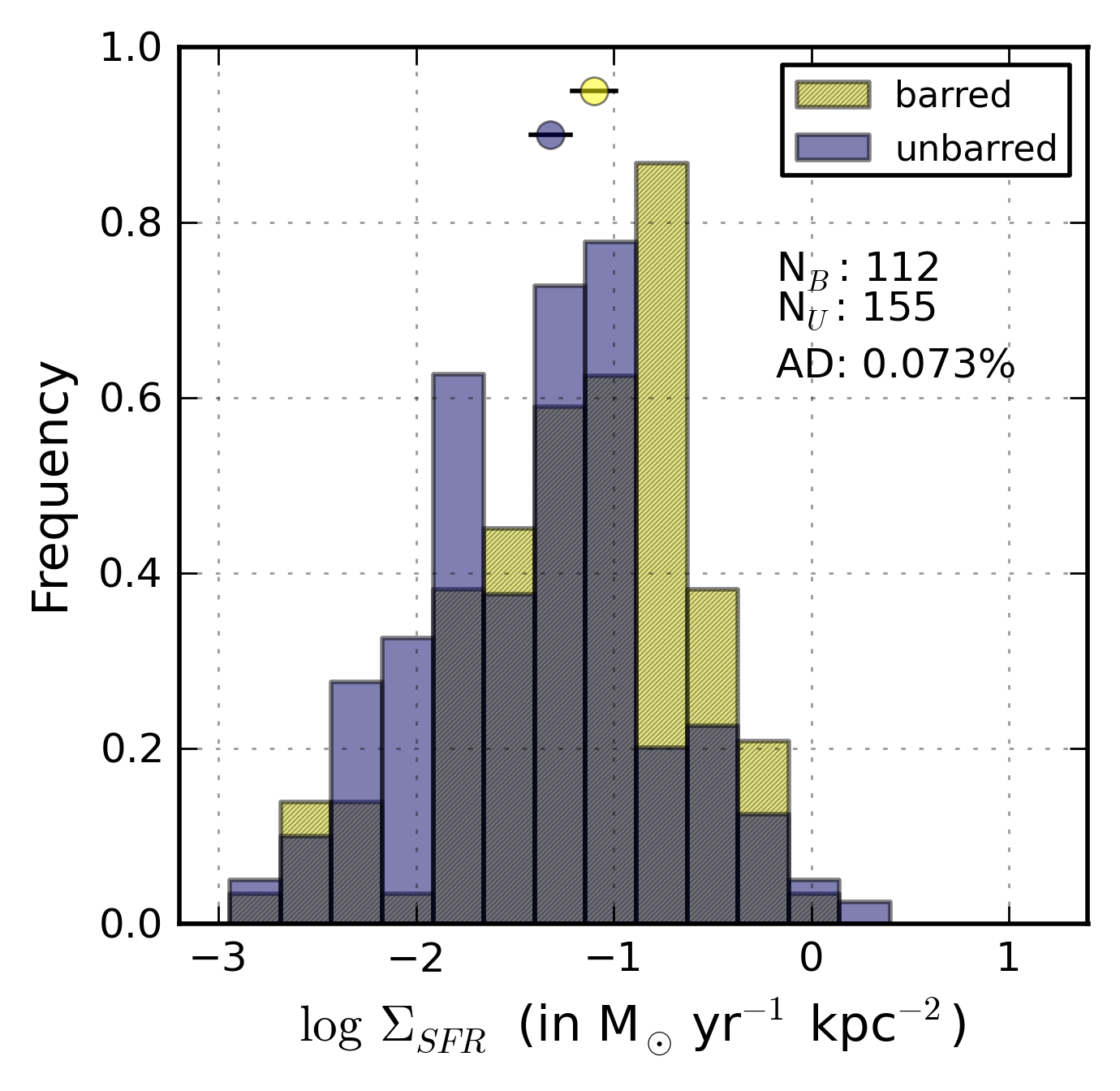}
\caption{\label{SFR_plot}Comparative histograms of star formation rate per unit area for all non-AGN barred (yellow hatched) and unbarred (purple) galaxies of the sample.}
\end{figure}

The distributions of the logarithm $\Sigma_{SFR}$ for barred and unbarred galaxies are shown in Fig.~\ref{SFR_plot}. Both distributions are different according to the A-D test ($P$-value $<0.1$\%), with barred galaxies showing marginally larger star formation rate per unit area. This difference is slightly larger when considering only star-forming galaxies, where barred galaxies show a median log~$\Sigma_{SFR}$ which is $\sim$0.3~dex larger than unbarred galaxies.
A larger SFR in the centres of barred galaxies was also reported by \cite{ellison} for galaxies of similar total stellar mass in a similar study, and by other authors \citep[see e.g.][]{hummel90,Ho_barras,wang2012,ooy,zhou}. 

\section{Electron density}
\label{edensity}
\begin{figure}
\centering
\includegraphics[width=0.75\columnwidth]{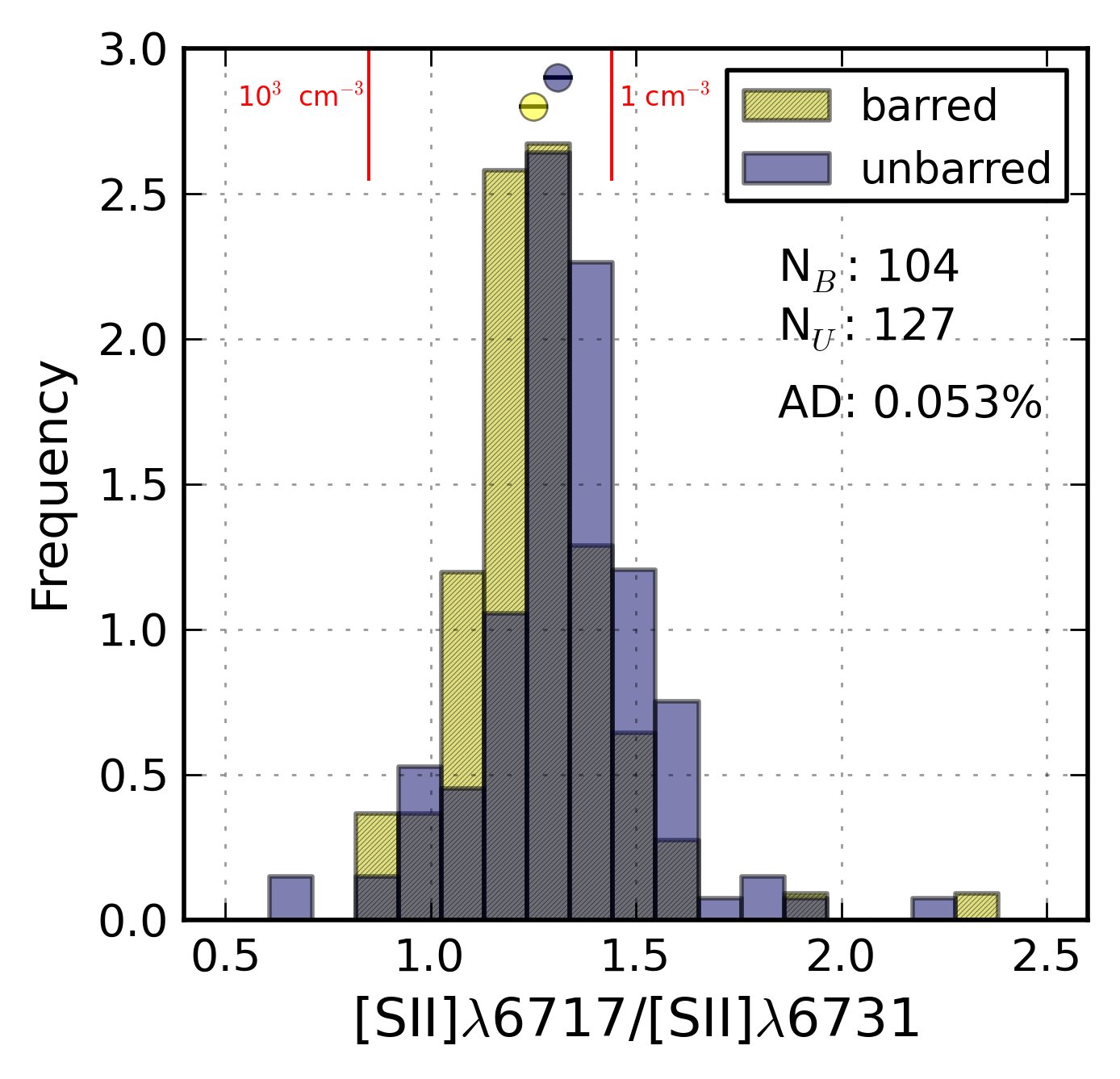}
\caption{\label{hist_sii_all}Comparative histograms of values of the [\sii]$\lambda$6717/[\sii]$\lambda$6731 line ratio for barred  (yellow hatched) and unbarred  (purple) galaxies separately. Only non-AGN galaxies are considered.}
\end{figure}

The [\sii]$\lambda$6717/[\sii]$\lambda$6731 line ratio is sensitive to the electron density (N$_e$). Nearly 8\% of all galaxies in our sample have a [\sii]$\lambda$6717/[\sii]$\lambda$6731 line ratio  which is above the low-density theoretical limit \citep[=1.43, implying N$_e\sim$1cm$^{-3}$, for T$_e$=10000K;][]{temden}. Most of these targets have the largest [\sii]$\lambda$6717/[\sii]$\lambda$6731  relative errors ($\sim$20-40\%). Here, we only consider galaxies where their  [\sii]$\lambda$6717/[\sii]$\lambda$6731 line ratio value is still compatible with being below the low-density theoretical limit, taking into account their corresponding error bars \citep{lopez-hernandez}.
The median value of this line ratio in the centres of the non-AGN galaxy sample is 1.29$\pm$0.02  (1.30$\pm$0.02 for pure SF galaxies), which corresponds to N$_e\sim100-150$~cm$^{-3}$. This is in agreement with typical values found  in centres of galaxies measured previously by \cite{KKB} and \cite{Ho_hii}.
In Fig.~\ref{hist_sii_all} we compare the distributions  [\sii]$\lambda$6717/[\sii]$\lambda$6731 for barred and unbarred non-AGN galaxies. The median value is marginally lower for barred (1.25$\pm$0.03) than for unbarred galaxies (1.31$\pm$0.03), indicating that, on average, barred galaxies tend to posses a larger central electron density than unbarred galaxies. The A-D test yields a very low $P$-value (0.05\%),  which indicates that both distributions are different.  Distributions are also different in the sub-sample of pure star-forming galaxies ($P$-value of 0.7\%,  see Table~\ref{stats}).

\begin{figure*}
\centering
\includegraphics[width=0.96\columnwidth]{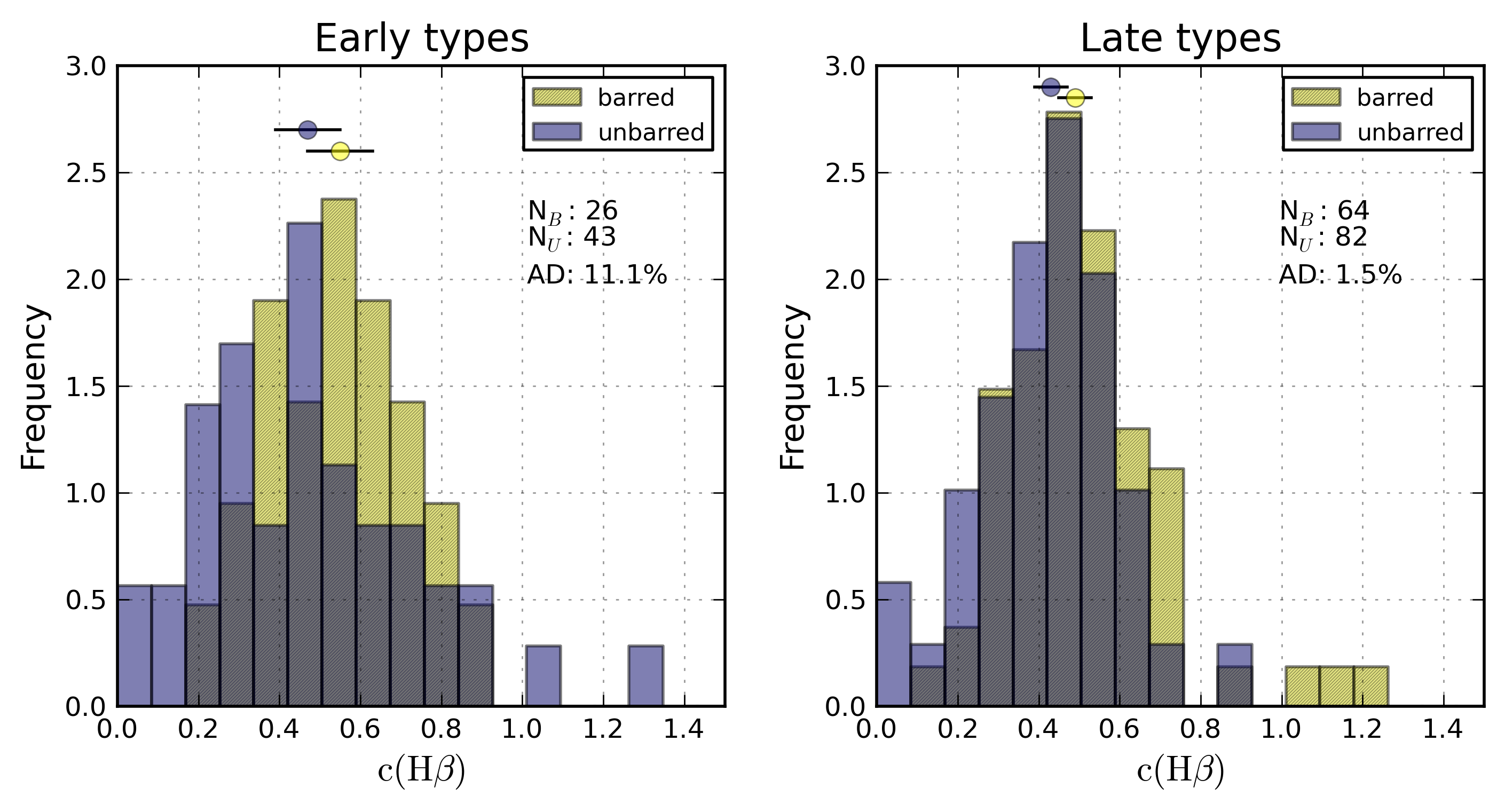}
\includegraphics[width=0.96\columnwidth]{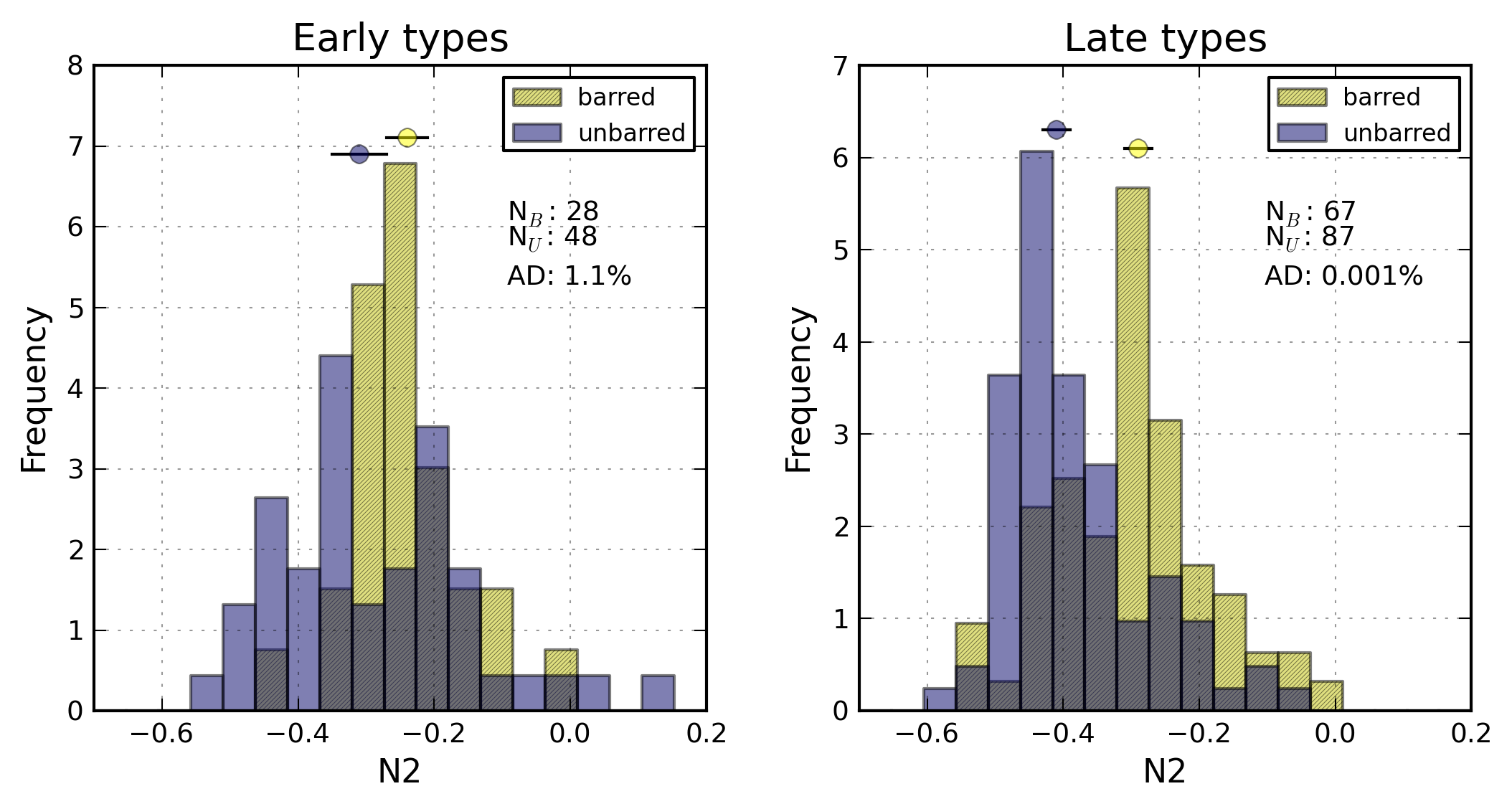}
\includegraphics[width=0.96\columnwidth]{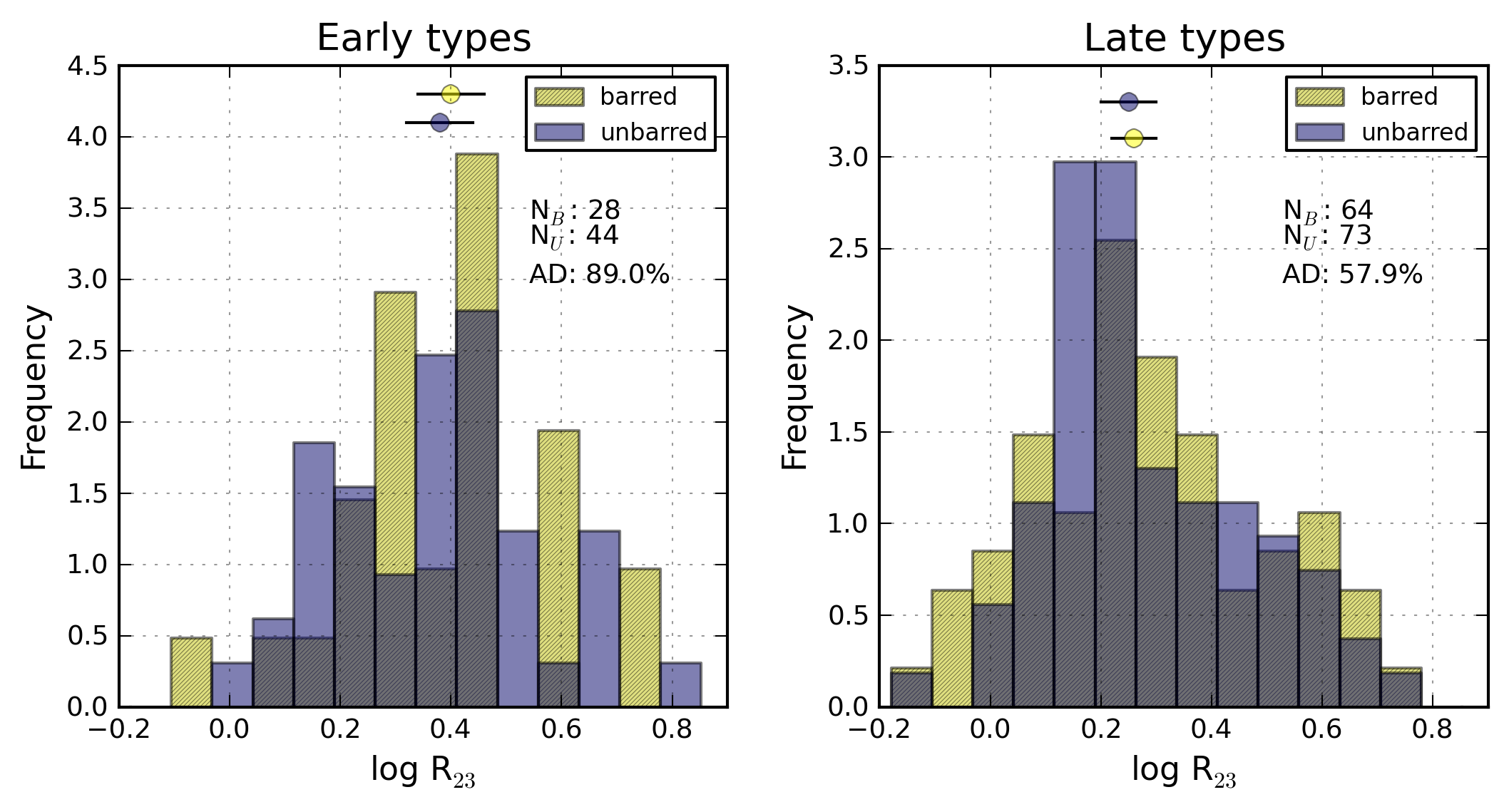}
\includegraphics[width=0.96\columnwidth]{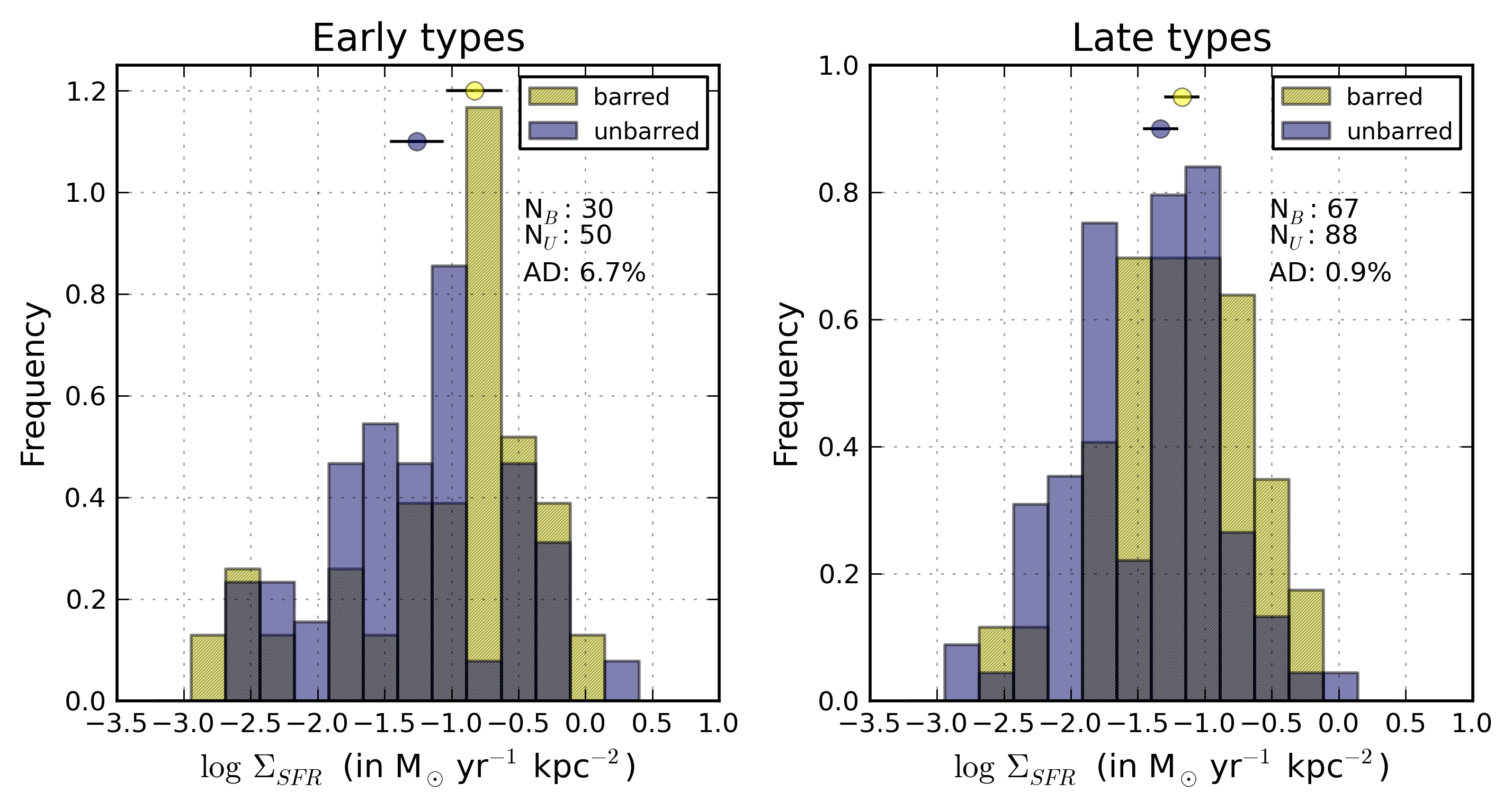}
\includegraphics[width=0.96\columnwidth]{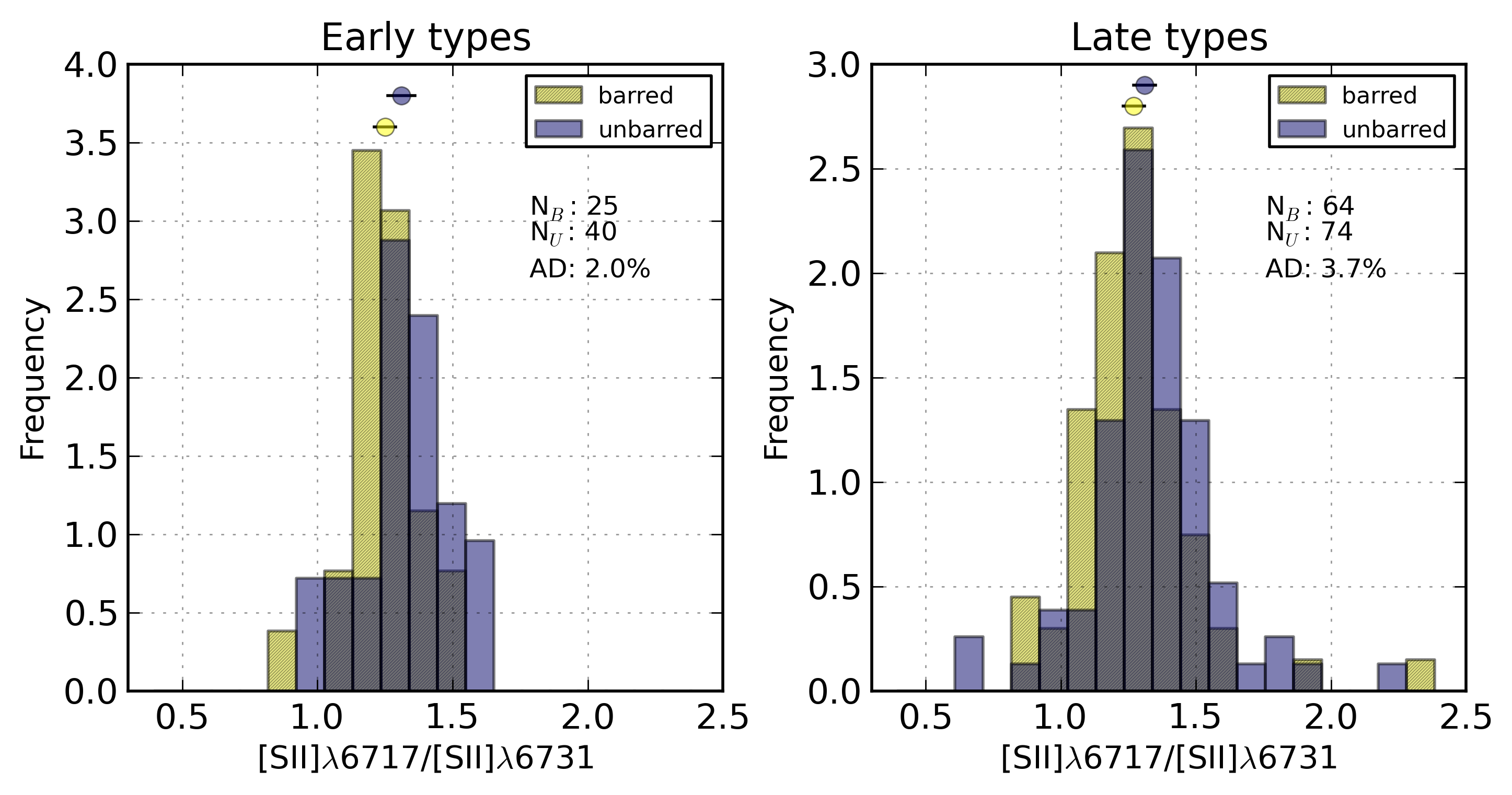}
\includegraphics[width=0.96\columnwidth]{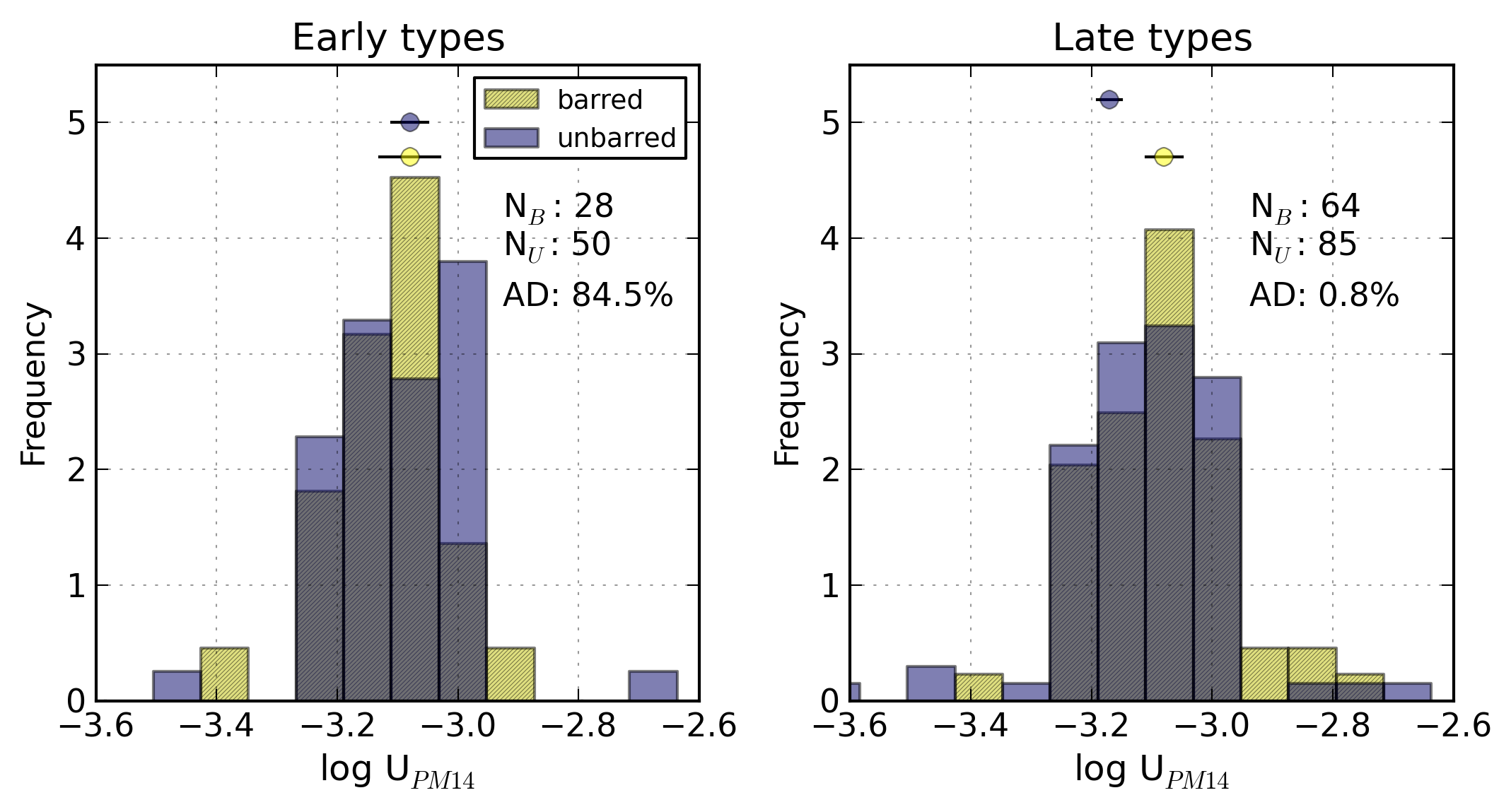}          
\includegraphics[width=0.96\columnwidth]{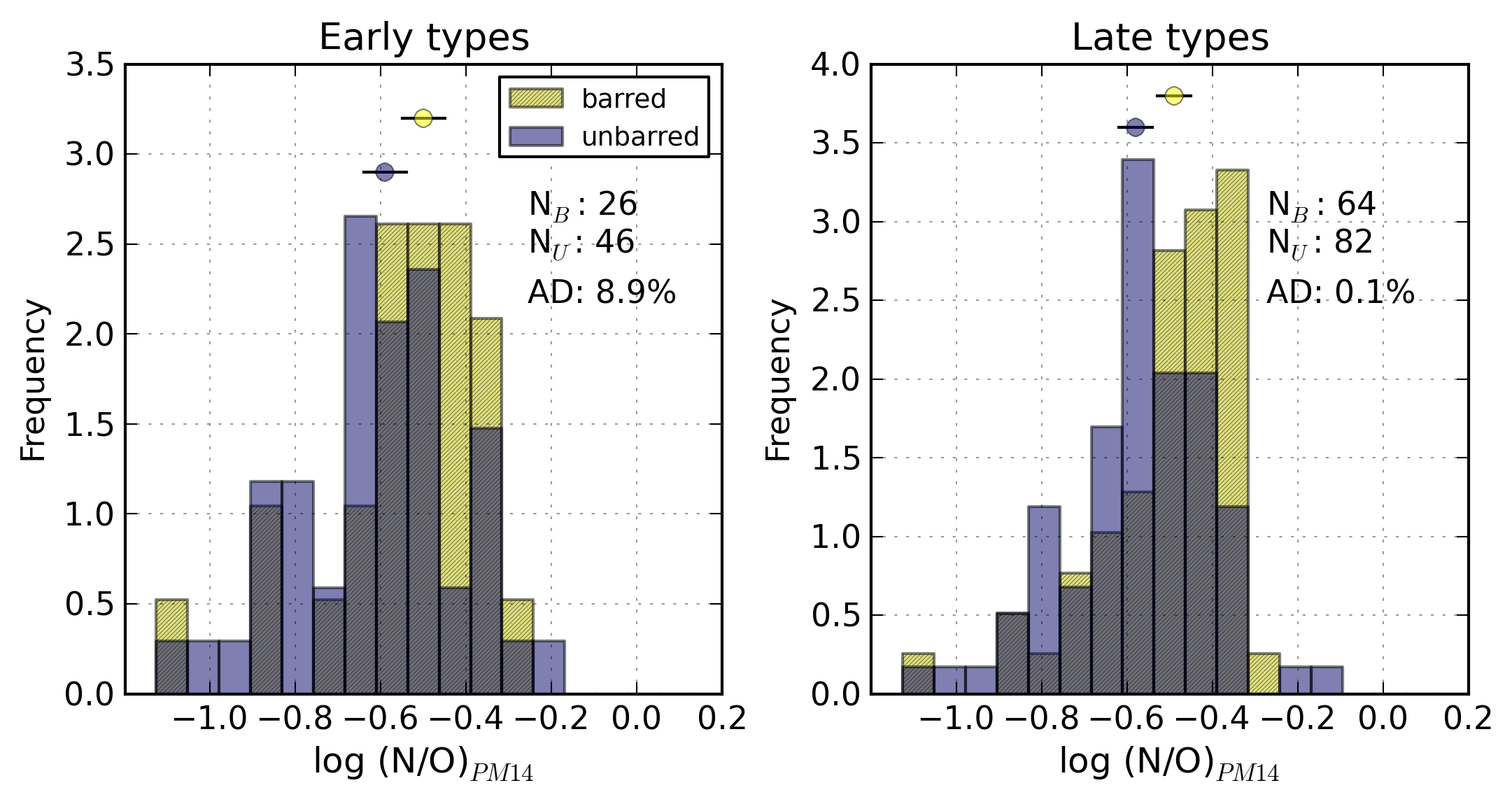}          
\includegraphics[width=0.96\columnwidth]{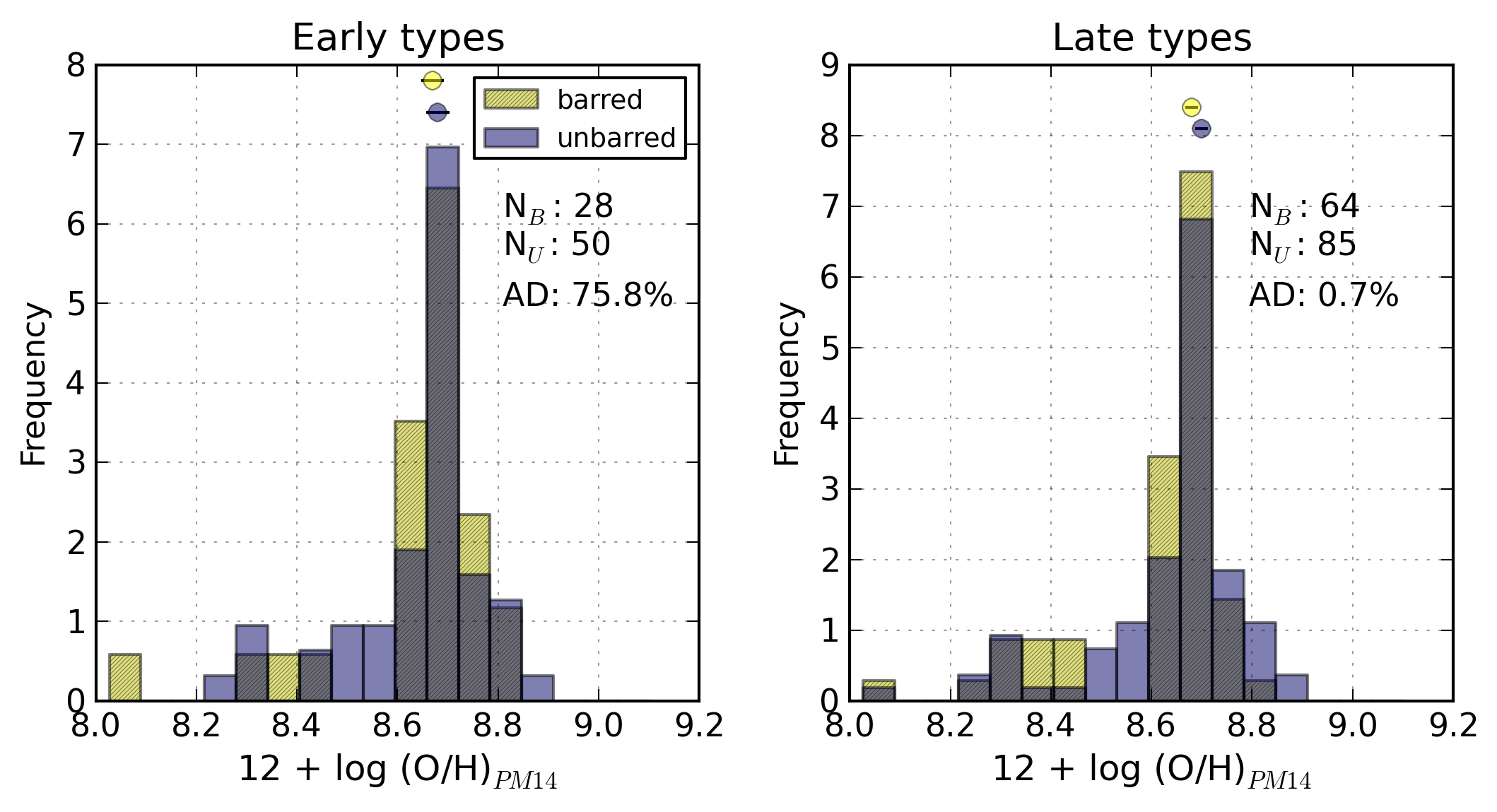}     
\caption{Comparative histograms of (from top to bottom and from left to right): the Balmer extinction at the \hb\ emission line, $c(\hb)$, N2, $\log$ R$_{23}$,  decimal logarithm of the SFR per unit area, [\sii]$\lambda$6717/[\sii]$\lambda$6731 line ratio, logarithm of the ionisation parameter, N/O abundance ratio and oxygen abundance for barred  (yellow, hatched) and unbarred  (purple) galaxies separately for  early- and  late-type galaxies. The separation between early- and late-type galaxies was carried out using the B/D luminosity ratio as described in Sect.~\ref{morfo} for the non-AGN sub-sample of galaxies.}
\label{early_late}%
\end{figure*}


\section{Dependence of central differences on Hubble type}
\label{morfo}
The amount of gas driven by a bar towards the galaxy centre might depend upon, amongst others, the availability of gas in galaxy disc and on the gravitational torque created by the bar-like mass distribution. Gas content varies across the Hubble sequence,  but also bars in early-type galaxies tend to be longer and stronger than in late-type spirals \citep{erwin,chapelon}. Our sub-samples of non-AGN galaxies are not similar in terms of B/D flux ratio and bulge mass  (Fig.~\ref{sample_noAGN}), which might indicate that both sub-samples are different in the relative population of morphological types. We next explore whether any of the differences between barred and unbarred galaxies found in previous sections could be due to a bias towards earlier types in the sub-sample of barred galaxies.

The fact that most of the differences between barred and unbarred galaxies of the non-AGN sub-sample are also observed in the pure star-forming sub-sample, where the distribution of bulge mass and bulge-to-disc flux ratio are equivalent for barred and unbarred galaxies (Fig.~\ref{sample_SF}),  seems to indicate that, even if a bias were present, bars tend to produce higher dust concentrations, larger N/O ratios, star formation, and electron densities in the centre of galaxies.  A deeper analysis is, in any case, necessary to clarify the nature of these differences.
\begin{table*}
\caption{\label{stats_early_late} Statistics for the sub-samples of early- and late-type barred and unbarred galaxies.} 
\centering
\begin{tabular}{l|c|cc|cc||c|cc|cc}
\hline \hline   
             &\multicolumn{5}{c||}{Early-type} &\multicolumn{5}{c}{Late-type}\\
             &\multicolumn{5}{c||}{(based on B/D)} &\multicolumn{5}{c}{(based on B/D)}\\
\hline
                     & $P$-value & \multicolumn{2}{c|}{barred}&  \multicolumn{2}{c||}{unbarred}& P-value  & \multicolumn{2}{c|}{barred}&  \multicolumn{2}{c}{unbarred}\\
                     &                          & median             & N     &     median           & N      &           &  median             & N    &  median         & N    \\
\hline
$c(\hb)$             & 11.1\%                   & 0.55$\pm$0.08      &26    & 0.47$\pm$0.08        &43     & 1.5\%      &  0.49$\pm$0.04      & 64    & 0.43$\pm$0.04    & 82  \\
$\log$~R$_{23}$       & 89.0\%                   & 0.40$\pm$0.06      &28    & 0.38$\pm$0.06        &44     & 57.9\%     &  0.26$\pm$0.04      & 64    & 0.25$\pm$0.05    & 73  \\
N2                   & 1.1\%                    & -0.24$\pm$0.03     &28    &-0.31$\pm$0.04        &48     & 0.001\%    & -0.29$\pm$0.03      & 67    & -0.41$\pm$0.01   & 87  \\
log~$\Sigma_{SFR}$    & 6.7\%                    & -0.8$\pm$0.2       &30    &-1.3$\pm$0.2          &50     & 0.92\%     &  -1.17$\pm$0.12     & 67    & -1.33$\pm$0.12   & 88  \\     
{[\sii]$\lambda$6717/[\sii]$\lambda$6731}&2.0\% &  1.25$\pm$0.04     &25    & 1.31$\pm$0.05        &40     & 3.7\%      &  1.27$\pm$0.04      & 64    & 1.31$\pm$0.04    & 74  \\ 
$\log$~U              &85\%                     & -3.08$\pm$0.05     &28    &-3.08$\pm$0.03        &50     & 0.8\%      &  -3.08$\pm$0.03     & 64    & -3.17$\pm$0.02   & 85  \\
$\log$~(N/O)          &8.9\%                    & -0.50$\pm$0.05     &26    &-0.59$\pm$0.05        &46     & 0.05\%     &  -0.49$\pm$0.04     & 64    & -0.58$\pm$0.04   & 82  \\
12+$\log$~(O/H)       &76\%                     &  8.67$\pm$0.02     &28    & 8.68$\pm$0.02        &50     & 0.7\%      &   8.68$\pm$0.01     & 64    &  8.70$\pm$0.01   & 85 \\

\hline
\end{tabular}
\tablefoot{Median values and errors (95\% confidence interval of the median) of the distributions of $c(\hb)$, R$_{23}$ and N2 gas metallicity indicators, star formation rate per unit area, [\sii]$\lambda$6717/[\sii]$\lambda$6731 line ratio, logarithm of the ionisation parameter, N/O abundance ratio and oxygen abundance in the centres of barred and unbarred galaxies in the sub-samples of early and late-type galaxies as defined in Sect.~\ref{morfo}  and number of galaxies in each sub-sample. Only non-AGN galaxies are considered. The $P$-values of the $k$-sample A-D test for the comparison of distributions for barred and unbarred galaxies is shown for all cases.}
\end{table*}
\begin{table*}
\caption{\label{stats_bulge_mass} Same as Tables~\ref{stats} and \ref{stats_early_late} except for the sub-samples of low and high stellar mass bulges.}
\centering
\begin{tabular}{l|c|cc|cc||c|cc|cc}
\hline \hline   
             &\multicolumn{5}{c||}{M$_{bulge}< 10^{9.7}$ M$_\odot$} &\multicolumn{5}{c}{ M$_{bulge} \geqslant 10^{9.7}$ M$_\odot$}\\
\hline
                     & $P$-value & \multicolumn{2}{c|}{barred}&  \multicolumn{2}{c||}{unbarred}& P-value  & \multicolumn{2}{c|}{barred}&  \multicolumn{2}{c}{unbarred}\\
                     &                          & median             & N     &     median           & N      &           &  median             & N    &  median         & N \\
\hline
$c(\hb)$              & 0.2\%                    & 0.46$\pm$0.03      &50    & 0.42$\pm$0.04        &68     & 58\%     &  0.52$\pm$0.07      & 54   & 0.50$\pm$0.06    & 72 \\
$\log$~R$_{23}$        & 49\%                     & 0.31$\pm$0.05      &49    & 0.25$\pm$0.05        &56     & 67\%     &  0.34$\pm$0.06      & 56   & 0.35$\pm$0.05    & 74 \\
N2                    & 0.001\%                  & -0.30$\pm$0.03     &54    &-0.43$\pm$0.02        &68     & 3.0\%    & -0.26$\pm$0.03      & 56   &-0.33$\pm$0.04    & 84 \\
log~$\Sigma_{SFR}$     & 0.45\%                   & -1.15$\pm$0.13     &54    &-1.37$\pm$0.12        &68     & 2.6\%    & -0.99$\pm$0.17      & 58   &-1.29$\pm$0.16    & 87 \\     
{[\sii]$\lambda$6717/[\sii]$\lambda$6731}&0.97\% &  1.26$\pm$0.03     &53    & 1.34$\pm$0.05        &57     & 2.0\%    &  1.25$\pm$0.05      & 51   &1.31$\pm$0.03     & 70 \\ 
$\log$~U              &2.9\%                      &  -3.08$\pm$0.03    &51    & -3.18$\pm$0.02       &65     & 54\%      &  -3.07$\pm$0.03     & 56   &  -3.09$\pm$0.03  & 86 \\
$\log$~(N/O)          &0.3\%                      &  -0.51$\pm$0.04    &52    & -0.63$\pm$0.04       &63     & 0.3\%     &  -0.47$\pm$0.04     & 53   & -0.56$\pm$0.03   & 80 \\
12+$\log$~(O/H)       &0.9\%                      &   8.69$\pm$0.01    &51    & 8.70$\pm$0.02        &65     & 43\%      &   8.67$\pm$0.02     & 56   &  8.67$\pm$0.01   & 86 \\
\hline
\end{tabular}
\end{table*}

Morphological classification is only available for $\sim$46\% of the total galaxy sample, i.e. 245 galaxies, of which 113 are barred and 132 unbarred. This classification comes from the  \cite{nair} catalogue. However, morphological decomposition of the galaxy sample has already been performed using the code BUDDA v2.1\footnote{http://www.sc.eso.org/$\sim$dgadotti/budda.html} \citep{buda,dimitri08,dimitri_morpho} that is able to fit up to four different galactic components (a bulge with a S\`ersic profile, a single or double exponential disc, a S\`ersic bar, and a Moffat  central component) to a galaxy image. The morphological decomposition of the galaxies from this sample was performed over SDSS $g$, $r$, and $i$-band images \citep{dimitri_morpho} and the  B/D flux ratios for the three photometric bands were therefore obtained.

The behaviour of the B/D flux ratio along the Hubble sequence is well known.  However, this relation has a high dispersion and  varies significantly, depending on the function used to fit the bulge ($R^{1/4}$ vs S\`ersic  $R^{1/n}$) on the photometric band or on the galaxy inclination and dust-extinction \citep[see][and references therein]{alister}. We have analysed the dependence of the logarithm of the $g$, $r$, and $i$-band  B/D  light ratios obtained with BUDDA  \citep{dimitri_morpho} with the T-type by  \cite{nair} for the galaxies in common between both samples. From these relations and additional information from the  Galaxy Zoo 2\footnote{http://data.galaxyzoo.org} catalogue \citep{GZ2}, we have separated all galaxies in our sample into early- (T-type$<$2) and late-types (T-type$\geqslant$2). For more information on this separation, please see Appendix~\ref{morpho-class}.

The bulge stellar light and, therefore, its mass has been claimed as a key parameter that might distinguish galaxies across the Hubble sequence \citep[e.g.][and references therein]{meisels,alister}. In addition to the  B/D light ratios we have also used the bulge mass \citep{coelho} to explore whether the observed differences between barred and unbarred galaxies are dependent on Hubble types. We have separated all non-AGN galaxies into two groups  according to their bulge mass. The dividing bulge mass ($10^{9.7}$~M$_\sun$) has been selected to ensure that the total stellar galaxy mass distributions for barred and unbarred galaxies are equivalent in both bulge mass bins (with A-D $P$-values greater than 10\%).

\begin{figure*}
\centering
\subfigure{\label{masa_bulbo_a}\includegraphics[width=0.98\columnwidth]{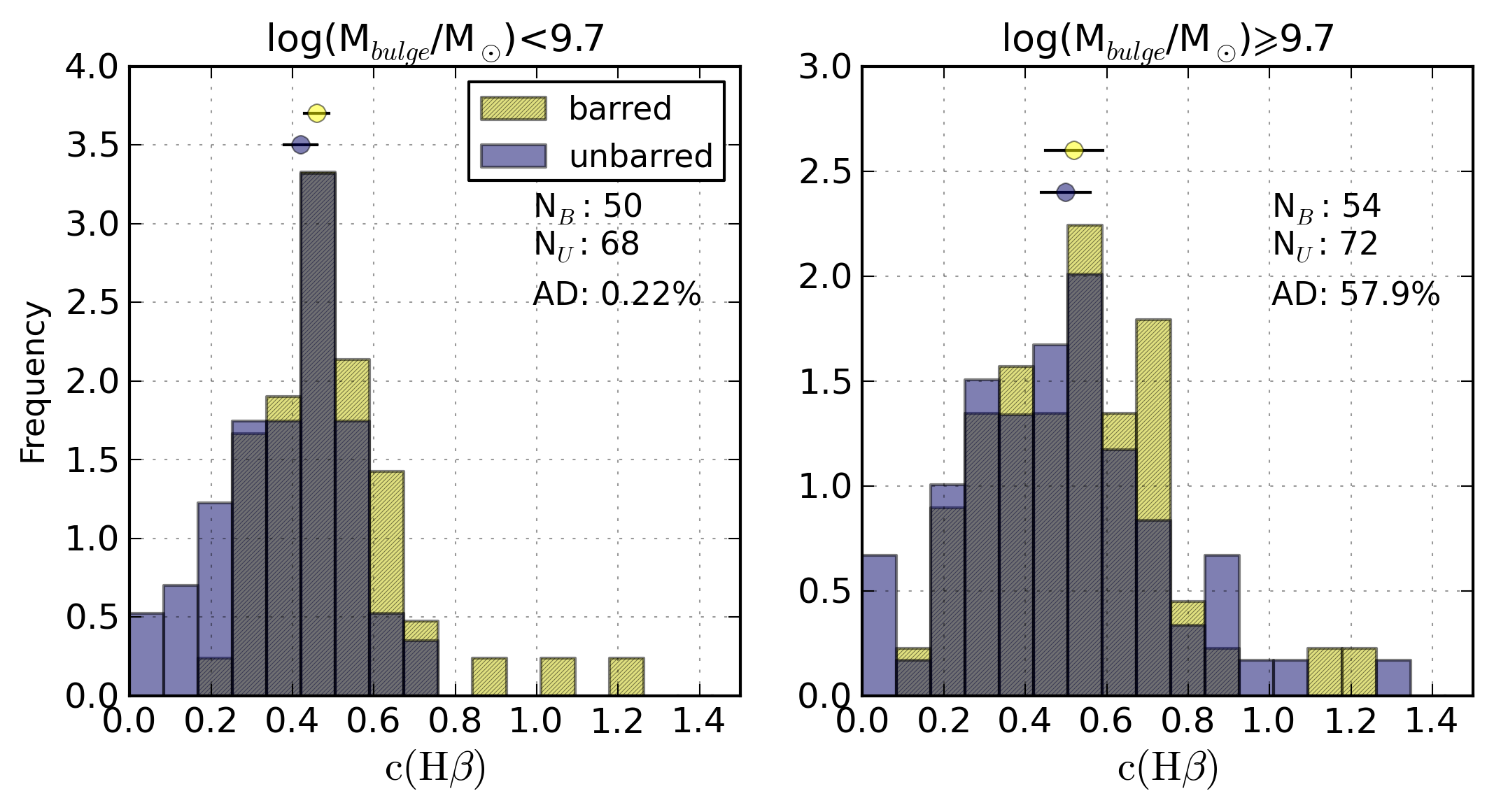}}        
\subfigure{\label{masa_bulbo_b}\includegraphics[width=0.98\columnwidth]{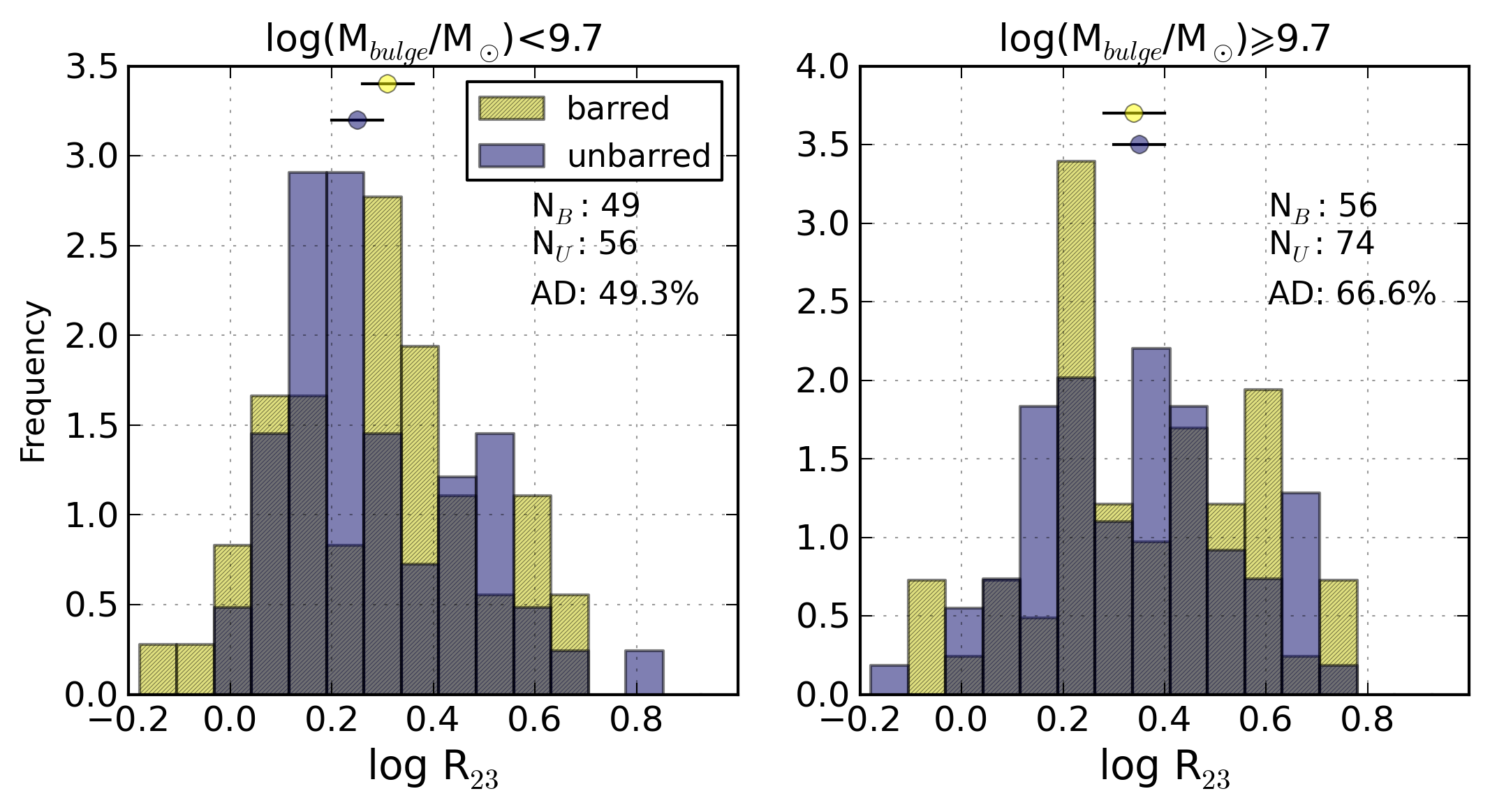}}              
\subfigure{\label{masa_bulbo_c}\includegraphics[width=0.98\columnwidth]{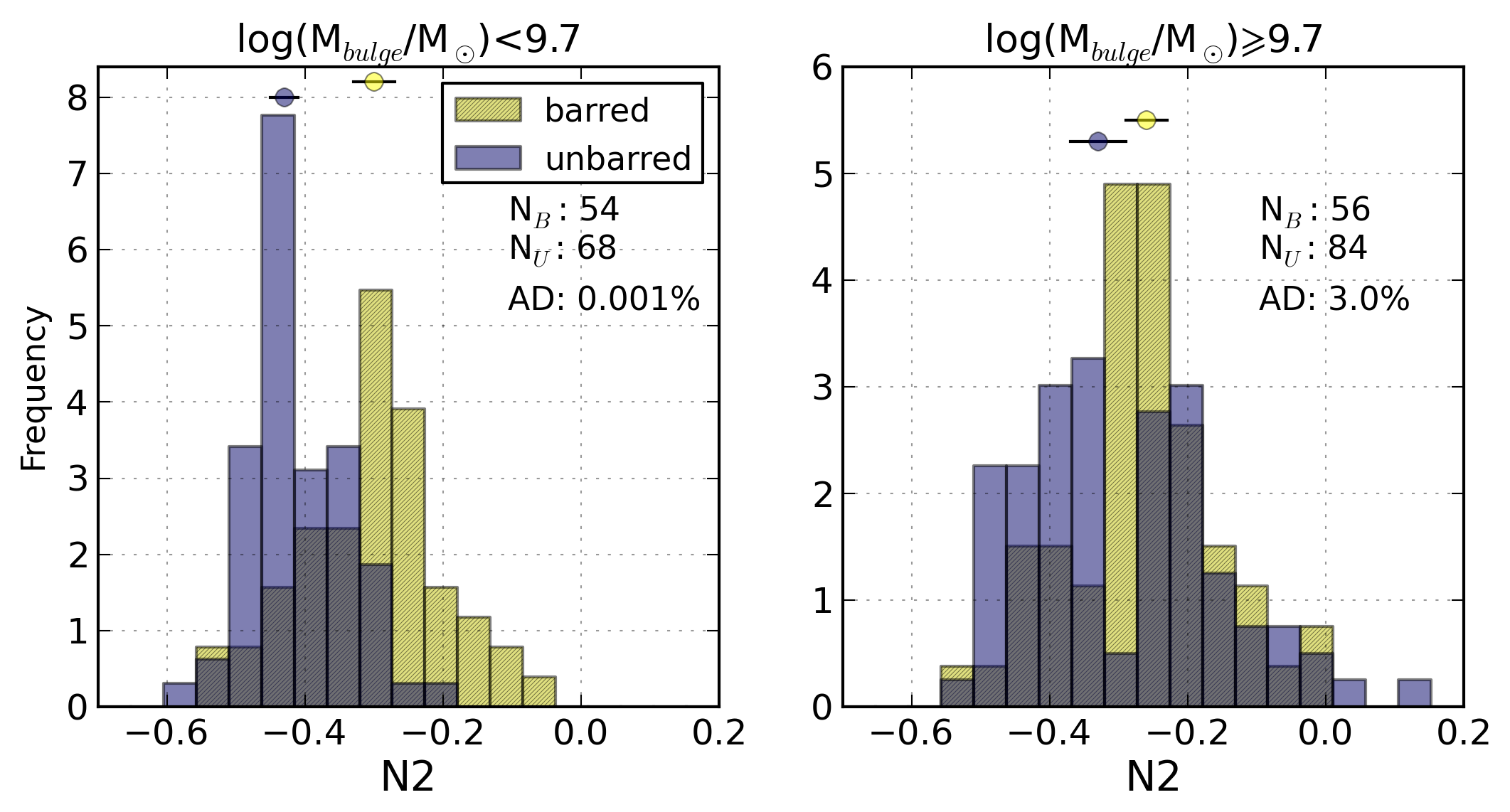}}               
\subfigure{\label{masa_bulbo_d}\includegraphics[width=0.98\columnwidth]{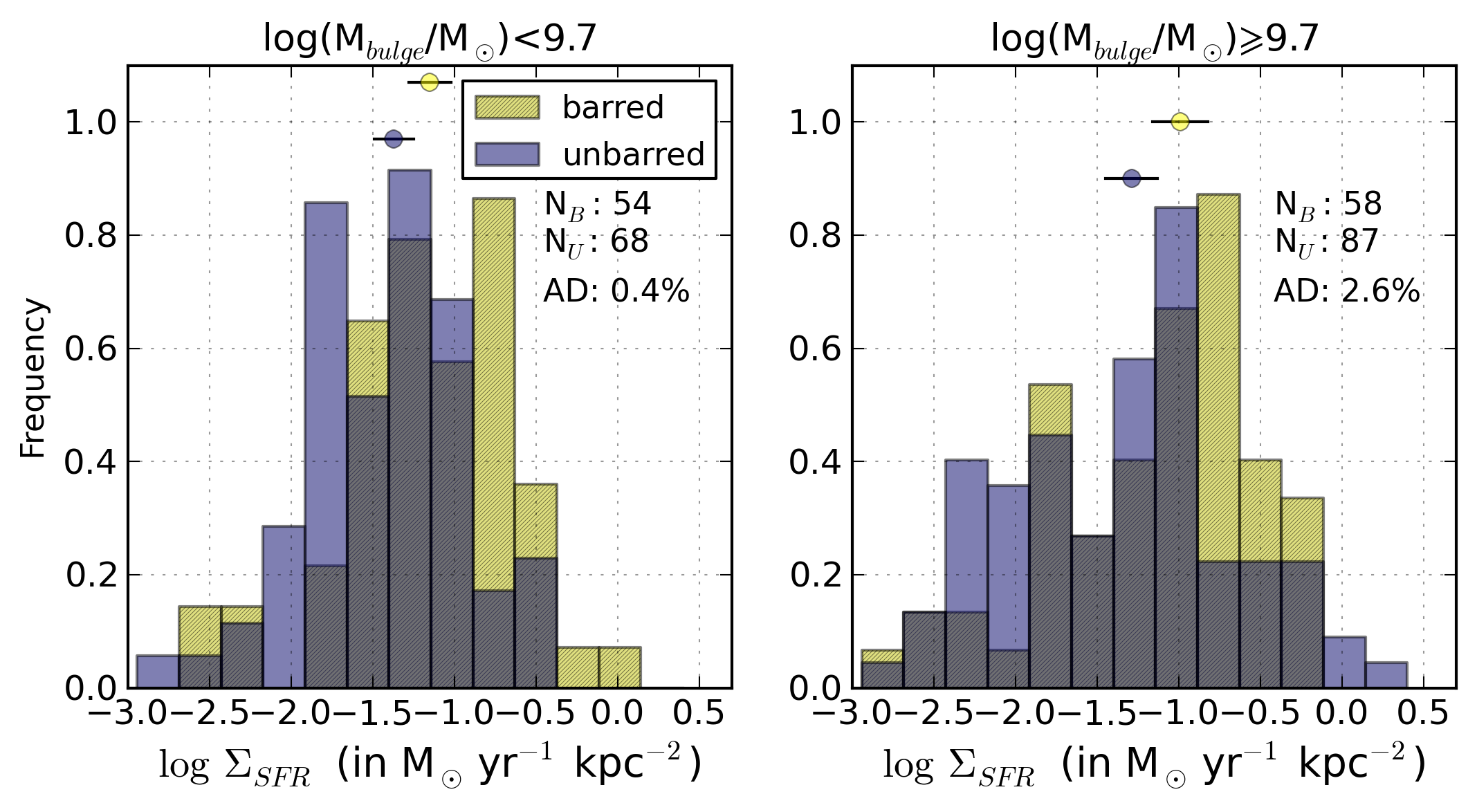}} 
\subfigure{\label{masa_bulbo_e}\includegraphics[width=0.98\columnwidth]{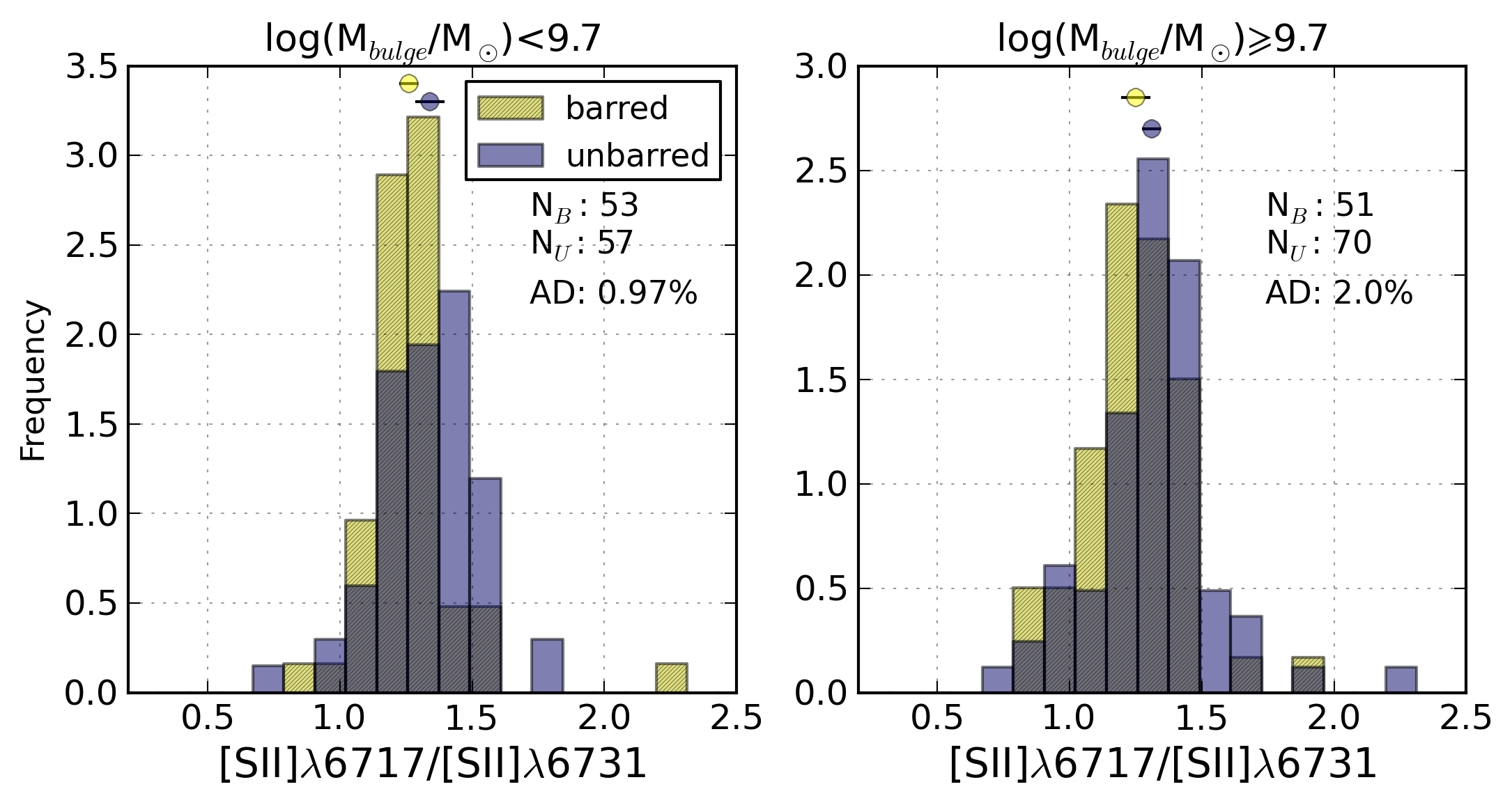}}     
\subfigure{\label{masa_bulbo_f}\includegraphics[width=0.98\columnwidth]{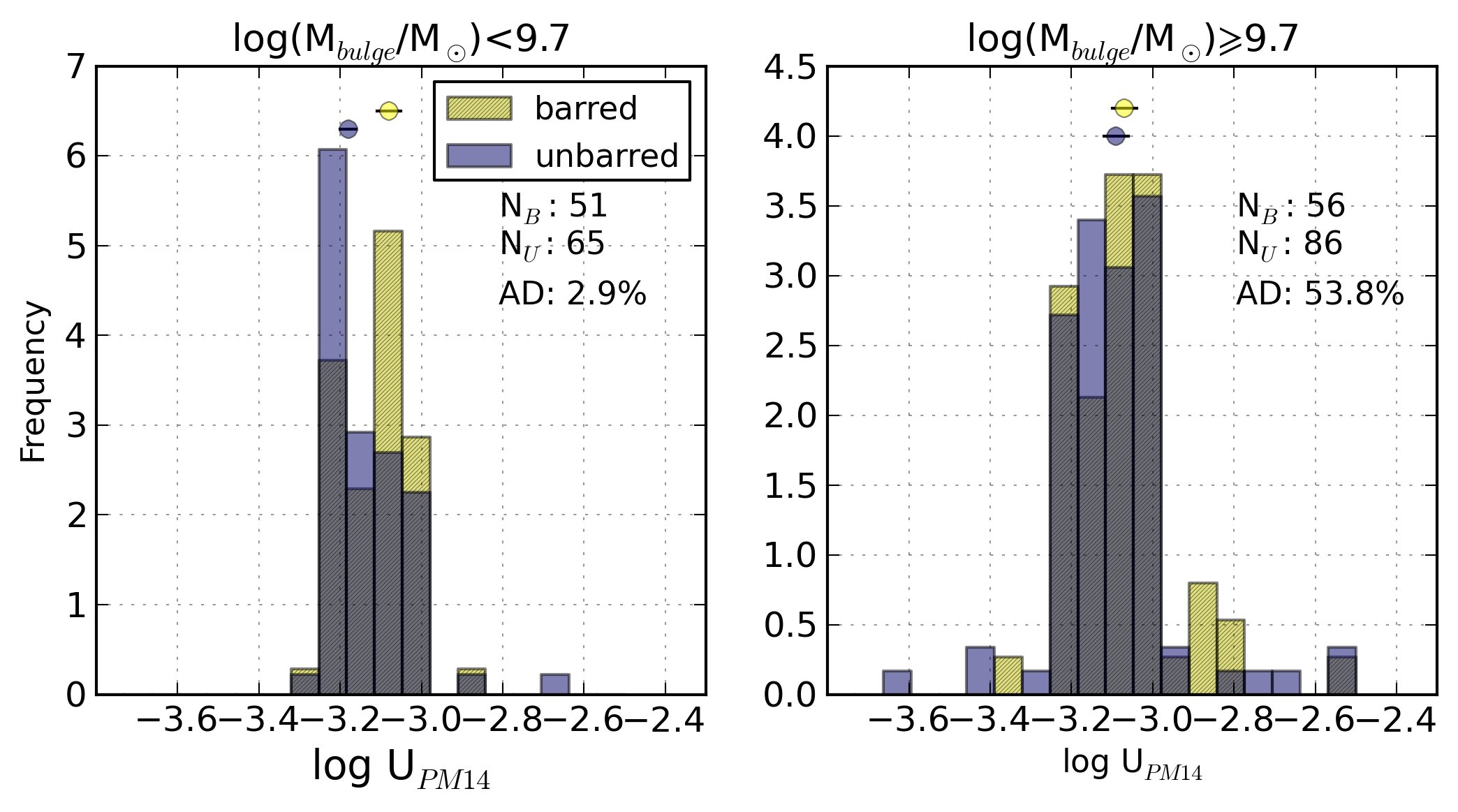}}           
\subfigure{\label{masa_bulbo_g}\includegraphics[width=0.98\columnwidth]{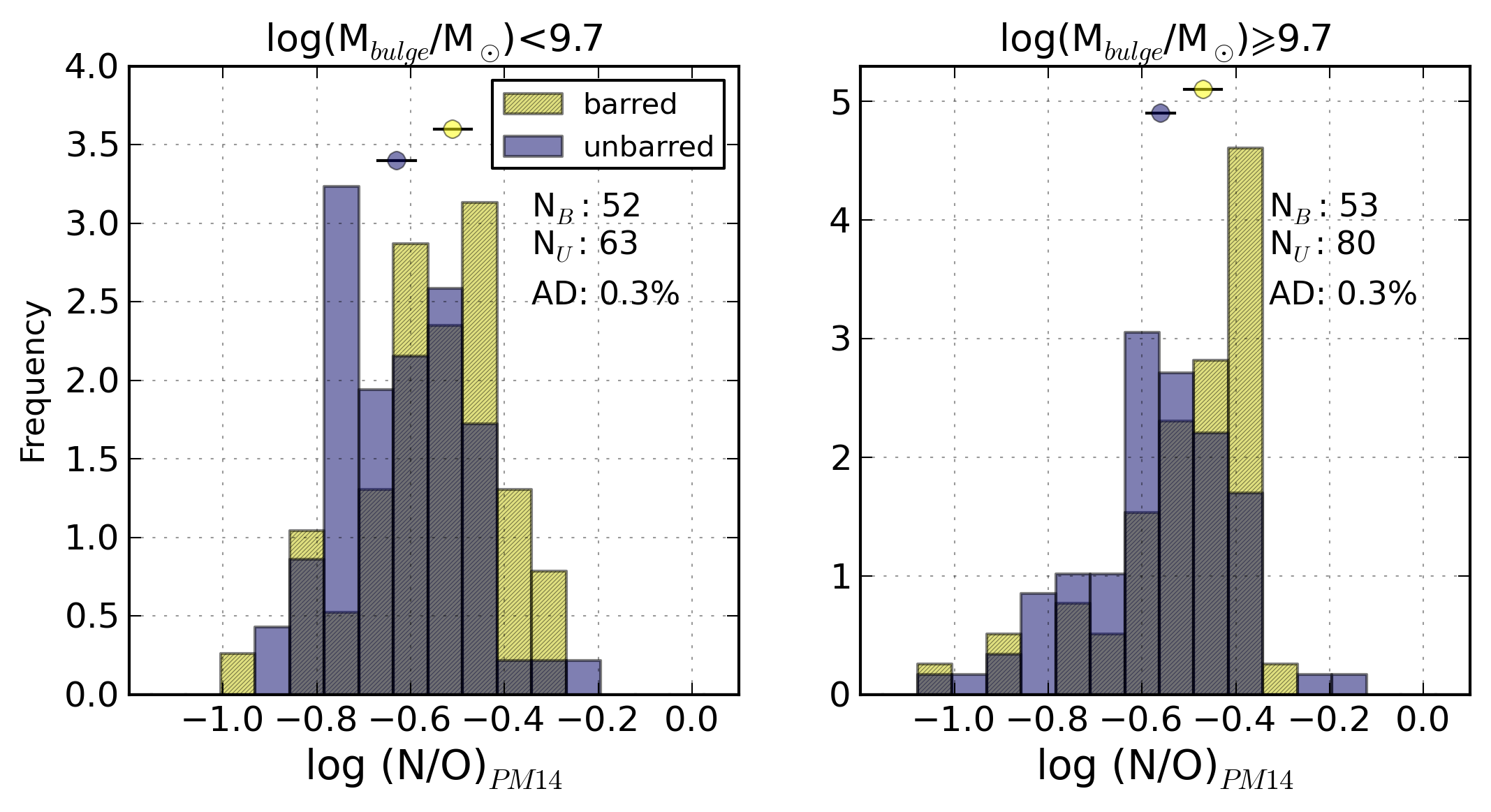}}          
\subfigure{\label{masa_bulbo_h}\includegraphics[width=0.98\columnwidth]{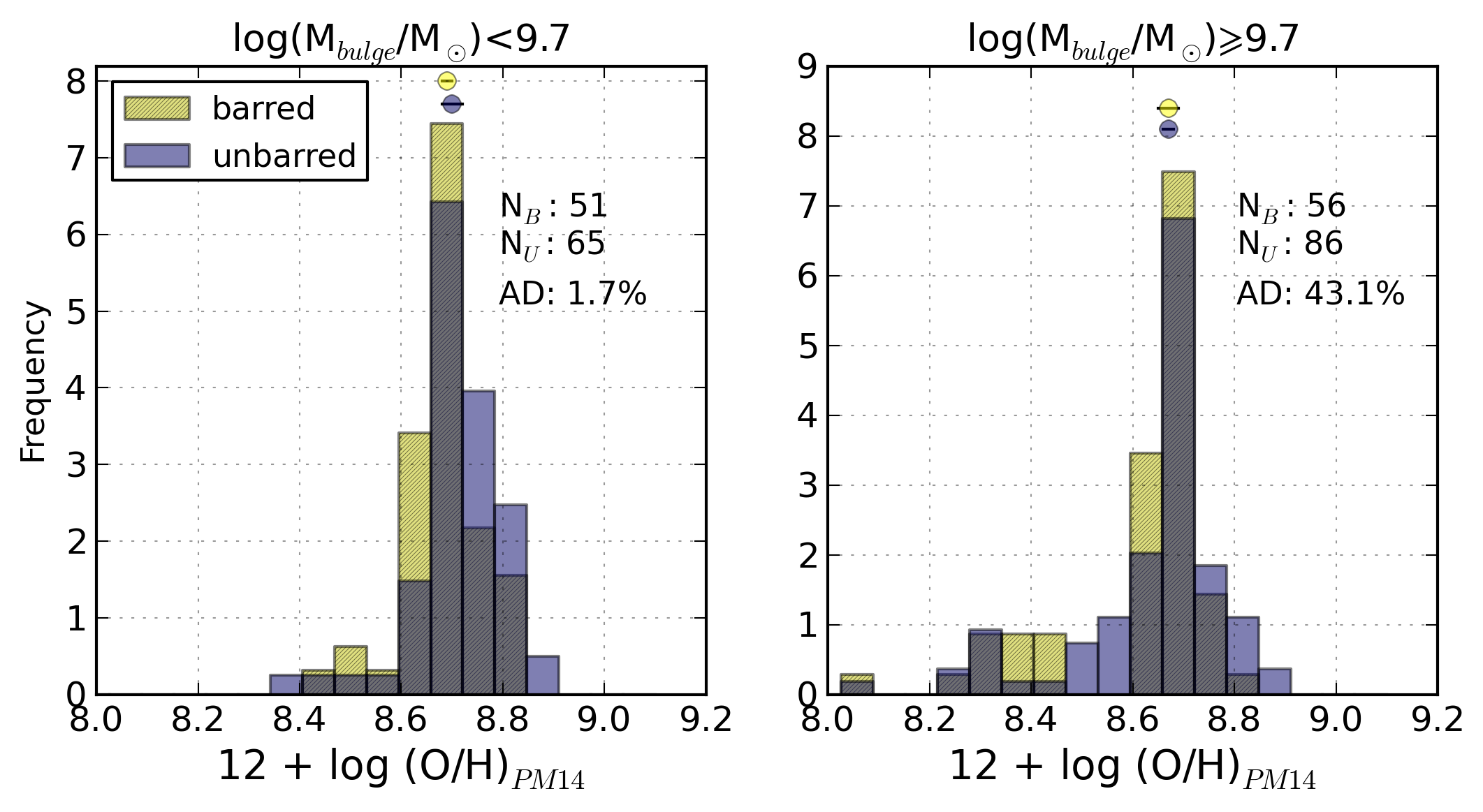}}               
\caption{Comparative histograms of (from top to bottom and from left to right): the Balmer extinction at the \hb\ emission line, $c(\hb)$, R$_{23}$, N2,  logarithm of the ionisation parameter, logarithm  of the SFR per unit area,  the [\sii]$\lambda$6717/[\sii]$\lambda$6731 line ratio, logarithm of the ionisation parameter, N/O abundance ratio, and oxygen abundance for barred  (yellow, hatched) and unbarred  (purple) galaxies separately for non-AGN galaxies with bulge mass lower than $10^{9.7}$ M$_\odot$ and for galaxies with heavier bulges ($\gtrsim 10^{9.7}$ M$_\odot$).}
\label{masabulbo}  
\end{figure*}

Figure~\ref{early_late} and Table~\ref{stats_early_late} show a comparison of central properties  between barred and unbarred galaxies  separately for  early- and  late-type disc galaxies (using  B/D light ratios, as explained above). The same comparison is made in Fig.~\ref{masabulbo} and Table~\ref{stats_bulge_mass} for galaxies with bulge mass below $10^{9.7}$M$_\odot$ and bulge mass larger than or equal to $10^{9.7}$~M$_\odot$. It can be seen that for galaxies with later types (or less massive bulges),  the distributions of all parameters except $\log$~R$_{23}$ are different between barred and unbarred galaxies, with A-D $P$-values lower than 4\% in all cases. 
For earlier type galaxies results depend on whether we split galaxies attending to  B/D light ratios  or bulge mass. Barred and unbarred earlier-type galaxies (according to B/D) only show different distributions for N2 and the [\sii]$\lambda$6717/[\sii]$\lambda$6731 emission-line ratio, while for galaxies with bulge mass above $10^{9.7}$M$_\odot$, in addition to those parameters, we also find  different distributions for $\Sigma_{SFR}$ and $\log$~(N/O).

Barred galaxies tend to have, on average, larger central Balmer \hb\ extinction, larger values of the N2 metallicity indicator, SFR per unit area, electron densities and N/O abundance ratio than unbarred galaxies in all sub-samples. However, taking into account the uncertainties in the median values, we cannot claim real differences in average values between these parameters in barred and unbarred galaxies, apart from  the N2 parameter and the N/O abundance ratio.  N2  is clearly larger in barred  late-type (by $\sim$0.12 dex) and lower bulge-mass (by $\sim$0.13 dex) galaxies compared with unbarred galaxies, while this difference is smaller ($\sim$0.07 dex), but still significant, in the early-type and higher bulge mass sub-sample. These differences between barred and unbarred galaxies in N2 translate into a larger N/O abundance ratio in barred galaxies by $\sim$0.12 dex in low bulge-mass galaxies, and $\sim$0.09 dex, in early-type, late-type and higher  bulge-mass sub-samples.

Finally, if we restrict our comparative analysis of barred and unbarred galaxies to pure star-forming galaxies, we find the same result as with the non-AGN galaxies, i.e.  late-type galaxies differ in central properties, depending on whether they have a bar or not, with barred galaxies exhibiting larger $\Sigma_{SFR}$, $c(\hb)$, N2, $\log$~(N/O) and $\log~U$, and lower [\sii]$\lambda$6717/[\sii]$\lambda$6731 than unbarred galaxies.  Also, similar to the non-AGN sample, the distribution of $12+\log$~(O/H) is different  in barred and unbarred  late-type galaxies and those with smaller bulge mass, with a marginally larger oxygen abundance in unbarred compared to barred galaxies. The results for late-type galaxies in both the pure star-forming and non-AGN samples, only differ for $\log$~R$_{23}$. This  shows a different distribution for barred and unbarred galaxies in the SF sample (although with identical median values). However, in the non-AGN sample, the distributions are identical in barred and unbarred galaxies ($P$-value 58\%). The number statistics are poorer for SF galaxies, but this reveals that the inclusion of
the transition objects does not determine our main results. We will  comment further on this in Sect.~\ref{discussion}.

In summary, we observe statistically significant differences in the distributions of parameters for barred and unbarred galaxies in our later-type galaxy sub-sample, regardless of whether we select them according to their predicted T-type (from their B/D light ratio) or their bulge mass.  These differences appear in all parameters except in the metallicity tracer R$_{23}$. 
Median values are only marginally different in most parameters (except in N2 and $\log$~(N/O)), but they point towards a larger  $\Sigma_{SFR}$, $c(\hb)$, N2, $\log$~(N/O) and $\log~U$, and lower [\sii]$\lambda$6717/[\sii]$\lambda$6731 in barred galaxies. Results for earlier-type galaxies depend on whether we select them according to B/D light ratio or bulge mass. However,  with both  criteria for dividing, the distributions of N2 and [\sii]$\lambda$6717/[\sii]$\lambda$6731 are significantly different between barred and unbarred galaxies. The difference in the median value of N2 for barred and unbarred galaxies is slightly smaller than in later-type galaxies but is still significant.


\section{Barred versus unbarred difference trends with bulge and total stellar mass}
\label{trends}
Previous authors find observational evidence that massive barred galaxies have a higher current central-star-formation rate than unbarred galaxies of the same stellar mass \citep{ellison}.  However, we  see above that the effect of bars in  galaxy centres seems to be stronger or more visible in galaxies with lower bulge mass, even when the barred and unbarred sub-samples have been created to have indistinguishable total stellar mass distributions.

To better understand our own and previous authors' results, we  created a series of boxplots that show the variation of all parameters as a function of bulge and total stellar mass. 
Boxplots are a useful representation in our case, as sample sizes are not big enough to explore parameter dependences using histograms. Figure~\ref{boxplot_totalmass} shows boxplots for $c(\hb)$, N2,  $\log$~R$_{23}$, $\Sigma_{SFR}$, [\sii]$\lambda$6717/[\sii]$\lambda$6731, $\log~U$, N/O, and oxygen abundance as a function of total galaxy stellar mass (see caption for a description on basic boxplots features). It is clear from these plots that, for the most massive galaxies, there is only a significant difference between barred and unbarred galaxies for the N/O abundance ratio.
The most massive galaxies are also the galaxies with more massive bulges, and this result simply confirms the results from Sect.~\ref{morfo}. For less massive galaxies (M$_{\star}\lesssim10^{10.8}$~M$_\odot$), N2 and  $\Sigma_{SFR}$ are clearly higher for barred than for unbarred galaxies, with non-overlapping  box-notches  between barred and unbarred galaxies in the two lowest mass intervals.  
Our results are, therefore, only partially in agreement with \cite{ellison}. We do find a larger fibre SFR in barred galaxies above 10$^{10}$~M$_{\odot}$, but we do not see differences in SFR  between barred and unbarred galaxies above $\sim$10$^{10.8}$~M$_{\odot}$.

The differences between barred and unbarred galaxies are undoubtedly better correlated with bulge stellar mass (see Fig.~\ref{boxplot_bulgemass}). All properties (except R$_{23}$, $\log~U$ and 12+$\log$~(O/H)) are significantly different between barred and unbarred galaxies for bulge masses $\lesssim10^{10.0}$~M$_\odot$, but differences normally increase towards lower bulge masses, except for $c(\hb)$ in the lowest mass bin, where the median values are similar within errors for both types of galaxies. The observed differences always suggest that barred galaxies present larger dust extinction, star formation rate per unit area, [\nii]/\ha\ line ratio, lower  [\sii]$\lambda$6717/[\sii]$\lambda$6731 (indicating a higher electron density), and larger $\log$~(N/O) and $\log~U$ in their centres.

\begin{figure}
\centering
\includegraphics[width=0.98\columnwidth]{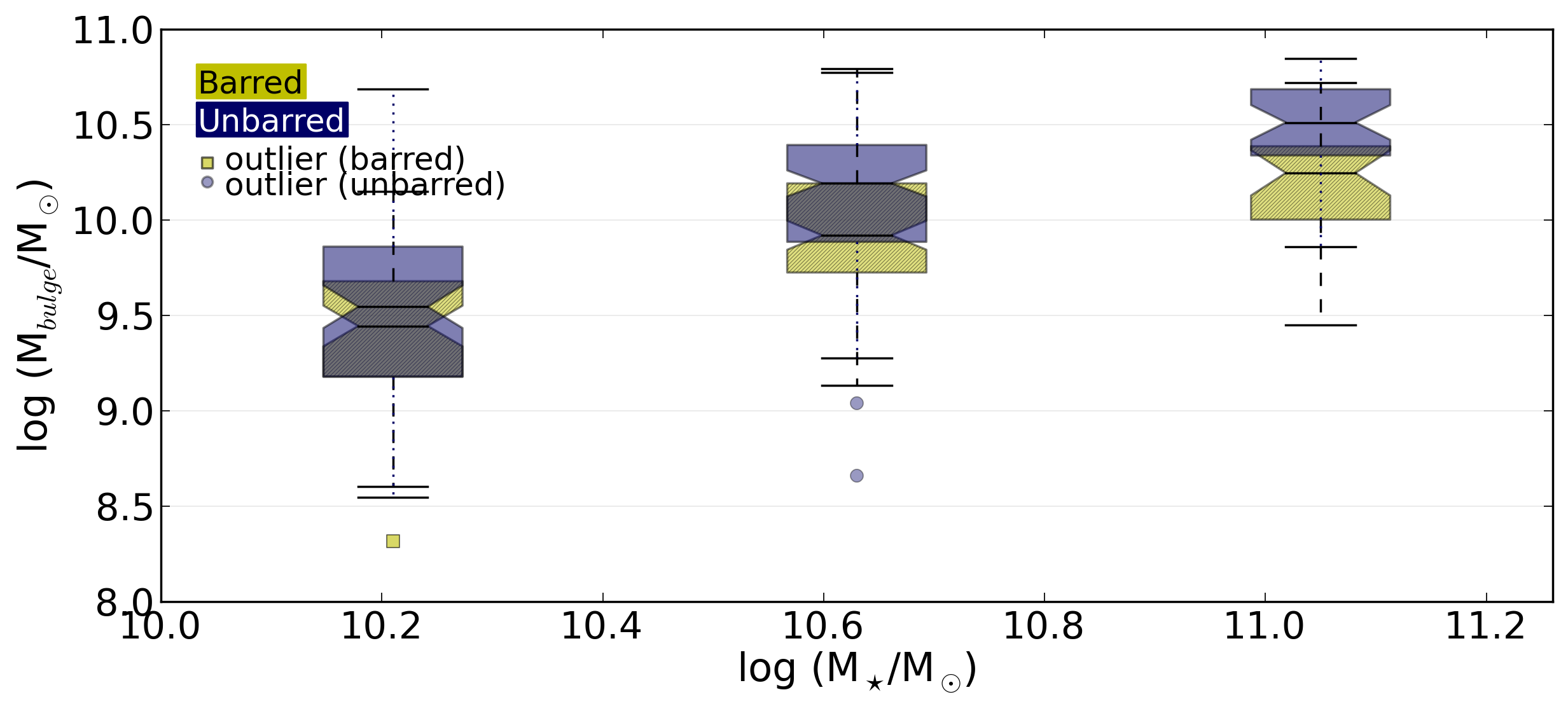}
\caption{\label{boxplot_Mbulge_total_mass}Boxplot showing median values of bulge stellar mass for three different bins in total galaxy stellar mass.}
\end{figure}

\begin{figure*}
\centering
\includegraphics[width=0.95\columnwidth]{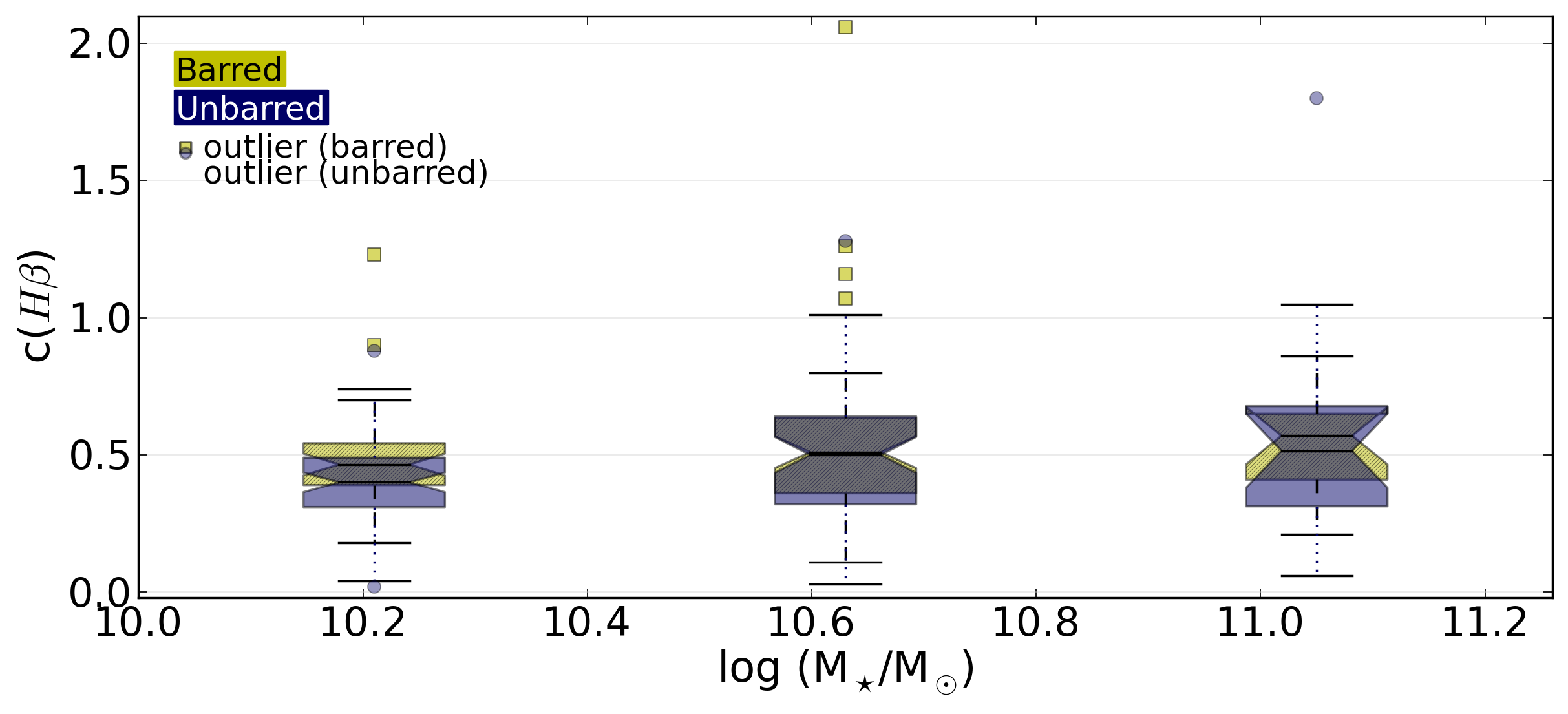}        
\includegraphics[width=0.95\columnwidth]{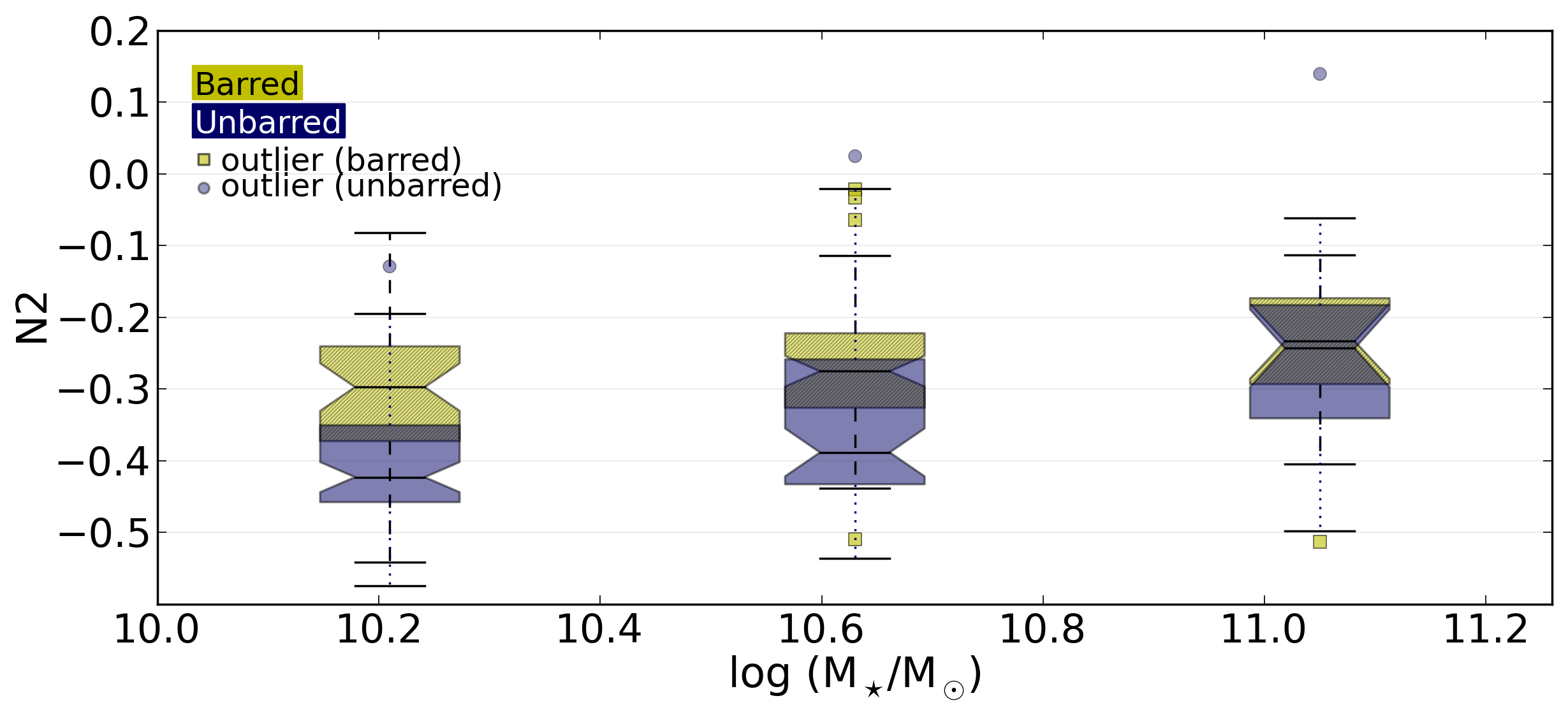}       
\includegraphics[width=0.95\columnwidth]{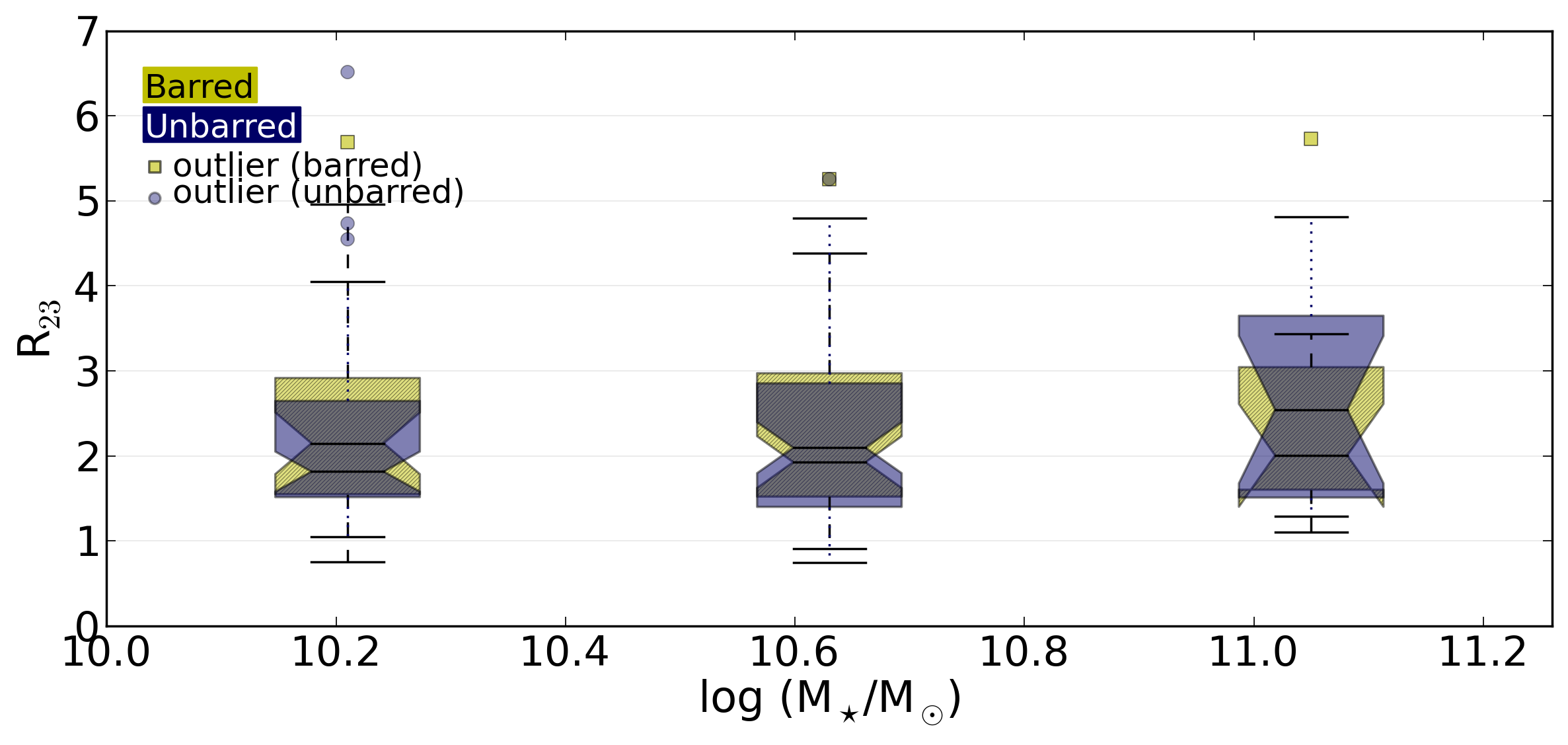}      
\includegraphics[width=0.95\columnwidth]{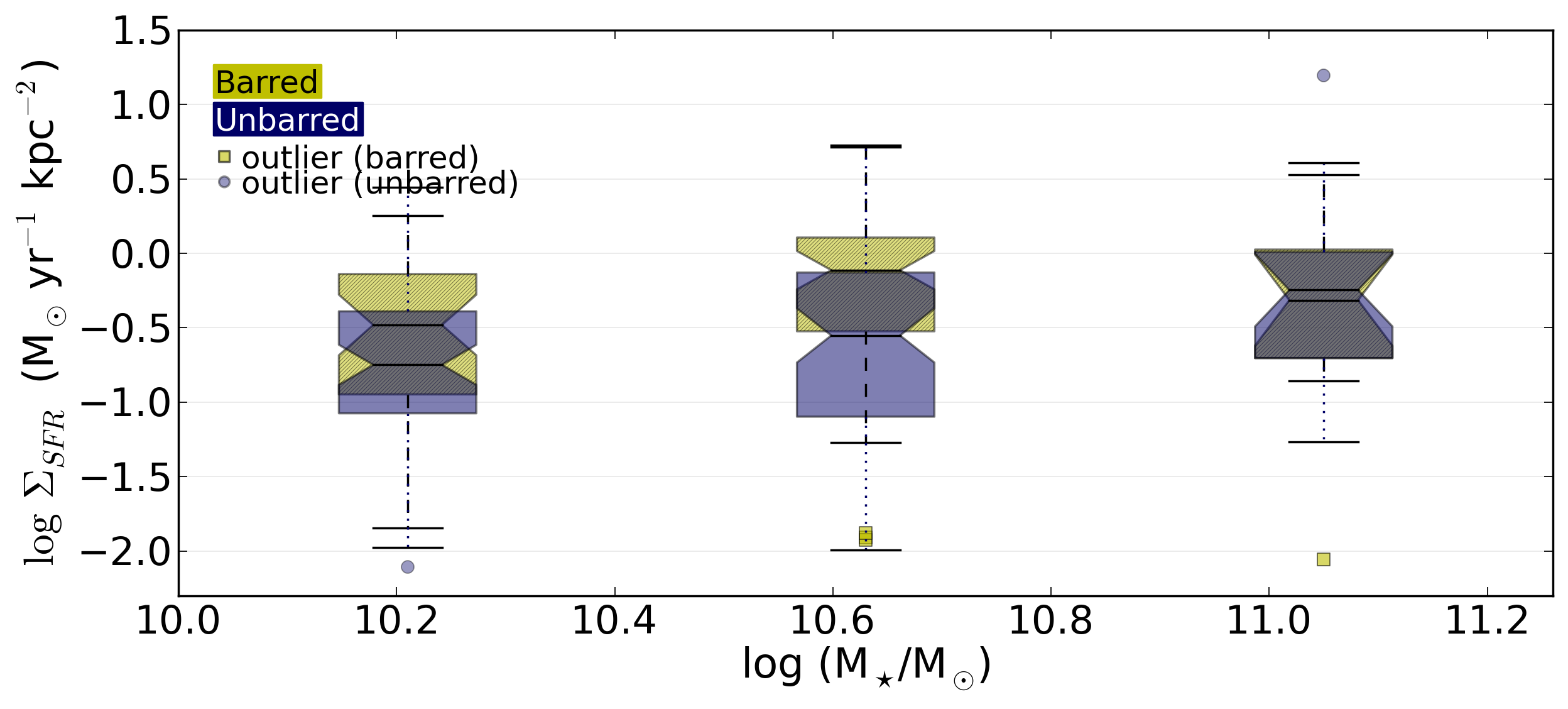} 
\includegraphics[width=0.95\columnwidth]{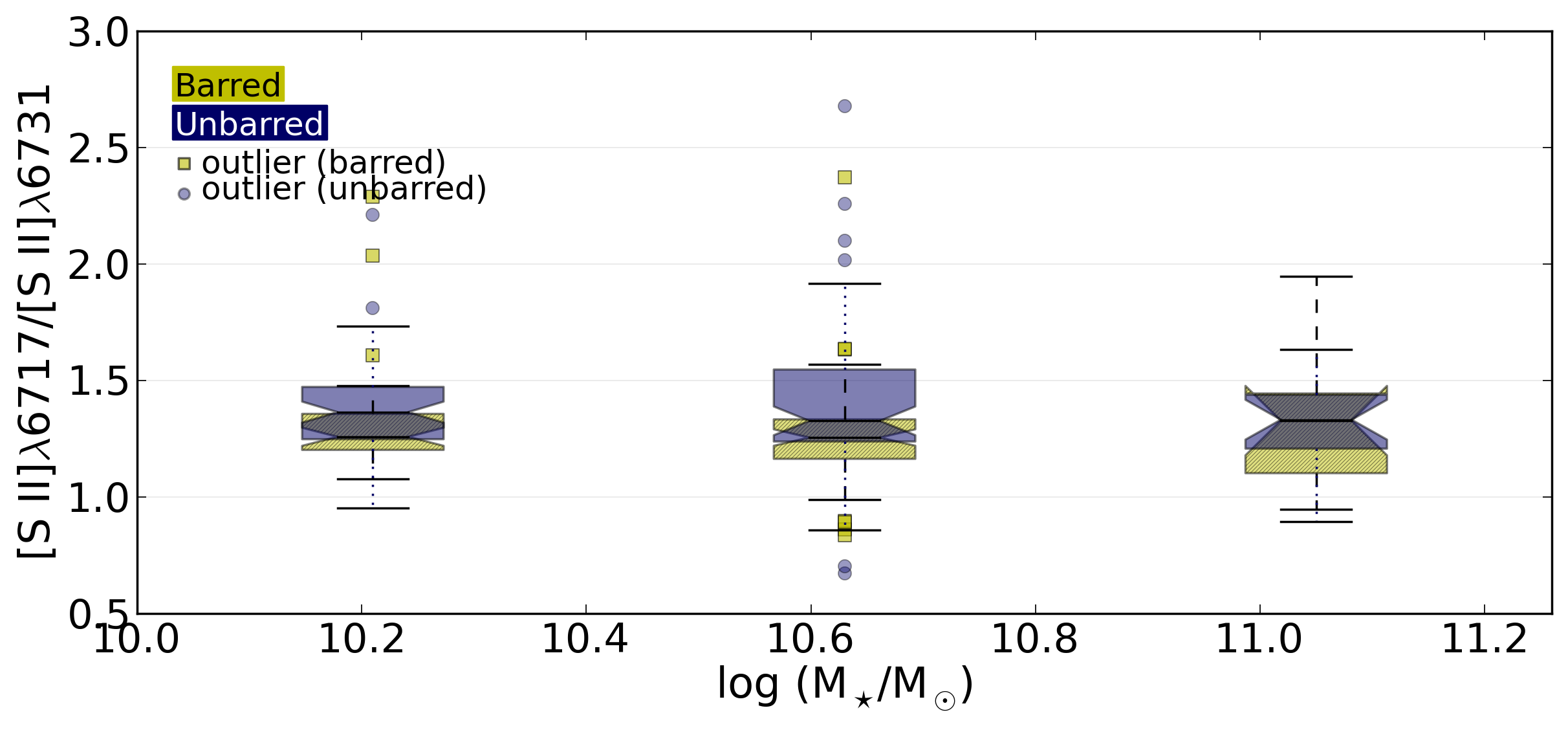}      
\includegraphics[width=0.95\columnwidth]{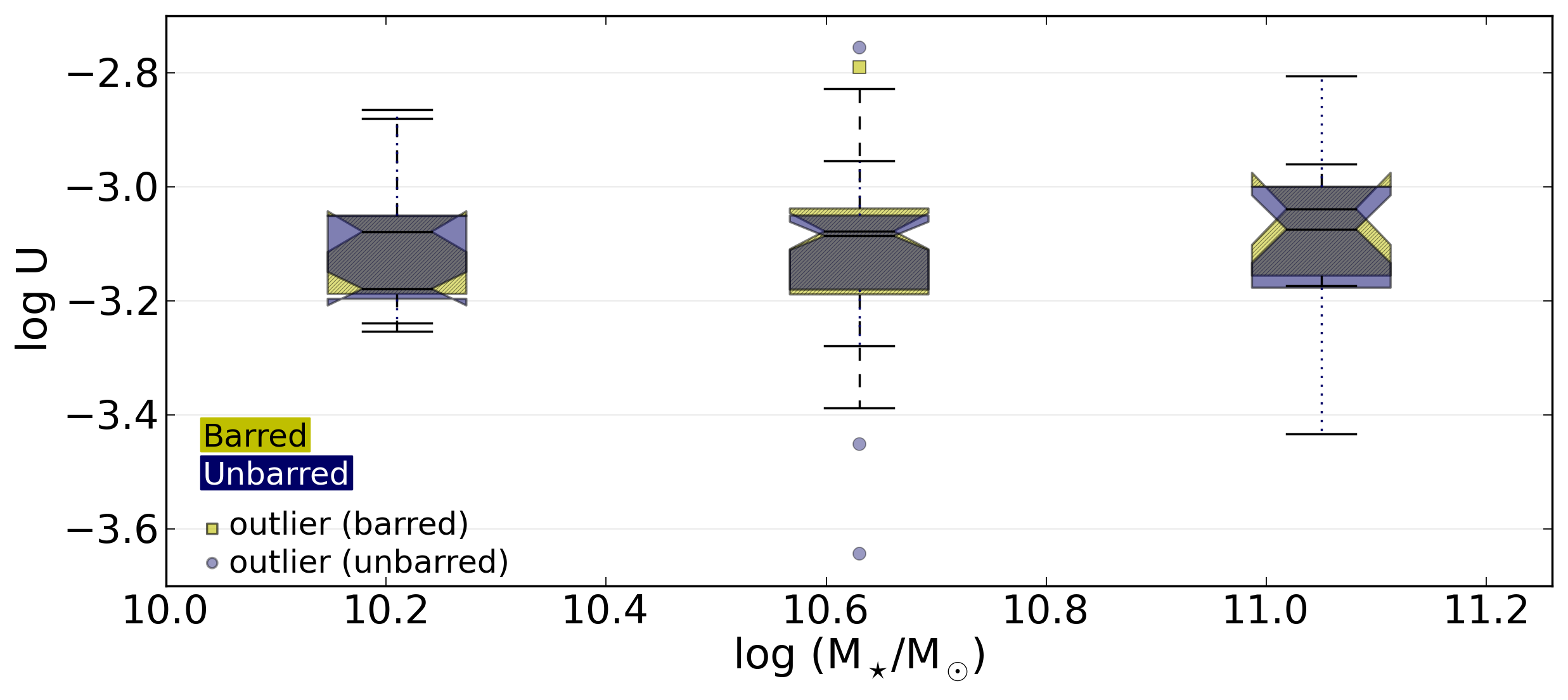}        
\includegraphics[width=0.95\columnwidth]{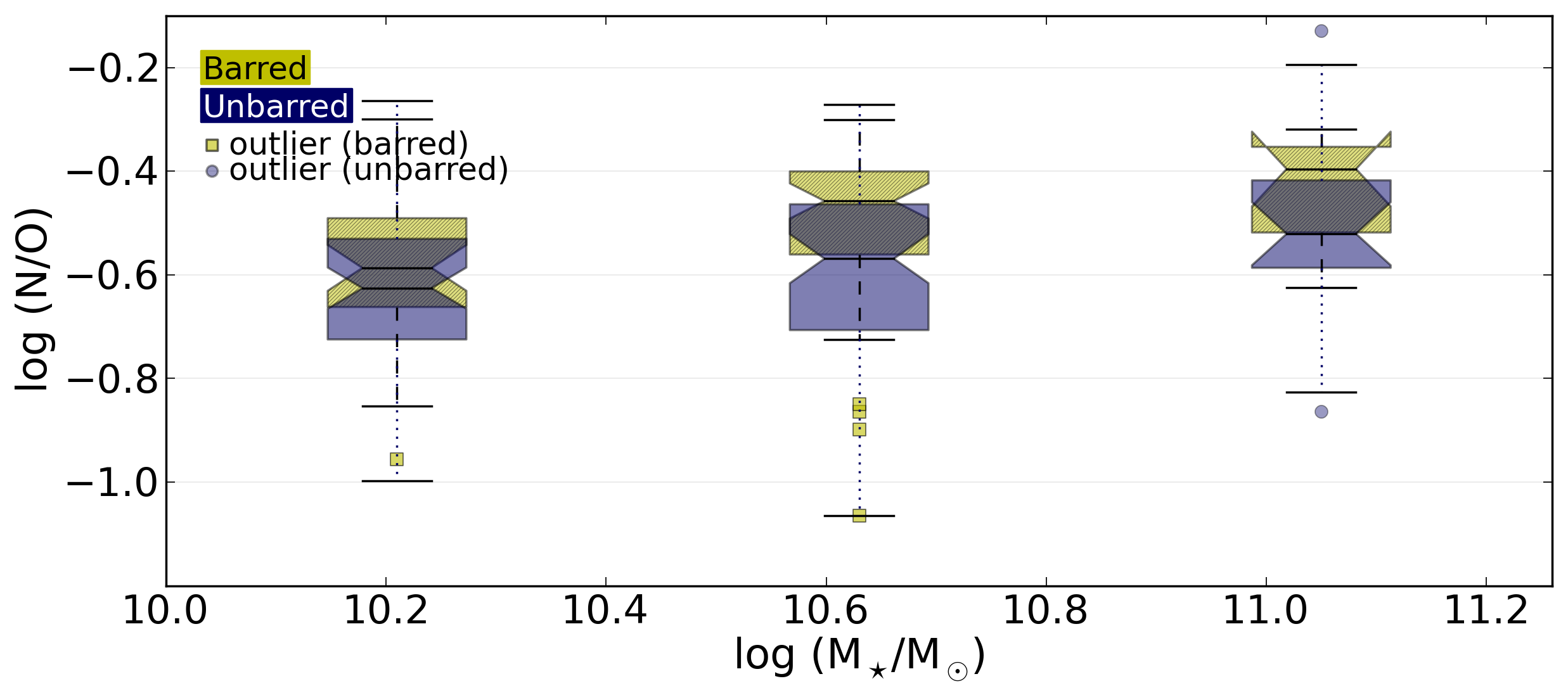}       
\includegraphics[width=0.95\columnwidth]{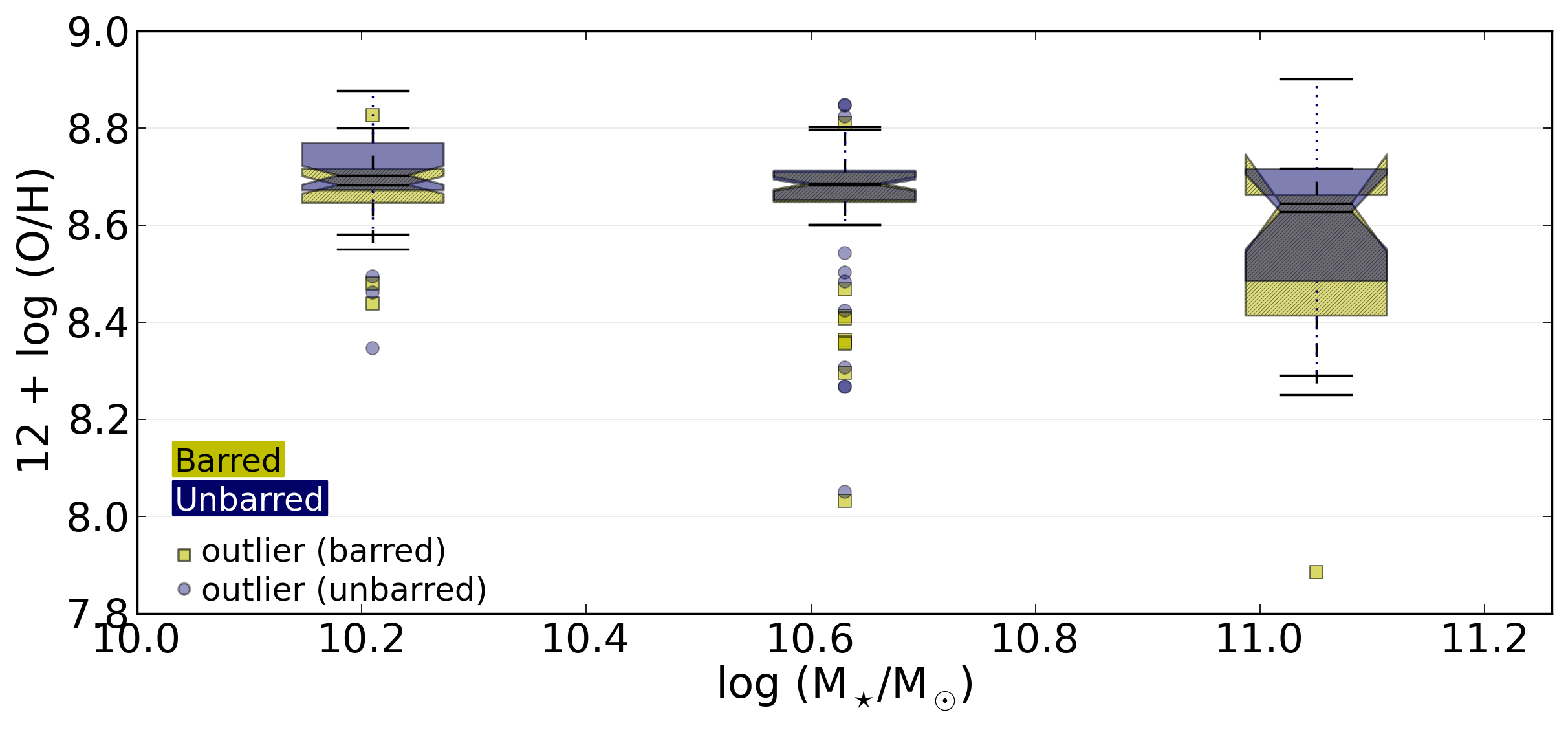}       
\caption{\label{boxplot_totalmass} Boxplots showing the distribution in $c(\hb)$, N2 =log([\nii]6583/\ha), R23,  SFR per unit area and [\sii]$\lambda$6717/[\sii]$\lambda$6731 line ratio, decimal logarithm of the ionisation parameter, N/O abundance ratio and oxygen abundance,  as a function of the total galaxy stellar mass (M$_\star$) for barred (yellow boxes) and unbarred (purple) galaxies. 
Each box represents the data distribution for galaxies in a 0.42~dex interval of logarithm in stellar mass centred at the abscissa axis values of the box position. Inside each box, the central horizontal black straight line marks the median value. The lower and upper quartiles are represented by the outer edges of the boxes, i.e. the box length is equal to the inter-quartile range, and therefore the box encloses 50\% of the data points. The notches mark the 95\% confidence interval for the median value. The whiskers extend to the most extreme data point within 1.5~times the inter-quartile range. Galaxies that do not fall within the reach of the whiskers are considered as outliers (yellow and purple circles for barred and unbarred galaxies, respectively).  The number of galaxies in each box, in order or increasing stellar mass, ranges from 35 to 39, 55 to 59, and 13 to 14 for barred galaxies, and from 57 to 64, 58 to 70, and 15 to 21, for unbarred galaxies.}
\end{figure*}

\begin{figure*}
\centering
\includegraphics[width=0.95\columnwidth]{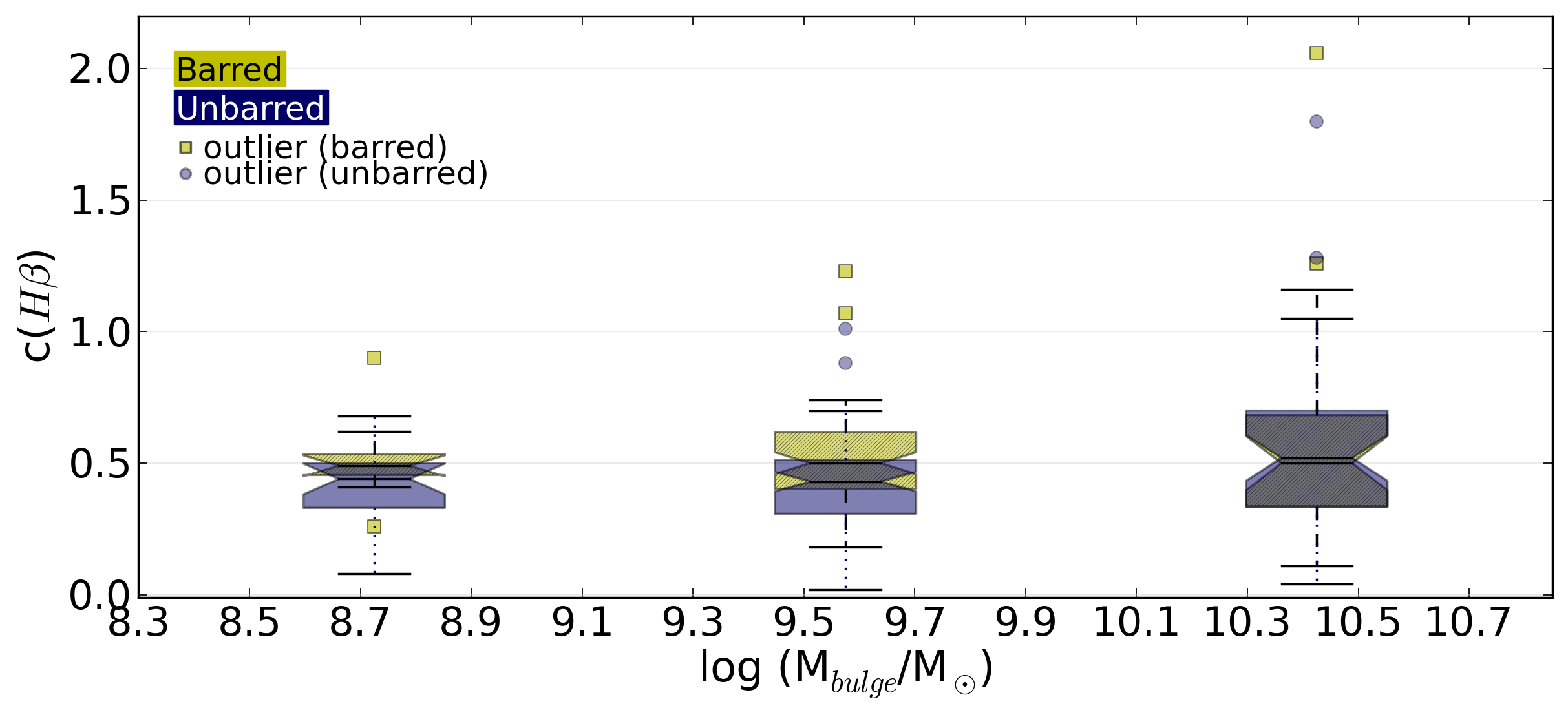}
\includegraphics[width=0.95\columnwidth]{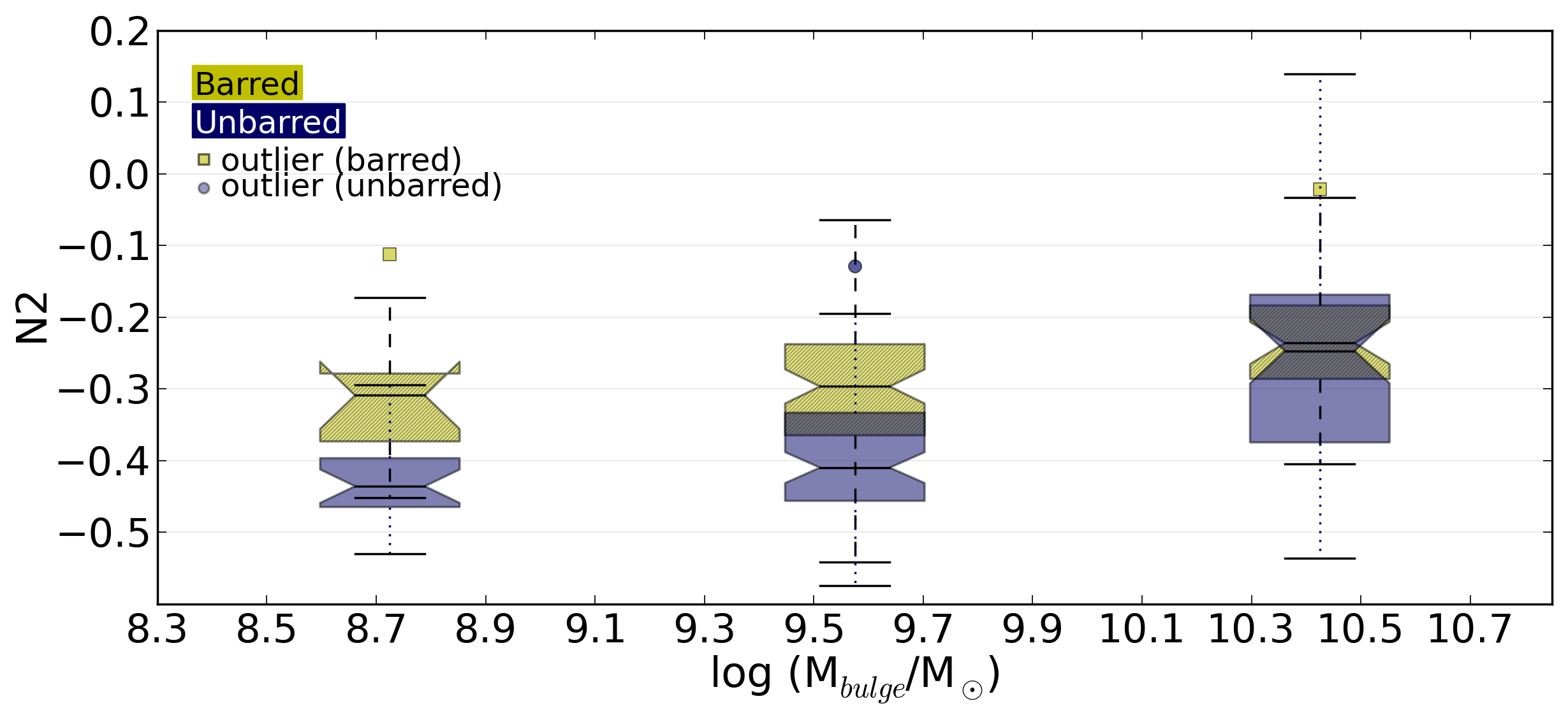}
\includegraphics[width=0.95\columnwidth]{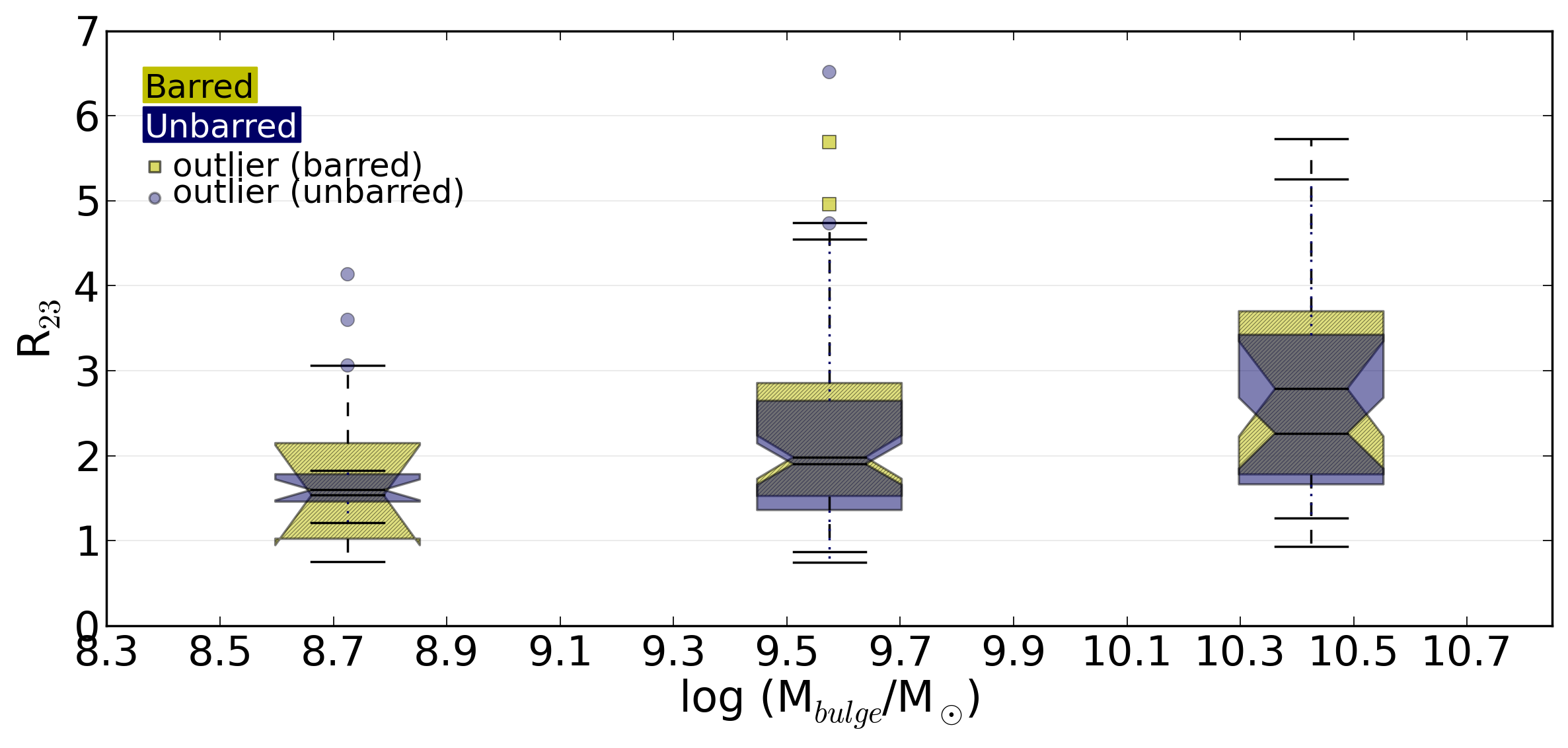}
\includegraphics[width=0.95\columnwidth]{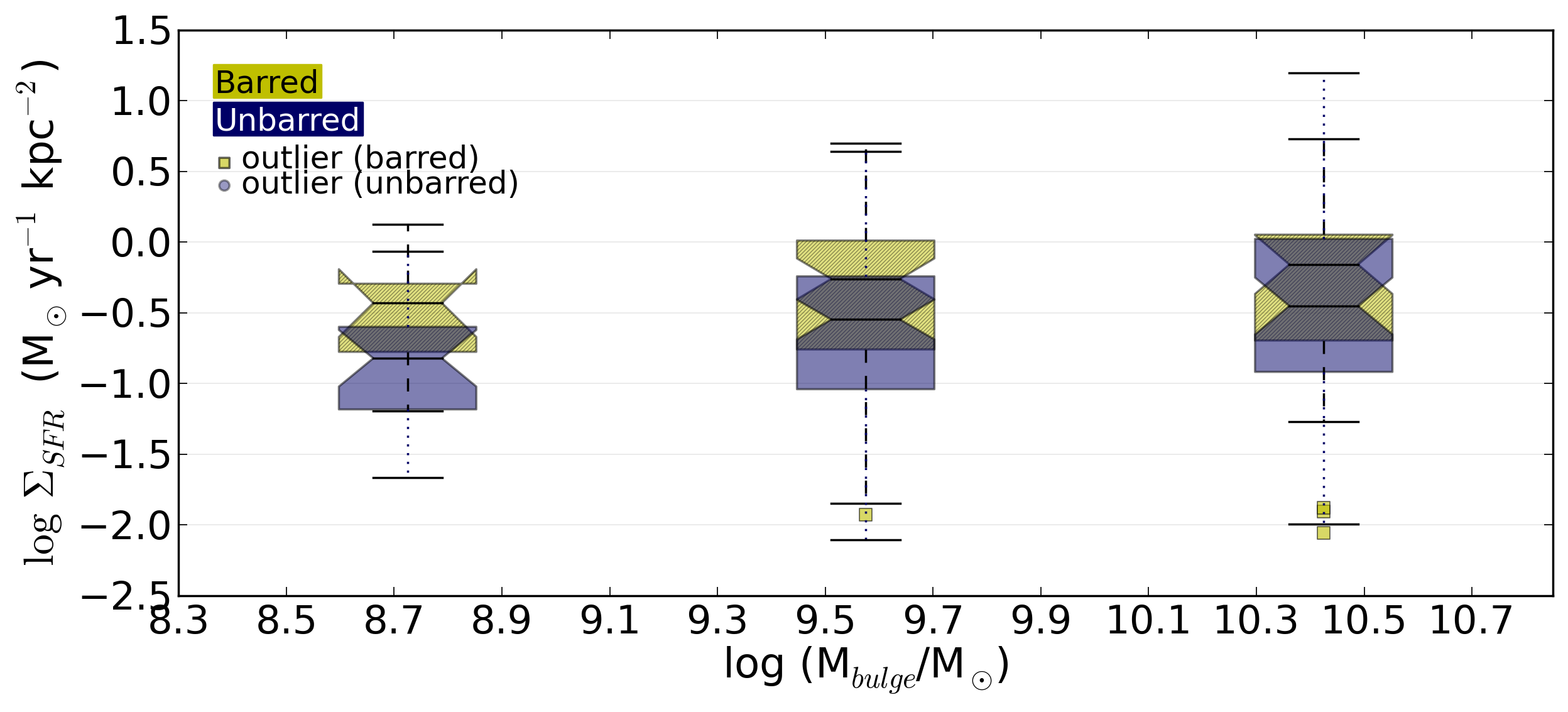}
\includegraphics[width=0.95\columnwidth]{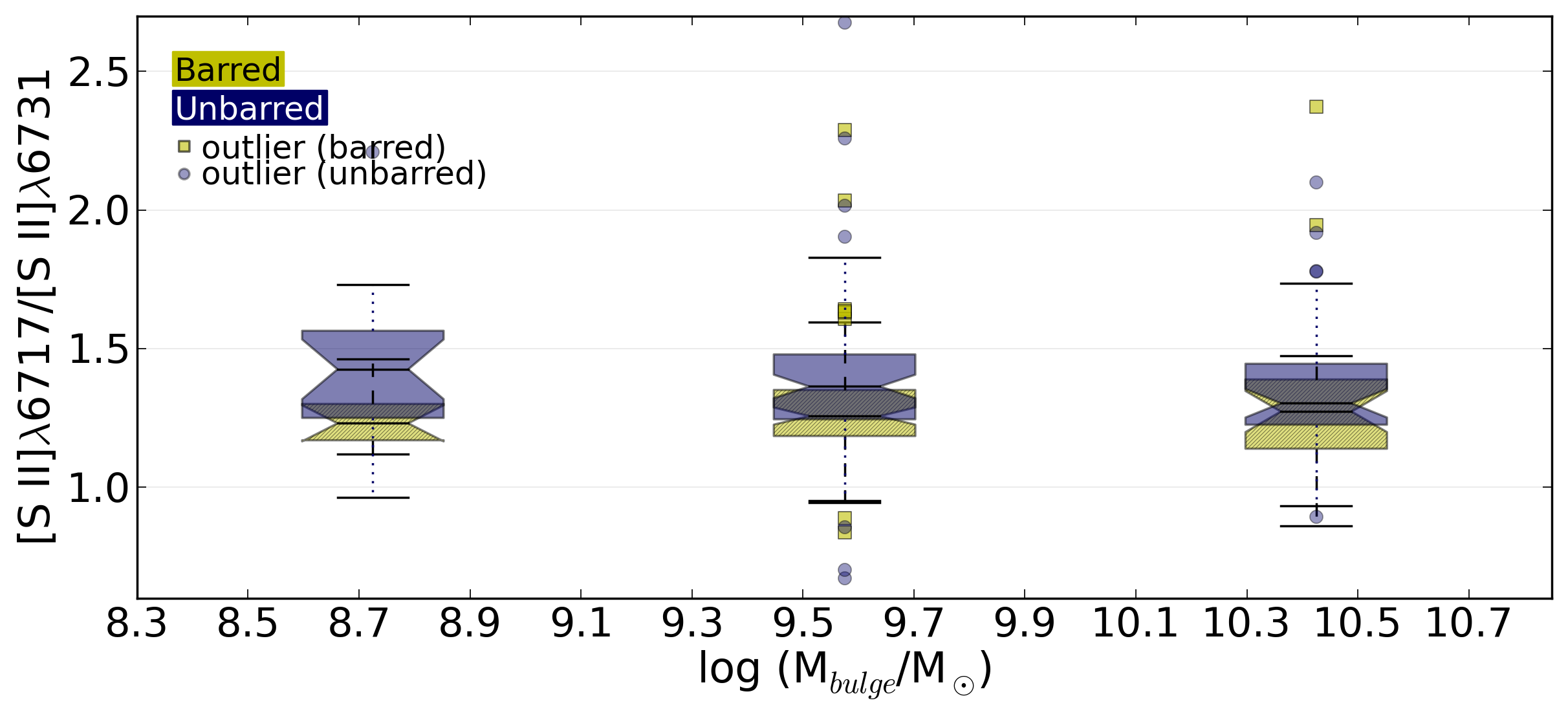}
\includegraphics[width=0.95\columnwidth]{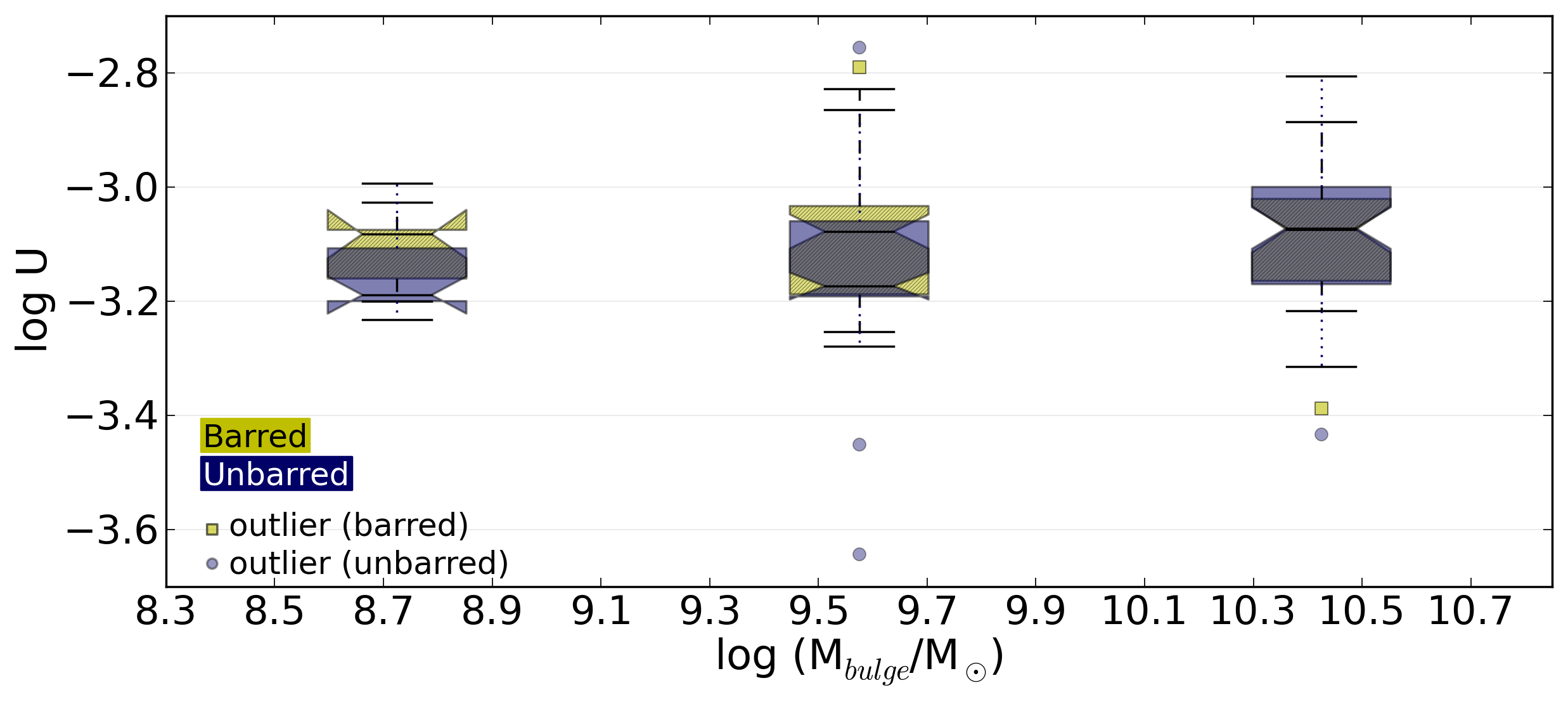}        
\includegraphics[width=0.95\columnwidth]{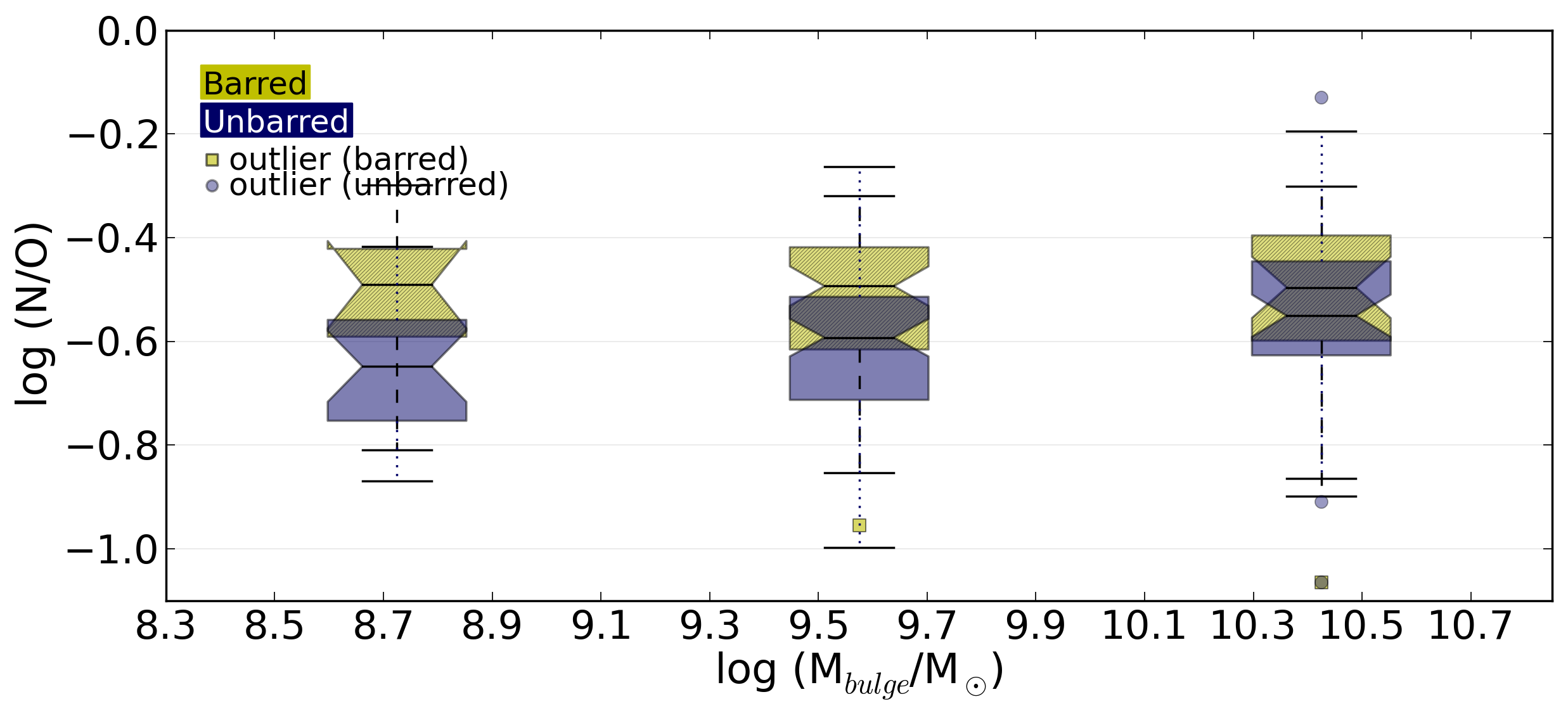}       
\includegraphics[width=0.95\columnwidth]{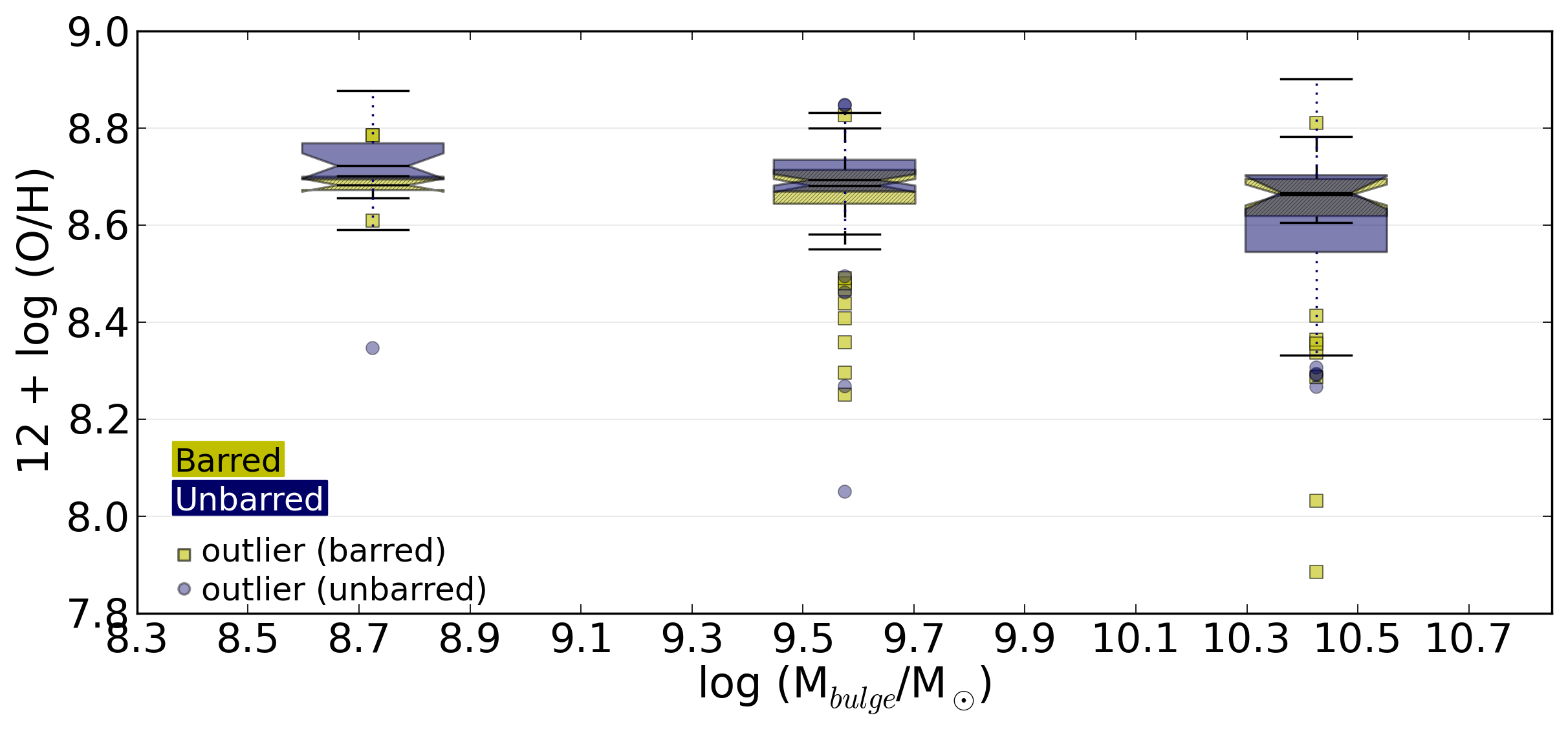}       
\caption{\label{boxplot_bulgemass} Same as Fig.~\ref{boxplot_totalmass} but as a function of the bulge mass.  Each box represents the data distribution for galaxies in a 0.85~dex interval of logarithm in bulge stellar mass centred at the abscissas values of the box position. The number of galaxies in each box, in order of increasing stellar bulge mass, ranges from 9 to 10, 66 to 71, and 28 to 32  for barred galaxies, and from 18 to 22, 70 to 81, and 28 to 32, for unbarred galaxies.}
\end{figure*}

We also note that the SFR per unit area is also larger on average for barred galaxies in the highest bulge mass bin.
However, as mentioned above, we do not see differences in  $\Sigma_{SFR}$ between barred and unbarred galaxies in the most massive galaxies (Fig.~\ref{boxplot_totalmass}). This is probably because bulge masses of barred galaxies are lower than in unbarred galaxies for the most massive galaxies  (Fig.~\ref{boxplot_Mbulge_total_mass}), as already pointed out by \cite{coelho}.

More massive bulges are typically classical bulges,  with S\`ersic indexes higher than $\sim$2.5, in contrast with the so called  pseudo-bulges, whose surface brightness profile is better fitted with lower S\`ersic indexes and which are typically less massive than classical bulges \citep{dimitri_morpho}. We have also analysed the barred and unbarred galaxy central properties as a function of S\`ersic index. Barred and unbarred galaxies with large values of the S\`ersic index  (n$\gtrsim$3.1) do not differ in their central ionised gas properties, apart from the SFR, which seems to be enhanced in barred galaxies,  in agreement with the result found for barred galaxies with the most massive bulges. However the global parameter dependences with the  S\`ersic index and the differences between barred and unbarred galaxies are not as clear as with bulge mass or even total stellar mass.

\section{Dependence on bar parameters}
\label{dependence_bar}
Previous sections show clear differences in the central ionised gas properties between barred and unbarred galaxies. These differences imply that the gas flows induced by bars are able to modify gas properties, especially in lower mass or lower bulge mass galaxies. Simulations predict a dependency between the gas inflow rate and the bar strength, with stronger bars supplying gas towards the centre at a higher rate \citep{athanassoula,ReganTeuben}.
Thus, it is natural to wonder whether these observational enhancements are related to dynamical or structural bar parameters.

Morphological decomposition of the galaxy sample was performed by \cite{dimitri_morpho} and a number of structural parameters for discs, bulges and bars (in the case of barred galaxies) are thus available. This allows for a comparison between parameters such as  bar ellipticity (as a proxy for bar strength) and bar length with central ionised gas properties. Our results are presented  and compared with previous work below.
The SFR per unit area ($\Sigma_{SFR}$) shows a weak tendency to increase with the effective radius of the bar, the bar semi-major axis normalised to the galaxy disc scale length, and the  bar ellipticity. However, the dispersion in these relations is high and the Spearman rank correlation coefficients ($\rho_S$) are 0.28,  0.20, and 0.12, respectively. For other parameters, such as the bar's S\'ersic index \cite[which might be related to the bar's  age;][]{kim} $\rho_S$ is lower than 0.1.
This  agrees with  previous works on central SFRs from  infrared emission \citep[e.g.][]{pompea, roussel}, which did not find evidence that stronger bars could produce a larger enhancement\  in SFRs. \cite{ellison} in turn, did not find a correlation between bar length and fibre SFR in the central region of galaxies from SDSS optical spectra. However, other authors have reported a positive relation between central star formation and bar parameters, but their results are, in some cases, conflicting:  \cite{ooy} report a higher effect of bars on central SF  but only for the reddest galaxies where the effects are more pronounced with increasing bar length. In contrast, other authors \citep{wang2012} find that only strong bars can enhance central SF, but the degree of enhancement depends solely on bar ellipticity, and not on bar size or bar mass. This  matches with \cite{zhou}, who find a weak positive trend between SFR in bulges and an ellipticity-based bar strength parameter.

Several studies have also tried to confirm observationally the effect of bars on the central oxygen abundance and on the radial oxygen abundance gradient \cite[e.g.][and references therein]{florido}, but results are also conflicting. For example, there is disagreement on whether bars produce a larger oxygen abundance in galaxy centres \citep[e.g.][]{ellison}, equal to that found in unbarred galaxies \citep[e.g.][]{chapelon,cacho} or whether O/H is lower in barred galaxies than in galaxies without a stellar bar \citep[][]{considere,Dutil_Roy}. However there seems to be a consensus that neither central nor the O/H abundance  gradient depends on bar strength \citep{chapelon,considere,cacho,ellison}. Our results agree with previous studies, since we also find no clear trend between 12+log(O/H) and bar parameters (effective radius, normalised length, ellipticity) with  correlation coefficients $\rho_S$ lower than 0.2, with a slightly better correlation with the bar boxiness ($\rho_S$=0.32). 

We also find no correlation between the N/O abundance ratio and any bar parameter, with $\rho_S$ lower than 0.1 in all cases, with the exception of a weak trend for the N/O to increase with the bar effective radius ($\rho_S$=0.15). The $c(\hb)$ and the [\sii]$\lambda$6717/[\sii]$\lambda$6731 emission-line ratio do not show any clear dependence on bar parameters either ($\rho_S\lesssim$0.2).


\section{Discussion}
\label{discussion}
In this paper we perform a comparative analysis of the properties of the ionised gas in the centres of disc galaxies that possess a stellar bar, with those without a bar structure. Our analysis is based on SDSS spectra.

The SDSS fibre size (3\arcsec) and our face-on galaxy sample redshift range (0.02 $\leqslant z \leqslant$ 0.07) implies that, in this work, we are analysing emission of the ionised gas located within galactocentric radii of between approximately 0.6 and 2.1~kpc. The ionised gas-emitting regions located at these galactocentric distances are usually termed as "nuclear \hii\ regions" or  "\hii\ region nucleus" when they are located in the immediate neighbourhood of a galactic centre, and "hotspots" or  "circumnuclear  \hii\ regions" when these regions surround the galaxy centre, frequently arranged in a ring or pseudo-ring shape \citep[see e.g.][]{KKB,Ho_hii,diaz07}. The spectroscopic studies of nuclear and circumnuclear spatially resolved \hii\ regions are still limited to a small number of targets.
These have been the focus of previous in-depth spectroscopic studies, mainly with the aim of comparing the physical and stellar content properties to those of normal disc \hii\ regions \citep[e.g.][]{KKB,Ho_hii,fabio02}, studying differences in properties with the Hubble-type of the host galaxy \citep[][]{Ho_hii} and/or deriving chemical abundances. These nuclear and circumnuclear  \hii\ regions are expected to be those with the highest metal content because of their location in the inner parts of spiral discs \citep{diaz07,dors2011}. To our knowledge  no  published work so far exists that compares the properties of resolved nuclear and circumnuclear \hii\ regions in barred and unbarred galaxies for a large number of galaxies in a systematic way.

Our derived values for internal Balmer extinction, [\sii]$\lambda$6717/[\sii]$\lambda$6731 ratio, oxygen abundance, and N/O ratio are within the values quoted by authors analysing individual, resolved nuclear, and circumnuclear \hii\ regions  \citep[e.g.][]{KKB,fabio02,diaz07, dors2008}.  
Our finding, that the current SFR and electron density (as traced by [\sii]$\lambda$6717/[\sii]$\lambda$6731) are also larger in barred systems, is also in agreement with  \cite{Ho_barras}, who made a spectroscopic survey of the nuclei (using a 2\arcsec$\times$4\arcsec aperture) of nearly 500 nearby galaxies. In particular, they only found a larger SFR in barred galaxies earlier than Sbc, compared with unbarred counterparts, with no difference between barred and unbarred late-type spirals. We also find a larger average for SFR per unit area in barred galaxies in our early-type sub-sample (see Fig.~\ref{early_late}) but the distributions for barred and unbarred galaxies are only slightly different, while the difference in both distributions is statistically different for the late-type galaxies, according to the A-D test. We note that the most significant differences between barred and unbarred galaxies (in all gas properties) are obtained for lower bulge or total stellar mass galaxies (i.e. later-type galaxies). The total stellar galaxy or bulge mass distribution for the galaxy sample of  \cite{Ho_barras}  is not shown and we cannot  go any further in comparing both results. 

\subsection{Enhanced [\nii]$\lambda$6583/\ha\  in centres of barred galaxies}

The most striking difference between barred and unbarred galaxies obtained in this work comes from the observed [\nii]$\lambda$6583/\ha\  emission-line ratio.
An enhanced  [\nii]$\lambda$6583/\ha\  emission-line ratio in the galaxy centres was first found by \cite{stauffer} and later by  different authors \citep[see, for example][]{KKB,Ho_hii,sanchez14}.  \cite{Ho_hii}  analysed the integrated spectra of a sample of 206 galaxy nuclei. They reported that early-type galaxies had the largest  [\nii]$\lambda$6583/\ha. According to these authors, it might be an indication that nitrogen is selectively enriched in the  centres of bulge-dominated galaxies. This agrees with the general trend found in this work (without distinction between barred and unbarred systems), since early-type galaxies are also, on average, more massive, and they host bulges with a greater mass than late-type galaxies (see Figs.~\ref{boxplot_totalmass} and \ref{boxplot_bulgemass}). In this work we go a step further and show that this line ratio is larger in the centres of barred galaxies and that the bar effect, in enhancing this line ratio, is more important in galaxies with less massive bulges.

A number of effects can enhance the [\nii]$\lambda$6583 to \ha\ emission-line ratio, namely the nitrogen excitation by a  source other than photoionisation (i.e. AGN or shocks) that enhances the [\nii] emission, compared with emission from recombination lines, and a larger abundance of nitrogen atoms compared with oxygen. 

\begin{figure}
\centering
\includegraphics[width=\columnwidth]{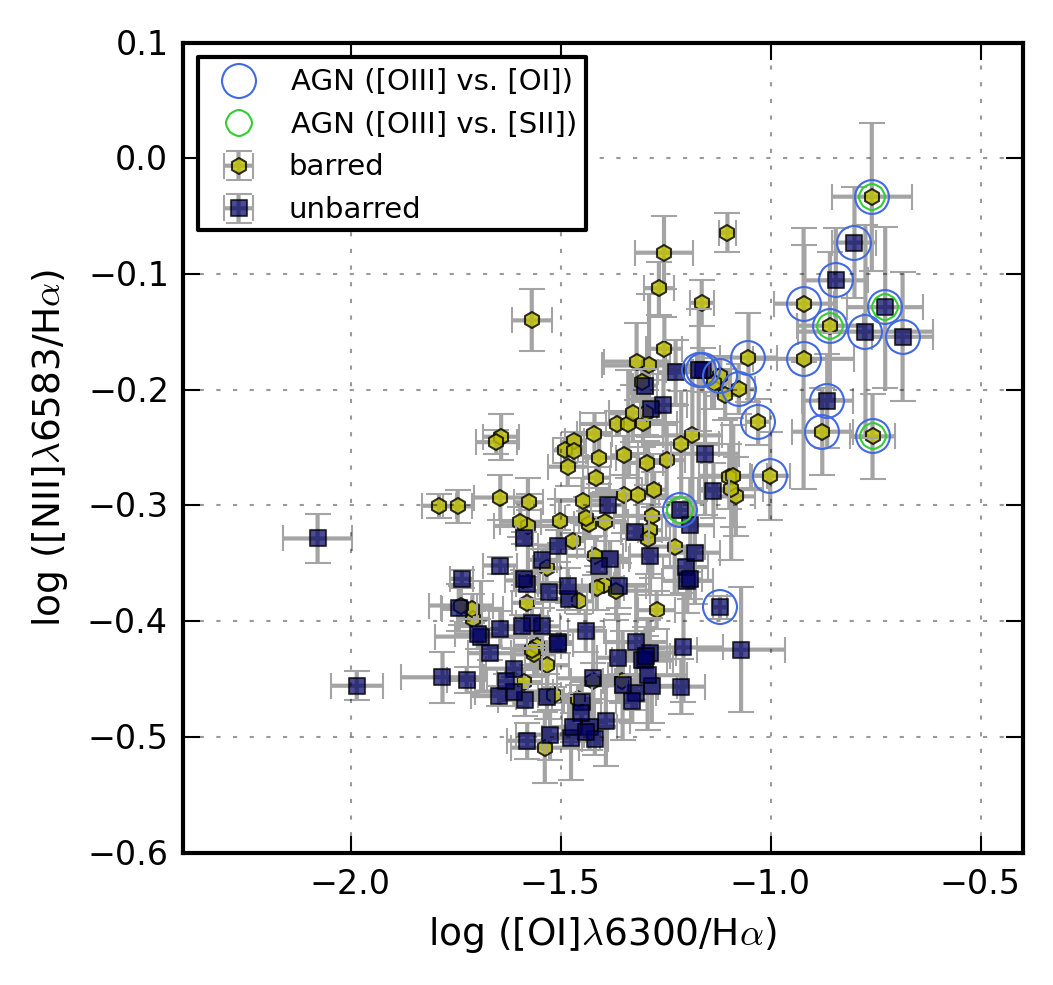}
\caption{\label{shocks}Comparison of shock- and AGN-sensitive emission-line ratios [\nii]$\lambda$6583/\ha\ and [\oi]$\lambda$6300/\ha\ for barred and unbarred galaxies. Galaxies that would be considered as AGN according to the [\oiii]/\hb\ {\em vs} [\oi]/\ha\ or [\oiii]/\hb\ {\em vs} [\sii]/\ha\ diagrams and the  \cite{kewley06} separation line are marked with blue and green circles, respectively.}
\end{figure}

We now analyse these cases separately. We have used the criteria of  \cite{kewley01}, based on the [\oiii]$\lambda$5007/\hb\ vs. [\nii]$\lambda$6583/\ha, to remove AGN from the galaxy sample. This separation criteria allows for objects of the composite area to be included, which may add a number of galaxies in which shocks or an AGN can enhance the [\nii]$\lambda$6583 emission, but also pure photoionised objects with high N/O ratio \citep{pmc09}. 
We note that pure SF galaxies also show a significant difference in [\nii]$\lambda$6583/\ha\ between barred and unbarred galaxies (see Table~\ref{stats}). However, to check whether the use of  \cite{kewley01} criteria  includes a high number of shock-ionised barred galaxies,  and enhances the observed differences (e.g. in the right-hand panel of Fig.~\ref{R23_N2}) between barred and unbarred galaxies, we  created Fig.~\ref{shocks}.
This figure shows [\nii]$\lambda$6583/\ha\ as a function of the shock and AGN sensitive [\oi]$\lambda$6300/\ha\ line ratio. We can see that galaxies with large log([\oi]$\lambda$6300/\ha) in their centres ($\gtrsim$ -1) are also galaxies with high  [\nii]$\lambda$6583/\ha\ and, according to the  [\oiii]$\lambda$5007/\hb\ vs  [\oi]$\lambda$6300/\ha\ BPT diagram, those galaxies would be classified as AGN \citep[][circled in blue]{kewley06}. However, there is roughly the same number of barred (11) and unbarred (10) galaxies  in this area of the plot. Of those galaxies, only three barred galaxies and two unbarred galaxies would  have been classified as AGN  with the  [\oiii]$\lambda$5007/\hb\ vs [\sii]$\lambda\lambda$6717,6731/\ha\ diagram and the separation line given by \cite{kewley06}, circled in green. The plot clearly shows that even by removing those galaxies, the average [\nii]$\lambda$6583/\ha\ is larger in barred than in unbarred galaxies.

\begin{figure*}
\includegraphics[width=0.62\columnwidth]{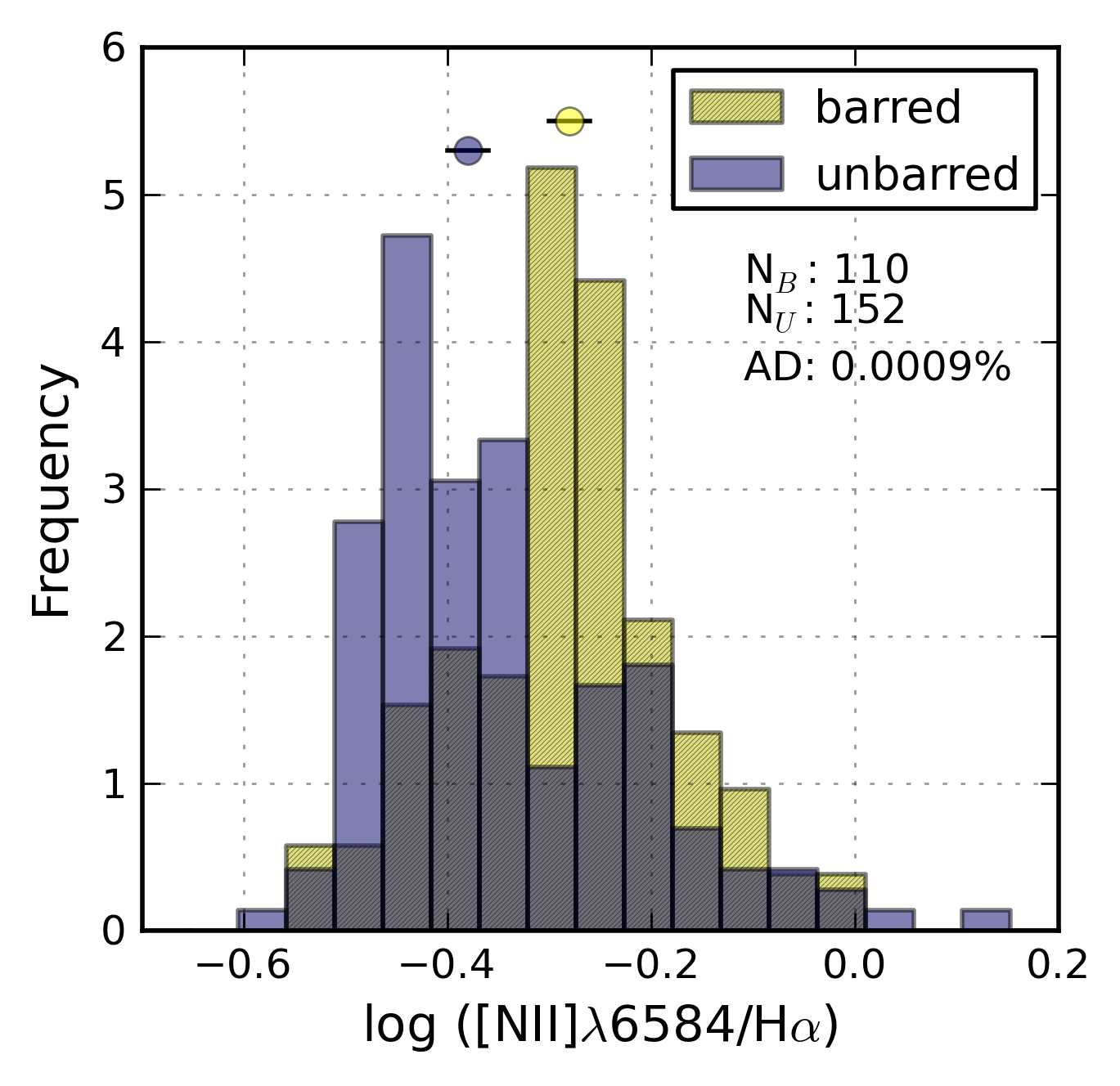}
\includegraphics[width=0.62\columnwidth]{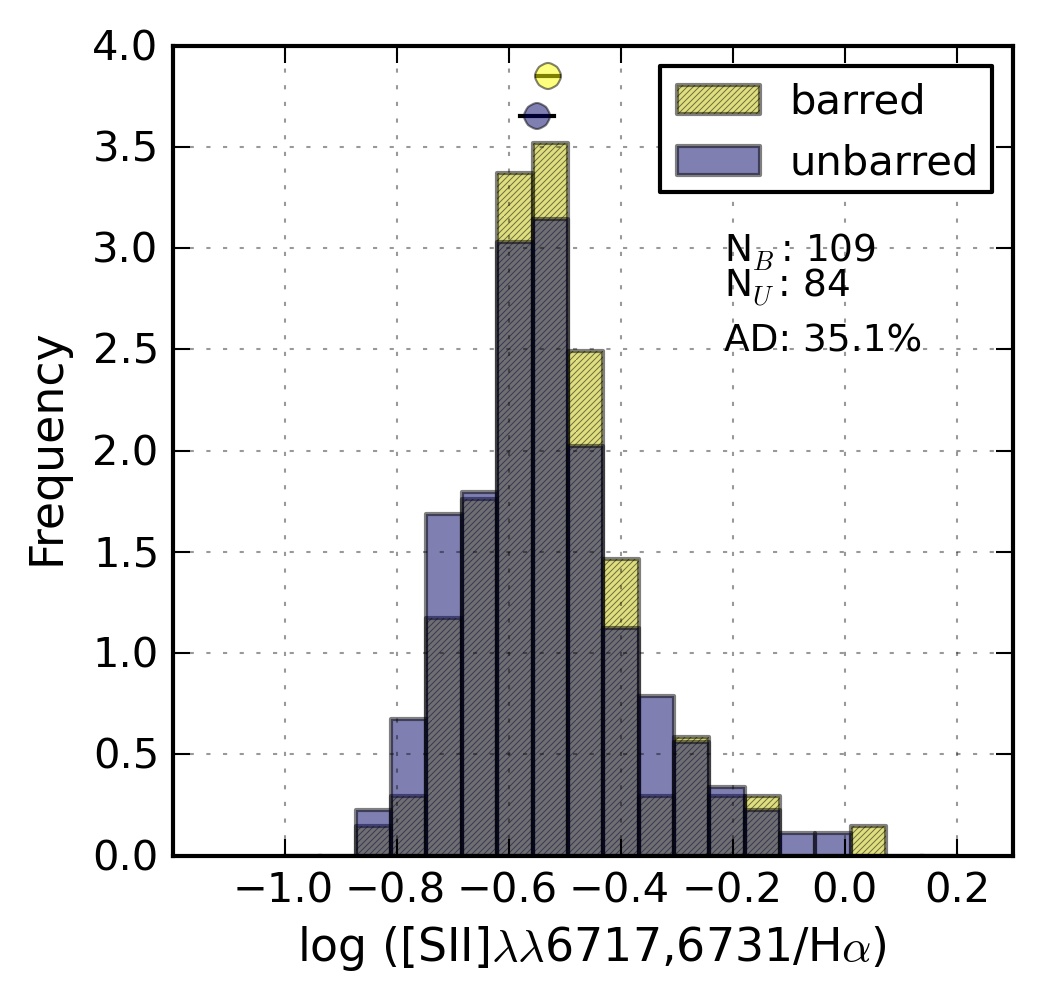}
\includegraphics[width=0.62\columnwidth]{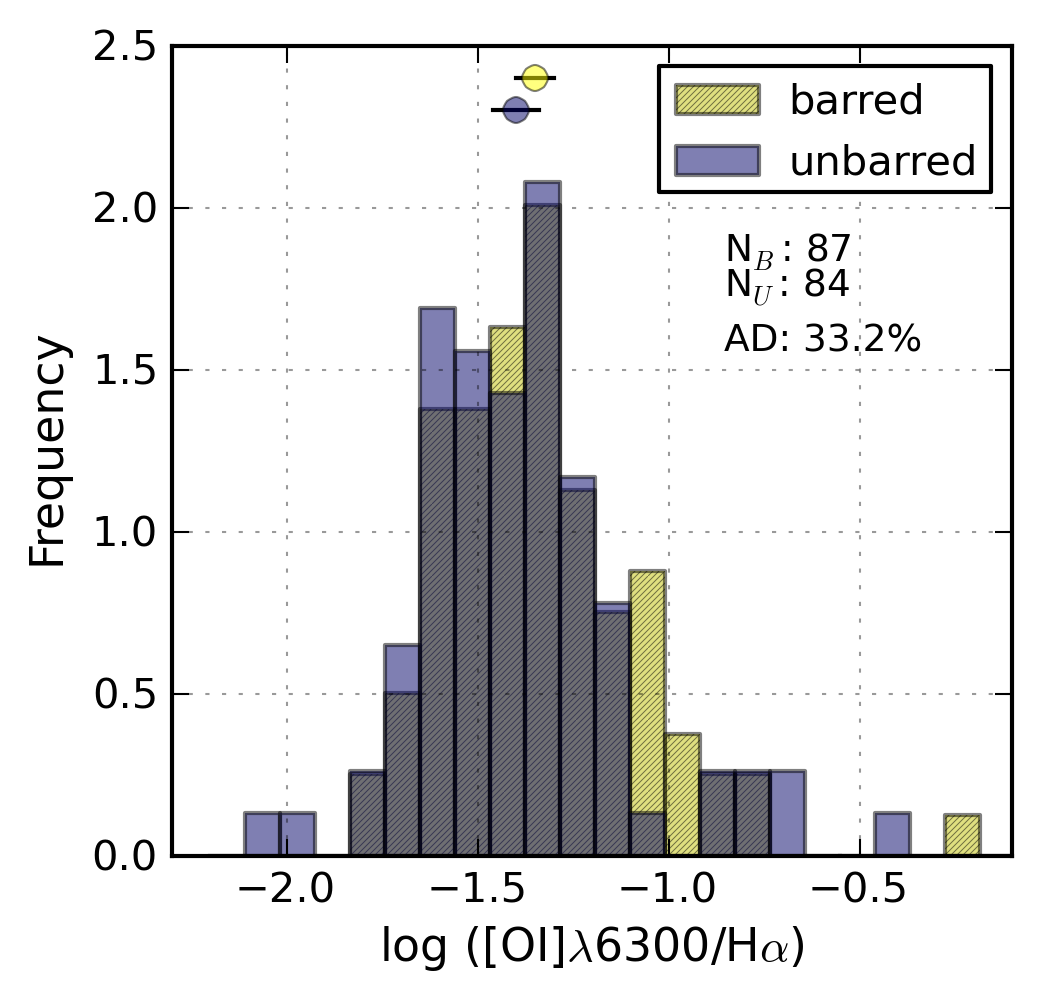}
\caption{\label{histog_shocks}Histograms showing the distribution of AGN and shock-sensitive emission-line ratios (from left to right: [\nii]$\lambda$6583/\ha,  [\sii]$\lambda\lambda$6717,6731/\ha,\ and [\oi]$\lambda$6300/\ha) for barred and unbarred galaxies separately. All galaxies classified as non-AGN are included (see Sect.~\ref{agn}). The barred/unbarred distributions are only different for [\nii]$\lambda$6583/\ha, and therefore the source of the differences in that line ratio must originate from something  other than shock ionisation
and/or AGN contamination alone.}  
\end{figure*}

In addition, if shocks or AGN excitation were responsible for a larger [\nii]$\lambda$6583/\ha\ in barred galaxies, we would also expect to see, on average,  larger values in other AGN and shock-sensitive emission-line ratios, such as [\oi]$\lambda$6300/\ha\ or [\sii]$\lambda\lambda$6717,6731/\ha, in barred galaxies when compared to their unbarred counterparts. The distributions of the three emission-line ratios are shown in Fig.~\ref{histog_shocks}. It is clear that the distributions are only significantly different in the case of [\nii]$\lambda$6583/\ha, while, in the case of [\oi]$\lambda$6300/\ha\ and [\sii]$\lambda\lambda$6717,6731/\ha, the mean values for barred and unbarred galaxies are identical, and that the distributions are not statistically different, with A-D test $P$-values greater than 30\%, compared to 0.0009\% for [\nii]$\lambda$6583/\ha.

A difference in N/O abundance ratio in the centres of barred and unbarred galaxies is thus the most probable explanation for the observed 
enhancement in the [\nii]$\lambda$6583/\ha\ emission-line ratio in barred galaxies.  But what is responsible for an over-abundance of nitrogen compared with oxygen in barred galaxies?

The N/O abundance ratio is a useful parameter to understand chemical evolution of galaxies \cite[see e.g.][]{vila-costas93,pmc09}, as both elements are produced by different mechanisms and by stars of different masses. 
While oxygen is always primary and created in massive stars, nitrogen is synthesised via the CNO and CN cycles in H burning stars, and  is  mostly secondary (the seed C and O were already present in the placental cloud when the star formed) for  metallicity higher than 12+$\log$~(O/H) = 8.3,   while for low metallicity very little secondary N is produced.
 Also, N  can be produced in stars of all masses, but mainly in  intermediate mass (between 4 and 8~M$_\sun$) stars \citep{henry}. Consequently, in a single burst of star formation, we expect that most of the N will be released to the ISM after O, with a time-delay roughly equal to the stellar lifetime of the main producers of N (intermediate mass stars), i.e. from $\sim$100 to 500~Myr after the oxygen was released. Star formation in spiral galaxies might be a continuous process \citep[e.g.][]{gavazzi} which complicates this scenario. In addition, gas flows (within the galaxy and from outside) can also, in theory, alter the initial abundance ratios. 

Both chemical evolution models \citep{molla} and observational works \citep{mallery07} show that the N/O abundance ratio depends on the star formation history. In simple terms this means that, in a closed-box model and at a fixed metallicity, all that is required for a galaxy or region to increase its N/O is for its current SFR to be less than its past average SFR, so that comparatively fewer high mass stars are now formed. Thus,  although newly synthesised oxygen is being released into the ISM from any  massive stars that are currently forming, intermediate-mass stars from previous generations (when SFR  was higher) have had time to evolve and a larger amount of N  has been released into the ISM, increasing the N/O abundance ratio.
However, metal-poor gas flows have a relatively small effect on the N/O abundance ratio according to some models \citep[e.g.][]{edmunds90}. While the oxygen abundance is significantly decreased  during infall, as a result of  dilution of the present gas, the N/O ratio is much less sensitive. The mass infall must be much larger than the galaxy mass (or bulge mass in our case) and the infall rate must be higher than the SFR to have any large deviations from the expected closed-box behaviour \citep{koppen05}.

A different history of SF in the centre of barred galaxies compared with unbarred galaxies is expected from numerical simulations and can, at least, qualitatively explain the higher N/O abundance ratio observed in bulges less massive than $\sim 10^{9.7}-10^{10}$~M$_\sun$. 
Estimating the efficiency of the bar in transporting material to the centre, using numerical simulations, is not easy because the inflow rate strongly depends on the simulation and the chosen parameters \citep[e.g.][]{kim}. However, it is expected that the efficiency of the bar, in driving material to the galaxy centre will depend on the  bar strength \citep[e.g][]{MartinFriedli,ReganTeuben}. The gas flow in the bar gravitational potential will form shocks along the bar's leading edges, giving way to gas flows towards the centre \citep[e.g.][]{athanassoula}. The availability of gas in the disc will also alter the amount of gas accreted into the bulge region and, as the bar evolves and the gas reservoir diminishes, the inflow rate will also decrease \citep{athanassoula05}.
Therefore strong gas flows can be expected soon after the bar forms, which presumably leads to  SF  in the inner parts \citep[e.g.][]{cole}. Later on, the bar weakens and gas flows continue to transport material towards the central regions, where it can form stars, but at a lower rate. Thus, according to chemical evolution models, this could explain the high N/O in barred galaxies compared with unbarred galaxies. Apart from bars, other mechanisms can produce gas flows towards the centre of the galaxies (holding or not a bar), for example tidal forces by an interacting companion  (e.g. Cox et al. 2008) or asymmetric structures like spiral arms  \citep[e.g.][]{kormendy}, but these might be more sporadic, and not produce a continuous flow, as in the case of barred galaxies.

Given that a considerably strong
bar is present, late-type galaxies have a large reservoir of gas and can, therefore, supply a large amount of gas towards the centre. This is not the case for bulge-dominated galaxies and this might be, at least partially, one of the reasons why differences between barred and unbarred galaxies are smaller or non-existent in galaxies with massive bulges ($\gtrsim 10^{10}$~M$_\sun$).
The fact that we do not find differences in central N/O abundance ratio in the most massive bulges could  indicate that bar effects on central gas properties are either negligible (maybe because there is not much gas in the disc to feed the inner parts) or invisible. This may be because there is another, more important and dominant, mechanism that masks the expected bar effects. 

In any case, for massive bulges (which are also the ones with larger N/O abundance ratio, see Fig.~\ref{boxplot_bulgemass}) it would be more complicated to detect an increase in N/O in barred galaxies. The reason being that, given this time-delay scenario in the ejection of  N, at high abundances the increase in  the N/O abundance ratio is smaller \citep{coziol99}. However, the fact that we do not see differences in other properties (c(\hb), 12+$\log$~(O/H), $\log~U$), or that these differences are smaller (as with $\Sigma_{SFR}$ or [\sii]$\lambda$6717/[\sii]$\lambda$6731) in the high bulge mass range, points towards a different bulge origin, formation, or evolutionary process, that is dependent on bulge mass (or total galaxy stellar mass).

Bulges are usually classified as two different types: disc-like bulges or pseudo-bulges, and classical bulges. The former are generally less massive than classical bulges and are more common among late-type galaxies. The formation of classical bulges is thought to be faster and normally externally driven, by means of gravitational collapse or hierarchical merging of smaller objects (see e.g. \citealt{dimitri_morpho} or  \citealt{athanassoula05_bulges} for details on the formation mechanisms of the different types), while disc-like bulges are thought to be formed by internal secular processes, predominantly driven by bars \citep[e.g.][]{kormendy,athanassoula05_bulges}.

There is observational evidence for a delay in the bar formation for less massive galaxies  \citep{sheth12}. Therefore, both bars and bulges in massive galaxies, as mentioned above, form at earlier times. These bulges might be classical as they form in more violent environments, with more gas and larger merger rates. Moreover, recent numerical simulations  \citep{kraljic} show that the formation of bulges in spiral galaxies is influenced by the bar presence. They find an over abundance of star formation in the bulge for barred galaxies after the bar forms.  It is thus possible that the difference observed in the central stellar metallicities of massive galaxies  \citep{coelho} is set at the time when the bar forms, and consumes and redistributes most of the gas. This would be compatible with the frequent co-existence of both type of bulges in the centres of galaxies as observed in previous works \citep{peletier,mendez-abreu,erwin2015}.
In this scenario, low mass systems do not develop a massive classical bulge, and the bar will form at later times.  Furthermore, the central gas accretion  lasts longer, secularly forming disc-like bulges in barred galaxies. This would generate a difference in the ionised gas properties between barred and unbarred low-mass systems as observed in the present work. 

Therefore, our result, that central properties of ionised gas  vary predominantly in galaxies with lower mass bulges, gives strength to  a different origin or evolutionary process for less and more massive bulges, with the gaseous phase of the former currently having a closer relation with bars.

\subsection{Comparison with recent works}
Several works have tried to find observational evidence for the effect of gas flows by analysing the gas content, SFRs and/or metallicities in  galaxy centres \citep[e.g.][]{Ho_barras,MartinFriedli} but their results were inconclusive. This was, in part, because these works were carried out with small data samples and sample selections may have introduced biased results caused by, now well-known, relations between the analysed gas properties with Hubble type and/or total stellar mass.
In an effort to side-step these problems, more recently \cite{ellison} and  \cite{cacho} have comparatively analysed gaseous-phase metallicities and SFRs in the centres of barred and unbarred galaxies in large samples (from $\sim$900 to 1500 galaxies, respectively) by using SDSS spectra. In both cases the galaxy samples have been taken from the catalogue of \cite{nair}\footnote{Bars are visually classified in the  \cite{nair} catalogue.}, requiring axial ratios $b/a \geqslant 0.4$ (i.e. $i \leqslant 68$\deg, compared to 26\deg\ in our sample) and redshift range $z\leqslant0.06$ and  $z<0.1$ in  \cite{ellison} and \cite{cacho}, respectively.
Special care was taken by the authors to avoid biases due to differences in inclination,  redshift and stellar mass between their barred and unbarred galaxy samples. However, and in spite of the similarity in their samples and in the data analysis, they arrive at differing results. While \cite{ellison} find that the SFR and the oxygen abundance in the galaxy centres are considerably  larger in  barred galaxies  more massive than 10$^{10}$~M$_\sun$ compared with unbarred galaxies of the same mass, \cite{cacho} do not find statistically significant differences between barred and unbarred galaxies, only a trend (which decreases with increasing mass) in their early-type sub-sample, for barred galaxies  to exhibit a larger 12+$\log$~(O/H) than unbarred galaxies. 

The study presented here is in apparent disagreement with both \cite{ellison} and \cite{cacho}, as we have found significant differences between  barred and unbarred galaxies, but we did not see differences in the oxygen abundance between barred and unbarred galaxies. However, our results  can explain, to a large extent,  the disagreement of \cite{ellison} and \cite{cacho} results, both with each others' and with ours. This mostly arises from the way in which star-forming galaxies are selected, and from the empirical calibration used to obtain oxygen abundances.  \cite{cacho} use \cite{kauffmann2003_AGN}  to separate  star-forming dominated galaxies from active galaxies, which rejects galaxies from the  transition area, where pure star-forming galaxies with high nitrogen abundance are expected to be located \citep[e.g.][]{pmc09}. \cite{ellison} and the present study use \cite{kewley01} and, therefore,  include transition objects in  our analysis. Our samples are thus expected to contain more nitrogen-rich galaxy centres.
In addition, \cite{ellison} use the [\nii]/[\oii] empirical calibration \citep{kewley02} to obtain 12+$\log$~(O/H), which depends on the N/O abundance. We believe their reported larger 12+$\log$~(O/H) in barred galaxies might  indicate a larger N/O abundance ratio in barred galaxies compared to unbarred galaxies.  \cite{cacho} use the R$_{23}$ parameter to obtain oxygen abundances, but later on, in an effort to explain the differences with Ellison's results, they apply the same empirical calibration as \cite{ellison}, [\nii]/[\oii].  With this calibration they also find larger 12+$\log$~(O/H) in barred galaxies compared to unbarred galaxies, but the difference is not as significant as in \cite{ellison} because their sample lacks a significant number of nitrogen-rich galaxies, as explained above. We also get larger 12+$\log$~(O/H) in barred galaxies when using empirical calibrations involving [\nii]$\lambda$6584, such as the ones given by \cite{pp04,kewley02} or \cite{b07}, but no difference when $R_{23}$ or the \cite{epm14} method is used, as described in Sect.~\ref{abund}.

Finally,  we would like to mention the fact that \cite{ellison} only find differences in the most massive galaxies of their sample  (M$_\star \gtrsim 10^{10}$~M$_\sun$)  does not contradict our results, since our galaxy sample contains  only galaxies more massive than 10$^{10}$~M$_\sun$. Our results do not match in that we do not find differences between barred and unbarred galaxies  more massive than $\sim$10$^{10.8}$~M$_\sun$, while \cite{ellison} finds larger SFRs and 12+$\log$~(O/H) up to their upper limit mass ($\sim$10$^{11.2}$~M$_\sun$).

We believe that the results presented here  show the footprints of bar-induced secular evolution of galaxies in the central gaseous component, which seems to be either more efficient or more relevant in galaxies with  less massive bulges (i.e. later-type galaxies), at least regarding the current central properties of the gas.

\section{Summary and conclusions}
\label{conclussions}
We  analysed the properties of the ionised gas in the centre of a sample containing 251 and 324  barred and unbarred face-on galaxies, respectively \citep[173 and 265 after removing AGN,][]{kewley01} of total stellar masses between 10$^{10}$ and 10$^{11.2}$~M$_{\odot}$, where both sub-samples have the same total galaxy stellar mass and redshift distributions. We compared the distributions of  internal extinction at \hb,  SFR per unit area ($\Sigma_{SFR}$), [\sii]$\lambda\lambda$6717,6731 emission-line ratio, empirical tracers of oxygen abundance ($R_{23}$ and N2=[\nii]$\lambda6854$/\ha), oxygen abundance, N/O abundance ratio, and ionisation parameters, for barred and unbarred galaxies separately. Our main conclusions are summarised below.

\begin{itemize}
\item  The whole non-AGN barred galaxy sample has statistically significant different distributions of $c(\hb)$, $\Sigma_{SFR}$,  [\sii]$\lambda\lambda$6717,6731, N2, and log(N/O) in their centres compared with unbarred galaxies. The differences are remarkably evident in N2 and log(N/O), which are $\sim$0.1 dex larger in barred than in unbarred galaxies. Barred galaxies also tend to have larger central internal extinction,  $\Sigma_{SFR}$, and electron density than unbarred galaxies, but in these cases the differences in median values between both distributions are within errors. We do not find differences between barred and unbarred galaxies, neither in the central value of the $R_{23}$ parameter, nor in 12+$\log$~(O/H), as determined from  the grids of photoionisation models by \cite{epm14}.

\item When we split our sample into earlier and later types  (according to their B/D light ratio or their bulge mass), the differences increase towards later-type galaxies. Especially significant is the difference in N2 that is $\sim$0.13~dex larger in late-type barred  galaxies compared to unbarred galaxies,  against a difference of $\sim$0.07~dex between barred and unbarred galaxies in the earlier-type sub-sample. This difference in N2 translates into a difference in N/O abundance ratio of 0.09 -- 0.12~dex from photoionisation models.

With this separation into earlier and later types, the oxygen abundance distribution appears to differ in barred and unbarred late-type galaxies, with a slightly larger median value for unbarred galaxies. There is no difference in 12+$\log$~(O/H) between barred and unbarred galaxies in the early-type sub-sample.
In addition to 12+$\log$~(O/H), barred and unbarred early-type galaxies seem to have similar central $c(\hb)$,  $R_{23}$ and log~U. The distributions of $\Sigma_{SFR}$ and N/O are  different in barred and unbarred early-type galaxies (indicating larger median values in barred galaxies) depending on the way we separate early- from late-type galaxies. 

\item  The parameter differences between barred and unbarred galaxies increase for galaxies with lower mass systems and seem to correlate better with bulge mass, except for the SFR per unit area, which seems to have a stronger dependence on total galaxy stellar mass.

\item We find no clear relation between the central ionised gas properties in barred galaxies and the bar structural parameters (effective radius, 
bar length, ellipticity or bar S\`ersic index),  but we do find weak trends for the central $\Sigma_{SFR}$ to increase with bar effective length.

\item The differences observed between barred and unbarred galaxies can  be explained qualitatively with current knowledge of  bar evolution simulations and chemical evolution models.

\end{itemize}

Taken all together, our results imply that the presence of bars does alter the properties of the central ionised gas of galaxies,  increasing their electron density, internal extinction, SFR per unit area, and, very notably, the N/O abundance ratio.  These bar effects on central ionised gas are more visible in later-type galaxies, or more precisely, in those galaxies with bulges that are less massive than $\sim 10^{10}$ M$_\odot$.

This work presents some of the clearest observational evidence so far for the effects of bars in galaxy centres. It also emphasises (1) the usefulness of the N/O abundance ratio, which gives important clues on  star formation history of galaxy centres, and (2) the risk of using certain empirical calibrations for oxygen abundance estimations, which can mask a high N/O under a false high O abundance.

\begin{acknowledgements}
 We thank the anonymous referee for his/her constructive comments which improved the paper. We acknowledge Joerg Dietrich for kindly sharing with us his python script to perform the $k$-sample Anderson-Darling test and William Schoenell for help in understanding STARLIGHT output spectra. 
This work has been supported by MICINN of Spain via grants AYA2011-24728 (EF, IP and AZ), FPA2010-16802 (EF) and AYA2010-21887-C04-01 (EPM), and from  the ``Junta de Andaluc\' ia'' local government through the FQM-108 project (AZ, IP and EF).
\end{acknowledgements}


\bibliographystyle{aa}
\bibliography{referencias}

\appendix

\section{Bulge-to-disc flux ratios for different galaxy morphological types.}
\label{morpho-class}
The decrease of the B/D flux ratio  along the Hubble sequence from early- to late-type galaxies is well known,  but this ratio, among others, depends on the structural component-fitting procedure of the galaxy, on inclination, and on dust-extinction \citep[see][and references therein]{alister}. Morphological classification is only available from the largest catalogue so far \citep{nair} for less than 50\% of our galaxy sample. Our aim here is to use the B/D flux ratios \citep{dimitri_morpho} and the morphological classification in T-types for our sub-sample of galaxies included in the \cite{nair} catalogue, to separate our galaxies into early- and late-types, according to their B/D flux ratios.  

\begin{figure*}
\centering
\includegraphics[width=1.6\columnwidth]{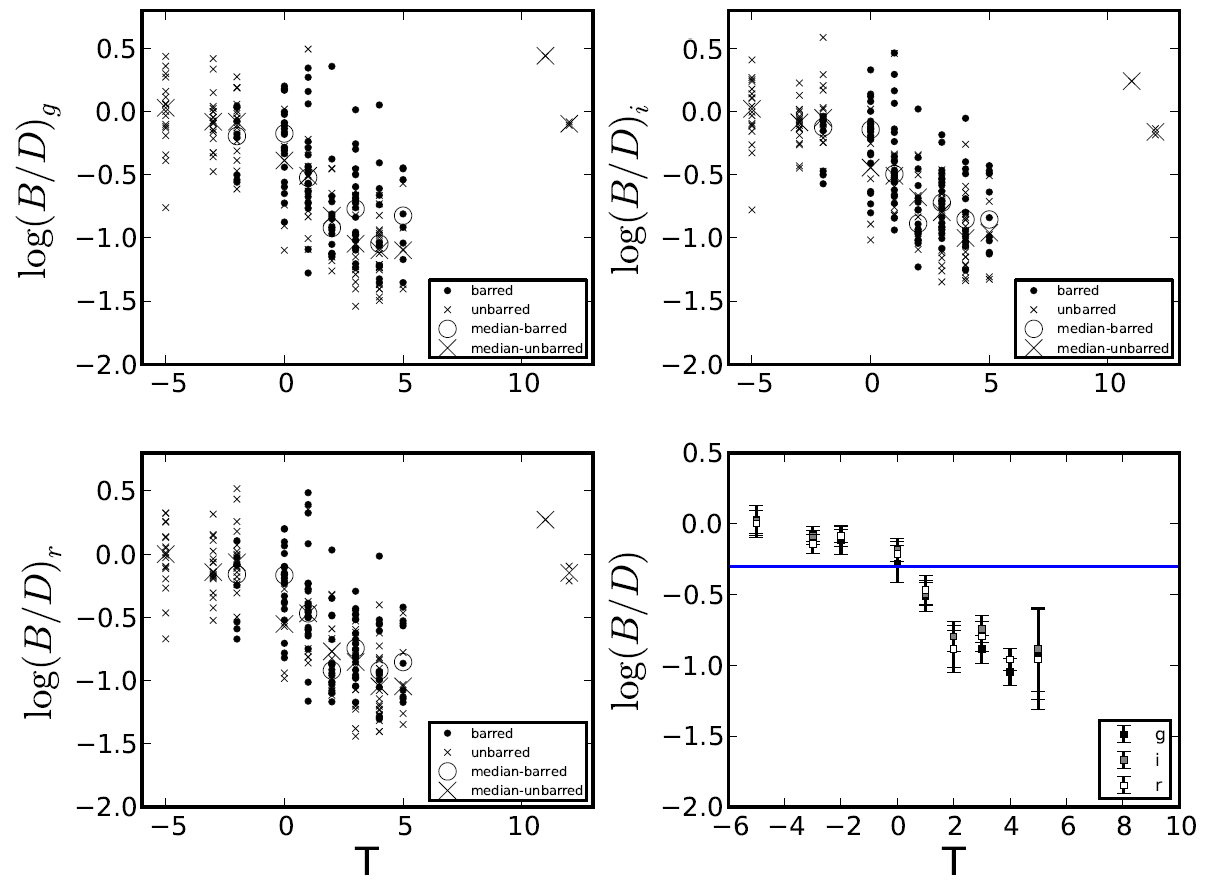}
\caption{\label{BDfrenteaT}Logarithm of the B/D flux ratio in the $g$, $r,$ and $i$-band  versus morphological type (expressed as T-type) for the galaxies of the sample contained in the \cite{nair} catalogue. Big crosses and big empty circles mark the median $\log$~(B/D)  for each T-type value for unbarred and barred galaxies, respectively. The bottom right-hand panel shows the median  $\log$~(B/D) in each band for each  T-type value. Error bars represent  the 95\% confidence level of the median value for the distribution of B/D values at each T-type value. The blue solid straight line marks B/D=0.5, that we have used to separate early from late-type galaxies.}
\end{figure*}

Figure~\ref{BDfrenteaT} shows the dependence of the logarithm of the $g$-, $r$-, and $i$-band B/D light ratios \cite[from BUDDA,][]{dimitri_morpho}, with the corresponding T-type \citep{nair} for barred and unbarred galaxies occurring in both samples simultaneously. 
The top and bottom left-hand panels show this relation for the three different photometric bands, with small crosses for unbarred galaxies and small, filled circles for barred galaxies. The large crosses and big empty circles indicate the median $\log$~(B/D) value for each T-type value for unbarred and barred galaxies, respectively. As  can be seen, there are no clear differences between barred and unbarred galaxies. The bottom right-hand panel shows the median  $\log$~(B/D)  values, for barred and unbarred galaxies together, for each T-type for all photometric bands. The differences in the median B/D light ratio between the different bands are, in general, smaller than the error bars.

\begin{figure}
\centering
\includegraphics[width=0.98\columnwidth]{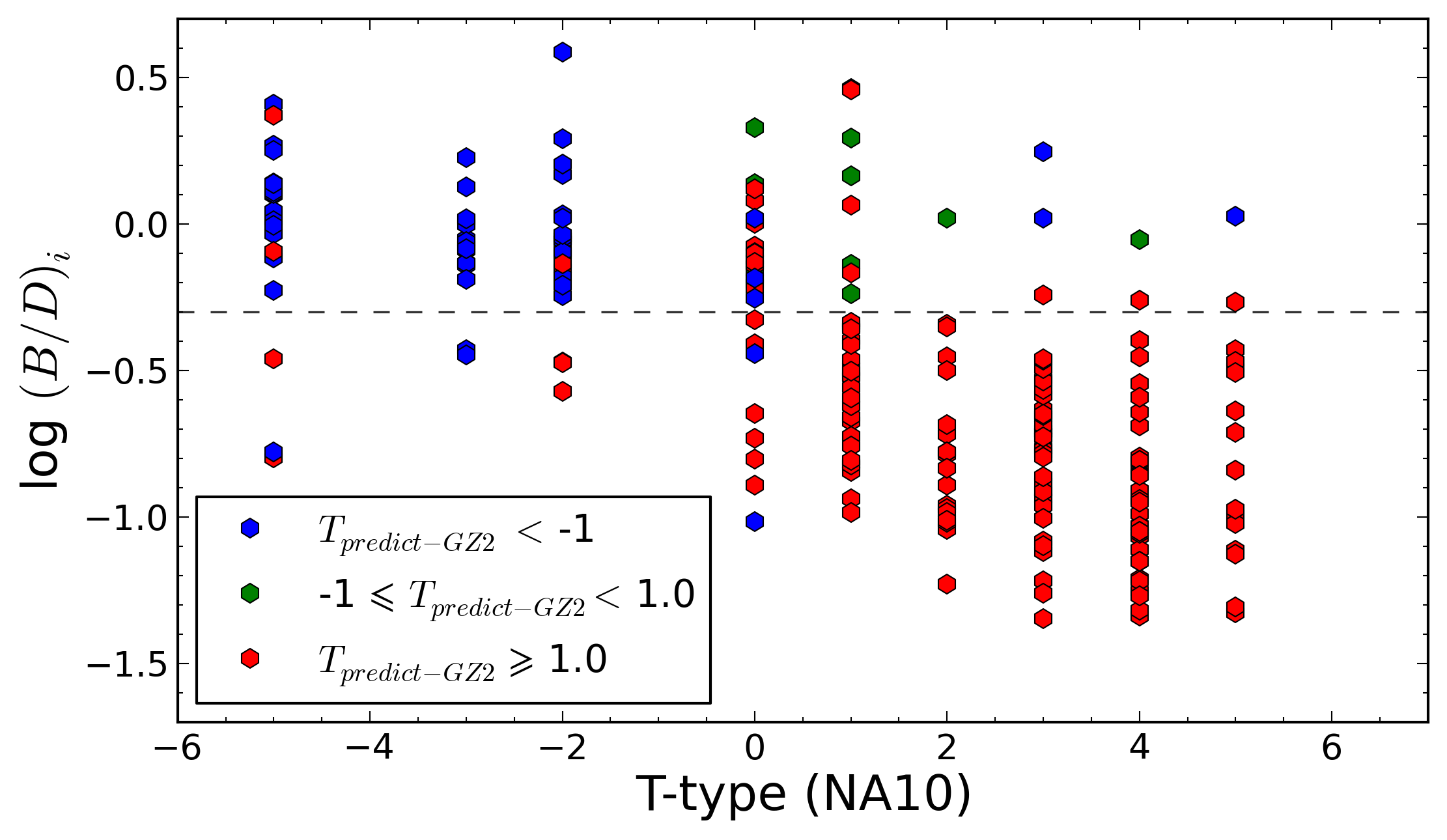}
\caption{Decimal logarithm of the bulge-to-disc flux ratio in the $i$-band as a function of the morphological T-type, as given in the \cite{nair} catalogue for the galaxies of our sample included in this catalogue. The colours represent different values of the predicted T-type Galaxy Zoo 2 parameter \cite{GZ2}.}
\label{BD_GZ}%
\end{figure}

Our median  values correspond well (within errors) to those from  \citet[][their Fig.~6]{alister} in the T-type range in common (between T-type 0 and 5), except for T=0, where we obtain a marginally larger  B/D ratio in the $r$ and $i$ bands. These differences could arise from poor statistics at T=0 in \cite{alister} (only three galaxies versus 30 in our sub-sample of galaxies are included in the \cite{nair} catalogue). 

We point out that a number of unbarred galaxies in our sample ($\sim$15), supposed to be disc galaxies, are classified with T-type$\leq$-3 by \cite{nair}, i.e. as ellipticals (see Fig.~\ref{BDfrenteaT} or \ref{BD_GZ}). These galaxies are very difficult to classify, since a face-on disc without spiral arms or a bar looks just like a round elliptical. \cite{dimitri_morpho}  classified galaxies after looking at isophotal contours and the profile very carefully, but as stated in the cited paper, the estimated  misclassification frequency is $\sim$5-10\%. With these uncertainties in mind, we have opted to keep these galaxies in our sample, and to consider them as S0 disc galaxies.

The B/D flux ratios decrease towards late-type spirals as expected. We have used this trend to create sub-samples dominated by early- or late-type spirals. In our sub-sample of galaxies included in the \cite{nair} catalogue, a value of $\log$~(B/D)=-0.30 (or  B/D=0.5) more or less separates galaxies with T-type$<$2 (earlier than Sa) from those later or equal to T-type=2 (Sab), but the contamination from late-types (with the previous definition) in the early-type sub-sample and vice versa is approximately 12\% and 30\%, respectively. The predicted T-type, as given in the Galaxy Zoo 2\footnote{http://data.galaxyzoo.org/} catalogue \citep[hereinafter GZ2][]{GZ2}, allows us to improve this separation.  We then consider early-type galaxies as those galaxies of the sample with either B/D$\geqslant$0.5, or those having simultaneously B/D<0.5 and T-type, as predicted by GZ2, lower than -1. Similarly, we consider as late-type galaxies those with B/D<0.5 and a T-type as predicted by GZ2, larger than or equal to one. In this way we are able to include $\sim$70\% and 90\% of the early (T-type$<2$) and late types (T-type $\geqslant$2) in the sub-samples of early- and late-type galaxies, respectively, and the contamination from late (early-types) in the early (late-type) sub-sample is $\sim$10\% (25\%). These numbers are the same in all three bands. However, we have used the $i$-band, as the B/D flux ratio is available for all sample galaxies in this filter.

\end{document}